\newif\ifShowKeys
\ShowKeystrue
\ShowKeysfalse

\documentclass[letterpaper,10pt]{article}

\pdfoutput=1
\usepackage[no-natbib-sort]{my-jheppub}

\ifShowKeys \usepackage[notcite]{showkeys} \fi

\usepackage{amsmath, amssymb}

\usepackage{xfrac}   

\usepackage{amsthm}

\usepackage{bm,environ,mathrsfs,array,arydshln,booktabs,float,slashed,hyperref}
\usepackage[mathcal]{euscript}
\usepackage{tensor} 	
\usepackage{esint}
\usepackage{physics}
\usepackage{braket}

\usepackage[nodayofweek]{date time}

\usepackage{graphicx,epsfig}
\usepackage{epic}
\usepackage{tikz} 
\usetikzlibrary{decorations.pathreplacing,decorations.markings,snakes}

\tikzset{middlearrow/.style={decoration={markings, mark= at position 0.5 with {\arrow{#1}} ,
}, postaction={decorate}}}

\usepackage{floatflt}

\usepackage{aurical}
\usepackage[T1]{fontenc}

\allowdisplaybreaks

\newcommand{\be}{\begin{equation}}
\newcommand{\ee}{\end{equation}}
\newcommand{\mc}{\mathcal }

\newcommand{\la}{\label}

\newcommand{\spek}{\notag \\ & }

\newcommand{\cN}{{\mathcal N}}
\newcommand{\cO}{{\mathcal O}}
\newcommand{\cA}{{\mathcal A}}
\newcommand{\cR}{{\mathcal R}}

\newcommand{\rf}{\text{f}}

\newcommand{\Vev}[1]{{\left\langle #1 \right\rangle}}

\makeatletter
\DeclareFontFamily{OMX}{MnSymbolE}{}
\DeclareSymbolFont{MnLargeSymbols}{OMX}{MnSymbolE}{m}{n}
\SetSymbolFont{MnLargeSymbols}{bold}{OMX}{MnSymbolE}{b}{n}
\DeclareFontShape{OMX}{MnSymbolE}{m}{n}{
<-6>  MnSymbolE5
   <6-7>  MnSymbolE6
   <7-8>  MnSymbolE7
   <8-9>  MnSymbolE8
   <9-10> MnSymbolE9
  <10-12> MnSymbolE10
  <12->   MnSymbolE12
}{}
\DeclareFontShape{OMX}{MnSymbolE}{b}{n}{
<-6>  MnSymbolE-Bold5
   <6-7>  MnSymbolE-Bold6
   <7-8>  MnSymbolE-Bold7
   <8-9>  MnSymbolE-Bold8
   <9-10> MnSymbolE-Bold9
  <10-12> MnSymbolE-Bold10
  <12->   MnSymbolE-Bold12
}{}

\let\llangle\@undefined
\let\rrangle\@undefined
\DeclareMathDelimiter{\llangle}{\mathopen}%
 {MnLargeSymbols}{'164}{MnLargeSymbols}{'164}
\DeclareMathDelimiter{\rrangle}{\mathclose}%
 {MnLargeSymbols}{'171}{MnLargeSymbols}{'171}
\makeatother

\newdimen\tableauside\tableauside=1.0ex
\newdimen\tableaurule\tableaurule=0.4pt
\newdimen\tableaustep
\def\phantomhrule#1{\hbox{\vbox to0pt{\hrule height\tableaurule
			width#1\vss}}}
\def\phantomvrule#1{\vbox{\hbox to0pt{\vrule width\tableaurule
			height#1\hss}}}
\def\sqr{\vbox{%
		\phantomhrule\tableaustep
		\hbox{\phantomvrule\tableaustep\kern\tableaustep\phantomvrule\tableaustep}%
		\hbox{\vbox{\phantomhrule\tableauside}\kern-\tableaurule}}}
\def\squares#1{\hbox{\count0=#1\noindent\loop\sqr
		\advance\count0 by-1 \ifnum\count0>0\repeat}}
\def\tableau#1{\vcenter{\offinterlineskip
		\tableaustep=\tableauside\advance\tableaustep by-\tableaurule
		\kern\normallineskip\hbox
		{\kern\normallineskip\vbox
			{\gettableau#1 0 }%
			\kern\normallineskip\kern\tableaurule}%
		\kern\normallineskip\kern\tableaurule}}
\def\gettableau#1 {\ifnum#1=0\let\next=\null\else
	\squares{#1}\let\next=\gettableau\fi\next}
\newcommand{\Yfund}{\tableau{1}}
\newcommand{\Ysymm}{\tableau{2}}
\newcommand{\Yasymm}{\tableau{1 1}}

\title{$\mc N=2$ conformal gauge theories at large R-charge: the $SU(N)$ case}

\author[a]{Matteo Beccaria,~}
\emailAdd{matteo.beccaria@le.infn.it}
\affiliation[a]{Dipartimento di Matematica e Fisica Ennio De Giorgi,
	Universit\`a del Salento \& I.\,N.\,F.\,N. - sezione di Lecce, 
	Via Arnesano 1, I-73100 Lecce,  Italy\\} 

\author[b]{Francesco Galvagno,}
\affiliation[b]{Dipartimento di Fisica, Universit\`a di Torino
	\& I.\,N.\,F.\,N. - sezione di Torino,
	Via P. Giuria 1, I-10125 Torino, Italy\\}
\emailAdd{galvagno@to.infn.it}

\author[a]{Azeem Hasan.}
\emailAdd{ahasan@gradcenter.cuny.edu}

\abstract{
Conformal theories with a global symmetry may be studied in the double scaling regime
where the interaction strength is reduced while the global charge increases.
Here, we study generic 4d $\mc N=2$  $SU(N)$ gauge theories with  
conformal matter content at large R-charge $Q_{\rm R}\to \infty$ with 
fixed 't Hooft-like coupling $\kappa = Q_{\rm R}\,g_{\rm YM}^{2}$. Our analysis concerns 
two distinct classes of natural scaling functions. 
The first is built in terms of  chiral/anti-chiral two-point functions. The 
second involves one-point functions of chiral operators in presence of 
$\frac{1}{2}$-BPS Wilson-Maldacena loops.
In the rank-1 $SU(2)$ case, the two-point sector has been recently shown to be captured by an
auxiliary chiral random matrix model. 
We extend the analysis to $SU(N)$ theories and provide an algorithm that computes  arbitrarily  
long perturbative expansions for all considered models, parametric  in the rank. 
The leading and next-to-leading
contributions are cross-checked by a three-loops computation in $\mc N=1$ superspace. 
This perturbative analysis identifies maximally non-planar Feynman diagrams as the
relevant ones in  the double scaling limit. 
In the Wilson-Maldacena sector, we  obtain  closed expressions for  the  scaling functions, 
valid for any rank and  $\kappa$. As an application, we analyze quantitatively the 
large 't Hooft coupling limit $\kappa\gg 1$ where we identify all  perturbative 
and non-perturbative contributions. The latter are associated with heavy electric BPS states
and the precise correspondence with their mass spectrum is clarified.
}

\begin{document}



\maketitle

\section{Introduction and summary of results}

The large charge limit of conformal quantum theories with a global symmetry is an interesting regime
where important simplifications may occur and novel exact results may be obtained
\cite{Hellerman:2015nra, 
Alvarez-Gaume:2016vff, Monin:2016jmo}. \footnote{The idea of a large charge/weak coupling compensation 
is closely related to the solvability 
of the BMN limit in AdS/CFT \cite{Berenstein:2002jq,Gubser:2002tv} and, more generally,
to the coherent-state effective theory description of ``semiclassical'' string states 
\cite{Tseytlin:2004xa,Minahan:2005mx} and its role in capturing the strong coupling 
regime \cite{Beccaria:2012xm}.} 
The simplest example is that of the 
$O(2)$ invariant scalar model in three dimensions, see e.g. \cite{Wilson:1973jj}, 
where an effective theory captures the dynamics of operators  with large 
$O(2)$ charge $\sim n$.
Exact results are obtained in the double scaling limit $n\to \infty$ with fixed
$\kappa\propto n^{2}\,g$ where $g$ is the quartic coupling of the Wilson-Fisher fixed point 
\cite{Arias-Tamargo:2019xld,Badel:2019oxl,Watanabe:2019pdh,Badel:2019khk}. For instance,
the anomalous dimension $\gamma_{n}$ of the composite operator $\varphi^{n}$ 
is exactly linear in $\kappa$, i.e. $\gamma_{n} -n\propto \kappa$, with a computable
coefficient \cite{Arias-Tamargo:2019xld}. In this model
higher order corrections in $\kappa$ are associated with suppressed diagrams 
in the double scaling limit. \footnote{An equivalent statement is that an exact saddle point analysis is possible in the double
scaling limit.}
Although the anomalous dimension $\gamma_{n}$ is inherently 
associated with a two-point function, similar results have been recently
extended to more general  
higher point functions with one anti-holomorphic insertion of $\overline\varphi^{n}$
\cite{Arias-Tamargo:2019kfr}.

In this paper we focus on another class of models where the 
large global charge limit is very interesting, i.e. four-dimensional $\mc N=2$ superconformal theories \cite{Howe:1983wj}.
The most common example is conformal super-QCD (SQCD) with gauge group $SU(N)$ and $2N$ hypermultiplets 
in the fundamental representation considered at large global charge in 
\cite{Hellerman:2017sur,Bourget:2018obm,Hellerman:2018xpi}. In this class of models the global symmetry 
is identified with the R-symmetry. Besides, thanks to $\mc N=2$ extended supersymmetry, it is possible 
to compute  non-trivial observables at high perturbative order using localization methods
\cite{Pestun:2007rz}. In some cases  non perturbative results may be obtained, as we shall illustrate.
In a typical setup, the large charge limit is approached with the Yang-Mills coupling 
$g\to 0$ while the R-charge $n$
grows as $1/g^{2}$. The corresponding 't Hooft-like double scaling limit 
is then 
\be
\la{1.1}
4\,\pi^{2}\,\kappa = n\,g^{2} = \text{fixed},\ \text{with}\ n\to\infty, 
\ee
where $\kappa$ is the new coupling.  
Perturbatively in $\kappa$ we can neglect instanton contributions because  we stay 
at weak-coupling for any finite $\kappa$.  Notice also that the gauge group rank $N$
is kept fixed in the double scaling limit (\ref{1.1}). Further
scaling regimes involving both 
$\kappa$ and $N$ have not been investigated yet.

\subsection{Large R-charge observables}

We shall consider two related but distinct sets of observables that will not trivialize at large R-charge. The first set (or  
{\em sector}) emerges naturally in the study of extremal correlators of chiral primaries, i.e.
higher point functions with only one anti-Coulomb branch operator. The simplest case is that of two-point functions
between a chiral primary and its antiholomorphic counterpart.
In conformal $\cN=2$ SQCD they have been computed in the double scaling limit (\ref{1.1}) 
in \cite{Bourget:2018obm,Beccaria:2018xxl}
by applying  localization methods \cite{Baggio:2014sna,Baggio:2015vxa,Gerchkovitz:2016gxx}. One 
considers the normalized ratio between the  two-point functions $\langle\cO_{n}\overline\cO_{n}\rangle$
in the  $\cN=2$ theory and in the $\cN=4$ SYM universal parent theory
when $\cO_{n}$ is a chiral primary with R-charge $\propto n\to \infty$.
This ratio 
is used to define the following
{\em scaling function} depending on the fixed coupling $\kappa$ and the gauge 
group rank (the position dependence is fully controlled by superconformal Ward identities 
\cite{Papadodimas:2009eu} and 
drops in the ratio)
\be
\la{1.2}
F^{\cO}(\kappa; N) = \lim_{n\to \infty}\left. F^{\cO}_{n}(g; N) \right|_{\kappa = \text{fixed}}\ ,\quad \text{with}\quad 
F^{\cO}_{n}(g; N) =  \frac{\langle\cO_{n}\overline\cO_{n}\rangle^{\cN=2}}
{\langle\cO_{n}\overline\cO_{n}\rangle^{\cN=4}}\ .
\ee
Technically, 
the explicit matrix model computation of $F^{\cO}(\kappa; N)$ is challenging because in the large $n$ limit
it becomes hard to disentangle the map from $S^{4}$ -- where the matrix model lives -- to flat space 
\cite{Gerchkovitz:2016gxx, Rodriguez-Gomez:2016ijh, Rodriguez-Gomez:2016cem, Billo:2017glv}. 
We remind that this step is  non-trivial because 
the preserved supersymmetry on $S^{4}$ is only $\mathfrak{osp}(4|2)\subset \mathfrak{su}(2|2)$ and mixing 
generically occurs, breaking the original flat space $\mathfrak{u}(1)_{\rm R}$.
Nevertheless,
for certain classes of chiral primaries $\cO_{n}$, it is possible to compute efficiently the perturbative expansion of 
the function $F^{\cO}(\kappa; N)$
by exploiting  the integrable structure of the $\cN=2$ partition function. In the simplest example,
$\cO_{n}$ is the maximal multi-trace operator
$\cO_{n} = \Omega_{n}\equiv (\tr\varphi^{2})^{n}$ where $\varphi $ is a complex combination of the two 
real scalar field belonging to the $\cN=2$ vector multiplet. The 
 two-point functions $\langle \Omega_{n}\,\overline\Omega_{n}\rangle$
 are then captured by an integrable Toda-chain \cite{Gerchkovitz:2016gxx,Baggio:2014ioa}.
 By exploiting this peculiar structure it is possible to  control the R-charge dependence and
evaluate the scaling function  (\ref{1.2})  at high perturbative order \cite{Beccaria:2018xxl}. 
Later, this approach based on decoupled semi-infinite Toda equations
has been  generalized to broader classes of primaries and is believed to 
be a general feature of  Lagrangian $\cN=2$ superconformal theories
\cite{Bourget:2018fhe}.

A second sector of observables arises in the study of one-point functions of chiral scalar
operators $\cO$ in presence of a circular $\frac{1}{2}$-BPS Maldacena-Wilson loop $\mc W$. 
For a circle of radius $R$ 
in $\mathbb R^{4}$ it reads
\be
\label{1.3}
\mc W = \frac{1}{N}\,\tr\,\mc P\,\exp\bigg\{g\,\int_{S^{1}}ds\,\bigg[i\,A(x)\cdot \dot x(s)
+\frac{R}{\sqrt 2}(\varphi+\overline\varphi)\bigg]\bigg\},
\ee
where $g$ is the Yang-Mills gauge coupling and  $A$ is the $SU(N)$ gauge field.
Exploiting conformal invariance and placing the chiral 
operator $\cO$ in the center of the loop, one can apply  localization methods to compute 
$\langle \cO\,\mc W\rangle$  \cite{Billo:2018oog}.
Again, one may consider the large R-charge limit by taking $\cO\to \cO_{n}$ as above and 
define the ratio \cite{Beccaria:2018owt}
\be
\la{1.4}
F^{\cO}_{\mc W}(\kappa; N) = \lim_{n\to \infty}\left. F^{\cO}_{\mc W, n}(g; N)\right|_{\kappa = \text{fixed}},
\quad \text{with}\quad 
F^{\cO}_{\mc W, n}(g; N) = \frac{\langle\cO_{n}\,\mc W\rangle^{\cN=2}}
{\langle\cO_{n}\,\mc W\rangle^{\cN=4}},
\ee
in analogy with (\ref{1.2}). 
For the sake of brevity, we shall name in the following  $\langle\cO\,\mc W\rangle$ 
a {\em one-point Wilson function} and $F_{\mc W}^{\cO}(\kappa; N)$ a {\em one-point Wilson 
scaling function}.

In this paper, we shall consider the observables
(\ref{1.2}) and (\ref{1.4}) in the  two (simplest and next-to-simplest \footnote{
Here, simplicity refers to the $S^{4}$ mixing, see for instance 
\cite{Beccaria:2018xxl,Bourget:2018fhe}.
}
) cases when the chiral primaries 
$\cO_{n}$ are the {\em towers} $(\tr\varphi^{2})^{n}$ or $\tr\varphi^{3}\,(\tr\varphi^{2})^{n}$ (the second choice is non-trivial
  for $N\ge 3$).
We shall denote the associated 
scaling functions by $F^{(2)}(\kappa; N)$ and $F^{(3)}(\kappa; N)$ respectively, 
and similarly for Wilson scaling functions. Their properties will be considered not
only in conformal SQCD, but also in a 
more general set of superconformal $\cN=2$
models with $SU(N)$ gauge group, obtained by imposing that the 1-loop coefficient of the beta function vanishes (see \eqref{RFSAb0}). These have a specific
matter content in the fundamental, symmetric or anti-symmetric 
representations \cite{Koh:1983ir}, see table \ref{tab:scft}.
\renewcommand{\arraystretch}{1}
\begin{table}[ht]
\begin{center}
\be
\def\arraystretch{1.3}
\begin{array}{cccc}
\toprule
\textsc{theory} & N_{F} & N_{S} & N_{A}  \\
\midrule
\mathbf{A} & 2N & 0 & 0 \\
\mathbf{B} & N-2 & 1 & 0 \\
\mathbf{C} & N+2 & 0 & 1  \\
\mathbf{D} & 4 & 0 & 2 \\
\mathbf{E} & 0 & 1 & 1  \\
\bottomrule
\end{array}\notag
\ee
%
%
\end{center}
\caption{The five families of $\cN=2$ superconformal theories with $SU(N)$ gauge 
group and matter in fundamental (F), symmetric (S) and anti-symmetric representations (A), cf. \cite{Koh:1983ir}.}
\label{tab:scft}
\end{table}
Theory $\mathbf{A}$ is $\cN=2$ conformal SQCD.
Theories $\mathbf{D}$ and $\mathbf{E}$ are quite interesting since they admit 
a holographic dual of the form $\mathrm{AdS}_5\times S^5/\Gamma$ 
with a suitable discrete group $\Gamma$  \cite{Ennes:2000fu}. 
Localization computations in these models have been recently fully discussed in \cite{Billo:2019fbi}.
For $SU(2)$ the only meaningful model is $\mathbf A$, while for $SU(3)$ 
we have the  identifications
\be
\la{1.5}
\mathbf{A}\equiv \mathbf{C}\equiv \mathbf{D},\qquad 
\mathbf{B}\equiv \mathbf{E},
\ee
so that we can  restrict to the $\mathbf{A}$ and $\mathbf{B}$ models. For $N>3$ there are no more accidental
identifications.
 
\subsection{Previous results and open questions}

Let us overview what is known about the large R-charge observables (\ref{1.2}) and (\ref{1.4})
and emphasize several open issues.

In the $SU(2)$ SQCD theory, i.e. the $\mathbf A$ model, 
the longest expansion of the two-point scaling function $F^{\mathbf A\,(2)}(\kappa; 2)$
has been computed in \cite{Beccaria:2018xxl}, while the one-point Wilson scaling function $F^{\mathbf A\,(2)}_{\mc W}(\kappa; 2)$
has been considered later in \cite{Beccaria:2018owt}. These explicit calculations show that 
-- at least up to order $\mc O(\kappa^{11})$ -- one has  (i) the equality
\be
\la{1.6}
F^{\mathbf A\,(2)}(\kappa; 2) = F^{\mathbf A\,(2)}_{\mc W}(\kappa; 2) \equiv F(\kappa; 2),
\ee
and (ii) a simple exponentiation structure in terms of simple $\zeta$-numbers
\begin{align}
\la{1.7}
& \log   F(\kappa; 2) = -\frac{9\, \zeta (3)}{2}\, \kappa ^2 +\frac{25\,  \zeta 
(5)}{2}\,\kappa ^3-\frac{2205\,  \zeta (7)}{64}\,\kappa ^4+\frac{3213\,  \zeta (9)}{32}\,\kappa ^5 -\frac{78771 \,
 \zeta (11)}{256}\,\kappa ^6 \\
 &+\frac{250965 \,
\zeta (13)}{256}\,\kappa ^7 -\frac{105424605  \,\zeta 
(15)}{32768}\,\kappa ^8 +\frac{265525975 \, \zeta 
(17)}{24576}\,\kappa ^9-\frac{12108123027  \,\zeta 
(19)}{327680}\,\kappa ^{10}+\mc O(\kappa ^{11}).\notag
\end{align}
Besides, (iii) the expansion (\ref{1.7})
has been conjectured in  \cite{Beccaria:2018owt} to admit the closed integral 
representation ($J_{n}$ are Bessel functions)
\be
\la{1.8}
\log  F(\kappa; 2) =4\, \int_{0}^{\infty}\frac{dt}{t^{2}}\ 
\frac{J_{0}(2t\sqrt\kappa)+2\,t\,\sqrt\kappa\,J_{1}(2t\sqrt\kappa)
-\kappa\,t^{2}-1}
{e^{t}+1}.
\ee
The relation (\ref{1.6}) shows that there is a puzzling connection between the two-point and one-point Wilson 
sectors.  Notice that the conjectured form  (\ref{1.8}) is very interesting because it gives access to the
non-perturbative (within the large R-charge limit framework) large $\kappa$ regime. 
Remarkably, (\ref{1.8}) has been proved for the two-point scaling function in \cite{Grassi:2019txd} (GKT)
by a {\em dual} description which is a chiral random matrix model of the Wishart-Laguerre type.  Such dual description involves matrices whose rank is related to the number of operator insertions $n$, so that the double scaling limit \eqref{1.1} corresponds to the usual 't Hooft limit for the random matrix model.

\medskip
In the higher rank $SU(N)$ SQCD theory, with $N>2$, things are less simple. 
The scaling function $F^{(\Delta)}$ has been computed by the Toda equation
in \cite{Beccaria:2018xxl} at $\mc O(\kappa^{10})$. 
In the $SU(3)$ case, the first orders of the weak-coupling
expansion read 
\begin{align}
\la{1.9}
\log F^{\mathbf A\,(2)}(\kappa; 3) &=-\frac{9\,\zeta (3)}{2}\,\kappa^{2}+\frac{425\,\zeta 
(5)}{36}\,\kappa^{3}-\frac{17885\,\zeta (7)}{576}\, \kappa ^4 +\frac{5565\, \zeta (9)}{64}\, \kappa ^5  \\
& +
\bigg(\frac{1925 \,\zeta (5)^2}{3456}-\frac{2668897 \,\zeta 
(11)}{10368}\bigg)\,\kappa ^6 +\bigg(\frac{32984237 \,\zeta 
(13)}{41472}-\frac{5005 \,\zeta (5) \,\zeta (7)}{864}\bigg)\,\kappa ^7+\mc O(\kappa^{8}),\notag
\end{align}
with similar results for $SU(N)$.
In the $N\ge 3$ theories, one can also consider the tower associated with 
$\cO = (\tr\varphi^{3})(\tr\varphi^{2})^{n}$, i.e. the function $F^{(3)}(\kappa; N)$. One finds for $N=3,4$
the expansions
\begin{align}
\la{1.10}
& \log F^{\mathbf A\,(3)}(\kappa; 3) = -\frac{9\,\zeta (3)}{2}\, \kappa ^2 +\frac{100 \, \zeta 
(5)}{9}\,\kappa ^3-\frac{15925\, \zeta (7)}{576}\, \kappa ^4+\frac{147\, \zeta (9)}{2}\, \kappa ^5\notag \\\
&+
\bigg(\frac{1925 \,\zeta (5)^2}{3456}-\frac{8599591 \,\zeta 
(11)}{41472}\bigg)\,\kappa ^6 + \bigg(\frac{3177031 \,\zeta 
(13)}{5184}-\frac{5005 \,\zeta (5) \,\zeta (7)}{864}\bigg)\,\kappa ^7+\mc O(\kappa^8),\notag \\
& \log F^{\mathbf A\,(3)}(\kappa; 4) = -\frac{9  \,\zeta (3)}{2}\,\kappa ^2+\frac{955 \,\zeta 
(5)}{92}\, \kappa ^3-\frac{429289 \,\zeta (7)}{17664}\, \kappa ^4+
\frac{78057  \,\zeta (9)}{1280}\,\kappa ^5 \\
&+\bigg
(\frac{293062 \,\zeta (5)^2}{475571}-\frac{68971014343 
\,\zeta (11)}{423464960}\bigg)\,\kappa ^6 + \bigg(\frac{387146868537 
\,\zeta (13)}{846929920}-\frac{28131103 \,\zeta (5) \,\zeta 
(7)}{4755710}\bigg)\,\kappa ^7+\mc O(\kappa ^8).\notag 
\end{align}
Now the exponentiation is no more in terms of simple $\zeta$-numbers with the exception
of the $\zeta(3)^{k}$ terms that are  fully resummed by the single $\zeta(3)$ term in the  above expansions.

For the higher rank one-point Wilson scaling functions the scenario is even more unsettled.
The only available result is the $SU(3)$ result for the $\mathbf A$ model with 
expansion \cite{Beccaria:2018xxl}
\be
\la{1.11}
\log F^{\mathbf{A}\,(2)}_{\mc W}(\kappa; 3) = 
-\frac{9\,\zeta(3)}{2}\,\kappa^{2}+\frac{175\,\zeta(5)}{18}\,\kappa^{3}
-\frac{12005\,\zeta(7)}{576}\,\kappa^{4}+\frac{1491\,\zeta(9)}{32}\,\kappa^{5}
-\frac{2247091\, \zeta(11)}{20736}\,\kappa^{6}+\mc O(\kappa^{7}).
\ee
Comparing (\ref{1.11}) with (\ref{1.9}) we see that  (\ref{1.6}) is certainly false in $SU(N)$
for $N>2$, i.e.
\be
\la{1.12}
F^{\mathbf A\,(2)}(\kappa; N) \neq  F^{\mathbf A\,(2)}_{\mc W}(\kappa; N),\qquad N>2.
\ee
Nevertheless,   (\ref{1.11}) strongly suggests an exponentiation similar to (\ref{1.7}).

Thus, in summary, at higher rank, one is led to ask the following list of open questions to 
be addressed in the generic $SU(N)$ case and depending on the specific tower $\cO_{n}$ 
and $\mathbf A$--$\mathbf E$ model:
\begin{itemize}
\item[Q1:]
Is there any relation between $F(\kappa; N)$ and $F_{\mc W}(\kappa; N)$, i.e. between the two-point and
one-point Wilson sectors ? 
Why does (\ref{1.6}) hold in $SU(2)$, but not in $SU(N>2)$ ? Is there any modified version of it that may 
work for higher rank ?

\item[Q2:]
Is it true that $\log F_{\mc W}(\kappa; N)$ 
may always be written as a series of simple $\zeta$-numbers ? 

\item[Q3:]
Is it possible to provide an all-order resummation, as in (\ref{1.8}), 
valid for any of $F(\kappa; N)$ and $F_{\mc W}(\kappa; N)$ ?

\item[Q4:] Does the GKT dual matrix model keep playing a role in answering the above questions, even at generic $N$ ? 

\end{itemize}

\subsection{Summary of results}

The analysis presented in this paper will consider and solve the previous open issues. 
In summary, our main results will be the following
\begin{enumerate}

\item
It is possible to compute $F^{(\Delta)}(\kappa; N)$ in any model and for both 
$\Delta=2,3$ by a suitable extension
of the GKT dual matrix model that captures the higher rank case.  This leads to an
efficient algorithm that computes
the perturbative expansion in $\kappa$ at any desired order with rather moderate (computational) 
effort.

\item
Using standard field-theoretical supergraph techniques on flat space we compute 
the two and three loops contributions to 
$F^{(\Delta)}(\kappa; N)$, i.e. the terms proportional to $\zeta(3)$ and $\zeta(5)$ respectively, and we give a hint of the generic $\zeta(2\ell-1)$ term.
This diagrammatical analysis of the double scaling limit matches the matrix model results, and is particularly useful to identify the class of diagrams contributing to that limit. These turn out to be specific {\em maximally non-planar}
insertions of certain polygonal loop diagrams. This is nicely opposite to what happens in the standard large $N$ limit. 

\item
There is indeed a close relation between the two-point and one-point Wilson sectors. For the $SU(2)$ theory we shall prove
the equality (\ref{1.6}). For $SU(N)$ with $N>2$ we shall prove the relation 
\be
\la{1.13}
F^{(2)}_{\mc W}(\kappa; N) = F^{(3)}_{\mc W}(\kappa; N) \equiv F_{\mc W}(\kappa; N).
\ee

\item 
In the Wilson sector, 
we shall also provide an efficient algorithm to compute the all-order expansion of $F_{\mc W}(\kappa; N)$ in powers of $\kappa$
based again on the higher rank extension of the dual matrix model. As a corollary of the construction, we shall prove
the exponentiation of $F_{\mc W}(\kappa; N)$ in terms of a series of simple $\zeta$-numbers.

\item
 Finally, we shall give very strong evidence for 
general resummations, similar to (\ref{1.8}), for all the five SCFT's and parametrical in $N$. 
From them, one can extract 
the perturbative ($\kappa\ll 1$) and non-perturbative ($\kappa \gg 1$) 
expansions $\log F_{\mc W}(\kappa; N)$.

\end{enumerate}

The last item in the above list will be our main result and is definitely non-trivial since 
such resummations are possible by a combination 
of (i) our proof that the higher rank dual matrix model may capture the Wilson scaling function together with (ii) our proof of exponentiation.
Such a result allows to explore the physics of the large $\kappa$ regime. The reason why it may be interesting is that
in the large $n$ limit (implicit at fixed $\kappa$) the path-integral
computing the scaling function is dominated by field configurations that are saddles of the modified $\mc N=2$
action taking into account the $\cO_{n}$ insertion.
As remarked in \cite{Grassi:2019txd}, 
this means that the relevant point in moduli space has  vacuum expectation
values growing like $g\,\sqrt{n}$. \footnote{The correct dimension is provided by a suitable infrared cutoff, like 
the inverse radius of the sphere in radial quantization.}
In the double scaling limit, this means that the hypermultiplet and short W-multiplet will have 
a mass $\sim \sqrt\kappa$, while magnetic BPS states, with mass $\sim g^{-2}\sqrt\kappa$,
will decouple.
At large $\kappa$, the electric BPS states will then lead to contact terms and exponentially suppressed contributions
vanishing like $\sim \exp(-c\,\sqrt\kappa)$. The non-perturbative contribution extracted from any 
resummation generalizing (\ref{1.8}) and parametric in $N$ will then be a direct probe into such a {\em heavy BPS} regime.

\paragraph{Plan of the paper}

In Sec.~\ref{sec:models} we briefly summarize the matrix model tools that are needed to discuss the 
large R-charge limit in the five superconformal theories in Tab.~\ref{tab:scft}. 
In Sec.~\ref{sec:2-point} we consider the extremal correlator sector and the two-point functions in the double scaling limit.
We review the GKT solution for the rank-1 $SU(2)$ gauge theory and generalize it to the higher rank case clarifying several technical 
issues. As an outcome, we provide specific results for the $\mathbf{ABCDE}$ models in terms of long expansions valid 
at weak coupling in the double scaling coupling $\kappa$. 
Sec.~\ref{sec:diagrams} is devoted to a diagrammatical check/interpretation of the results obtained in Sec.~\ref{sec:2-point}.
By using conventional Feynman diagram analysis in $\mc N=1$ super space, we identify the precise loop diagrams that 
give the leading order and next-to-leading order expansion of the scaling functions. In particular, we show that it is an insertion
characterized by the maximally non-planar topology.
Sec.~\ref{sec:wilson-exp} moves to the second sector of observables, i.e. one-point functions of chiral operator in presence 
of a $\frac{1}{2}$-BPS Wilson-Maldacena loop. We begin by collecting explicit data for the $SU(3)$ and $SU(4)$ theories
in order to extend the amount of explicit calculations and explore new features of the higher rank case. Then, in  
Sec.~\ref{sec:wilson-th}, we prove such features and obtain closed expressions for the one-point Wilson scaling functions 
that are valid in all the treated cases, i.e. for all $\mathbf{ABCDE}$ models at generic $N$ and for both types of 
large R-charge towers. Finally, Sec.~\ref{sec:resum} is devoted to the analysis of the resummations presented in 
Sec.~\ref{sec:wilson-th} in the above {\em heavy BPS} regime, i.e. for $\kappa\gg 1$, where we give a full account of the 
computed  non-perturbative corrections. Their detailed structure allows to match the spectrum of heavy BPS states
relevant in this regime.

\section{Matrix model description of the five $SU(N)$ theories}
\la{sec:models}

In this brief section, we summarize the matrix model description of the five $SU(N)$ theories
with matter content as in Tab.~\ref{tab:scft}. \footnote{The analysis of 
\cite{Koh:1983ir} identifies additional three cases, but they exist only for specific values of $N$.
Notice that they involve matter fields in the rank-3 antisymmetric representation. Although we do not consider them in this paper, all of our methods are applicable to them without any additional complications.}
Aspects of these theories related to the properties of their extremal correlators have been recently discussed in \cite{Bourget:2018fhe}.

\paragraph{Action}
For a general $\mathcal{N}=2$ theory with gauge group $SU(N)$, the partition function on a four sphere obtained by localization can be written as \footnote{Sometimes it may be convenient to rescale the matrix $a$
in order to make the Gaussian part of the action read simply $e^{-\tr a^{2}}$.}
\begin{align}
  \label{2.1}
  Z_{S^{4}} &= \int [\dd a] \exp (-\frac{8\pi^{2}}{g^{2}} \tr a^{2}) Z_{\text{1-loop}}\abs{Z_{\text{inst}}}^{2},
\end{align}
where $[\dd a]$ is the standard measure over the conjugacy classes of traceless Hermitian matrices reading
($a_{\mu}$ are the $N$ eigenvalues of $a$ ) 
\begin{align}
    \label{2.2}
    [\dd a] = \prod_{\mu =1}^{n} \dd a_{\mu} \prod_{ \nu < \mu }(a_{\mu} - a_{\nu})^{2} \delta\bigg(\sum_{\mu}a_{\mu}\bigg).
\end{align}
For a $\mc N=2$ 
superconformal theory with matter in the representation $\mathcal{R}$ of $SU(N)$, the interacting action in $Z_{\text{1-loop}}=e^{-S_{\rm int}}$ is 
conveniently written as 
\begin{align}
\la{2.3}
  S_{\rm int}(a) = -\log Z_{\text{1-loop}} = -\sum_{m=2}^{\infty}(-1)^{m}
  \frac{\zeta(2m-1)}{m}\left(\Tr_{\mathcal{R}} a^{2m} - \Tr_{\text{adj}} a^{2m}\right).
\end{align}
The difference of traces in (\ref{2.3}) stands for the
replacement of 
 $\mc N=4$ virtual exchanges of adjoint hypermultiplets by similar exchanges of 
matter hypermultiplets transforming in $\mathcal R$ \cite{Billo:2019fbi}. For the fundamental representation, 
we shall simply write $\Tr_{\rm fund} \equiv \tr$. The combination of traces appearing in $S_{\rm int}(a)$ can be expressed as 
\begin{equation}
	\label{tra2ntoC}
		\Tr_{\mathcal{R}} a^{2m} - \Tr_{\text{adj}} a^{2m} = \Tr^\prime_{\mathcal{R}} a^{2m} = C^\prime_{(b_1\ldots b_{2m})} \, a^{b_1}\ldots a^{b_{2m}}~,	
\end{equation}
where
\begin{equation}
	\label{defC}
		C^\prime_{b_1\ldots b_{2m}} = \Tr^\prime_{\mathcal{R}} T_{b_1}\ldots T_{b_{2m}}~.
\end{equation}
The indices $b_i = 1\dots N^2-1$ run over the gauge algebra. See appendix \ref{app:rules} for our conventions and for a systematic discussion of how to express the differences in (\ref{2.3})
in terms of traces in the fundamentals (see also \cite{Billo:2019fbi}). This procedure yields the explicit form of (\ref{2.3}) for the $SU(N)$ models with the matter content listed in Tab.~\ref{tab:scft}. 

Finally, the factor $Z_{\text{inst}}$ in (\ref{2.1}) 
takes into account the instanton corrections. In this paper they will not play any role while studying the double scaling limit. For this reason,
we simply drop $Z_{\text{inst}}$.

\paragraph{Observables}

As we mentioned in the Introduction, we shall primarily be interested in two classes of flat space correlation 
functions in such superconformal field theories. The first are two-point functions between a chiral primary $\mathcal{O}(x)$, with 
conformal weight $\Delta(\mc O)$, and its conjugate.
From conformal invariance we have (rank dependence is understood) \footnote{Notice that $\mc N=1$ superconformal invariance
is enough to protect the dimension $\Delta(\mc O)$ of chiral operators against radiative corrections, 
as first discussed in full generality in \cite{Conlong:1993eu}.}
\begin{align}
\la{2.4}
     \ev{\mathcal{O}(x) \,\bar{\mathcal{O}}(y)} = \frac{G_{\mathcal{O} \bar{\mathcal{O}}}(g)}
     {(x-y)^{2\Delta(\mathcal{O})}}.
\end{align}
The other class of observables will be one-point function of a chiral primary operator $\mathcal{O}$ in the presence of Wilson loop \eqref{1.3}.
In general it is given by:
\begin{align}
  \ev{\mathcal{O}(x) \, \mathcal{W}} &= \frac{A_{\mathcal{O}}(g)}{(2\pi\Vert x \Vert_{C})^{\Delta(\mathcal{O})}},
\end{align}
where $\Vert x\Vert_{C}$ is a distance between $x$ and the circle $C$, invariant under the $SO(1, 2) \times SO(3)$ subgroup of the 
conformal symmetry preserved by the Wilson loop, see App. A of \cite{Billo:2018oog}.

Both  $G_{\mathcal{O} \bar{\mathcal{O}}}(g)$ and $A_{\mathcal{O}}(g)$ are non-trivial
coupling dependent functions. They encode the information about the above correlation functions that is 
not fixed by conformal symmetry and from henceforth, with a little abuse of language,  we will refer to them
simply as two-point  and one-point Wilson functions.

For a $\mathcal{N}=2$ theory on $S^{4}$, they can be evaluated using the partition function \eqref{2.1}. 
Our focus will be on the Coulomb branch operators. For $SU(N)$ gauge group, 
they are generated by $\tr \varphi^{n}$ with $2 \le k \le N$ with $\varphi$ 
being one of the two complex combinations of the two scalars in the 
vector multiplet. Ignoring instanton corrections,
the recipe for computing $G_{\mathcal{O} \bar{\mathcal{O}}}(g)$ is simple. 
Given two Coulomb branch operators $\mathcal{O}(\varphi)$ and $\bar{\mathcal{O}}(\varphi)$ 
we can compute $G_{\mathcal{O} \bar{\mathcal{O}}}(g)$ on $S^{4}$ by inserting $\mathcal{O}$ at the north pole and $\bar{\mathcal{O}}$ at the south pole. This corresponds to inserting $\mc O(a)\,\bar{\mc O}(a)$ in the sphere partition function \eqref{2.1}
\begin{align}
\la{2.6}
    \ev{\mathcal{O}(N) \,\bar{\mathcal{O}}(S)}_{S^{4}} &= \frac{1}{Z_{S^{4}}}\int [\dd a]\, \mc O(a) \bar{\mc O}(a)\,\exp (-\frac{8\pi^{2}}{g^{2}} \tr a^{2}) Z_{\text{1-Loop}}.
\end{align}
The naive operators $\mc O(a)$ are not correct to reproduce flat space correlators due to conformal anomalies inducing a peculiar mixing 
on the sphere \cite{Gerchkovitz:2016gxx, Rodriguez-Gomez:2016ijh, Rodriguez-Gomez:2016cem}. 
\footnote{See for instance \cite{Gomis:2015yaa} for a detailed discussion of the mixing with the identity operator. Notice also that, as we 
remarked in the Introduction, there is not enough supersymmetry on $S^{4}$ to protect from this mixing.} 
In general, 
the matrix model chiral operator $\mc O$ has to be replaced by its {\em normal ordered} version defined in 
$:\mc O:$ \cite{Billo:2017glv}, i.e.
\begin{align}
   \label{2.7}
   :\mathcal{O}: \ =  \mathcal{O} + \sum_{\Delta(\mathcal{O}^{\prime}) < \Delta(\mathcal{O})}c_{\mathcal{\mathcal{O},\mathcal{O}^{\prime}}}(g)\mathcal{O}^{\prime}, 
\end{align}
where the coefficients $c_{\mathcal{\mathcal{O},\mathcal{O}^{\prime}}}(g)$ are determined by requiring the orthogonality 
condition with smaller dimensional operators $\ev{:\mathcal{O}:\,\mathcal{O}^{\prime}}_{S^{4}} = 0$.
%
Writing the explicit form of (\ref{2.7}) is clearly a major complication in the double scaling limit where the dimension of the considered
operators grows arbitrarily. Indeed, apart from some simple cases, the mixing coefficients are not known in closed form.
Nevertheless, we will see that a suitable  dual matrix model description can be used to overcome this technical difficulty.  

For observables involving the $\frac{1}{2}$-BPS Wilson loop, we have to supplement (\ref{2.6}) and (\ref{2.7})
with the correct replacement rule for the unit radius Wilson loop (\ref{1.3}), i.e. 
 \cite{Pestun:2007rz} \footnote{Several explicit  field-theoretical  verifications of the matrix model map can be found in \cite{Billo:2017glv,Billo:2018oog,Billo:2019job,Billo:2019fbi}.}
\begin{align}
  \label{2.8}
  \mc W(a) =\frac{1}{N} \tr e^{2\pi a} \ .
  \end{align}

\section{Extremal two-point functions at large R-charge in $SU(N)$ theories}
\la{sec:2-point}

In this section, we generalize the GKT dual matrix model approach \cite{Grassi:2019txd} in order to go beyond the rank-1 $SU(2)$ case and 
compute the large R-charge limit of extremal two-point functions in the general $SU(N)$ superconformal theories discussed in Sec.~(\ref{sec:models}).
Our main results (\ref{3.28}) will lead to a computational algorithm able to produce  long perturbative expansions of the scaling functions. These, in
principle, may be useful to derive (or check proposed) all-order resummations.

\subsection{Review of the $SU(2)$ Grassi-Komargodski-Tizzano solution}

We are interested in evaluating the  two-point function  (\ref{2.4}) in the special case
$\mc O = (\tr\varphi^{2})^{n}$. To this aim, 
we want to determine $G_{2n}(\tau, \bar\tau)$ in ($\tau$ is the complexified gauge coupling, $\Im \tau = \frac{4\pi}{g^{2}}$)
\begin{align}
\la{3.1}
\ev{(\tr\varphi^{2}(x))^{n}(\tr\overline{\varphi}^{2}(y))^{n}} = \frac{G_{2n}(\tau, \bar\tau)}{(x-y)^{4n}}.
\end{align}
To apply localization methods \cite{Pestun:2007rz,Gerchkovitz:2016gxx} one starts by considering the  
infinite matrix $\mathcal{M}$ defined by 
\begin{align}
\la{3.2}
\mathcal{M}_{nm} = \frac{1}{Z_{S^{4}}}\frac{\partial^{n+m}Z_{S^{4}}[\tau,\bar{\tau}]}
{\partial^{n}\tau \partial^{m}\bar{\tau}}, 
\end{align} 
where $Z_{S^{4}}$ is given by \eqref{2.1}. We shall denote by $\mc M_{(n)}$ the $n\times n$ truncation 
with matrix indices running in the range $0, \dots, n-1$.
As shown in \cite{Gerchkovitz:2016gxx}, it is possible to prove that 
\begin{align}
\la{3.3}
    G_{2n}(\tau, \bar\tau) = \frac{\det\mathcal{M}_{(n+1)}}{\det\mathcal{M}_{(n)}},
\end{align}
where the determinant ratio disentangles the mixing that occurs on $S^{4}$. 
In the $SU(2)$ case, the large R-charge limit of (\ref{3.3}) may be determined by the approach 
in \cite{Grassi:2019txd}. However,
the derivation cannot be naively 
extended to the higher rank $SU(N)$ case. As a preparation to the necessary changes, we 
now briefly summarize the GKT strategy. 

The first step is to use the so-called Andr\'eief identity, see for instance Lemma 3.1 in \cite{baik2003products}, 
which converts $\det \mathcal{M}_{(n)}$ from the determinant of a matrix with each elements defined as an integral to an integral of 
determinants: 
\begin{align}
\det_{kl} \int \dd\mu(y)\,f_{k}(y)g_{l}(y) = \frac{1}{N!}\int\prod_{j=0}^{N-1}
\dd\mu(y_{j})\,\det_{kl}(f_{k}(y_{l}))\det_{mn}(g_{m}(y_{n})),
\end{align}
where $f_{k},g_{k}$ with $k\in\{0,\cdots,N-1\}$ are two sets of $N$-functions and $\dd\mu(y)$ 
is the measure of integration. The relevant measure for $\det \mathcal{M}_{(n)}$ is $d\mu(a) = [\dd a]Z_\text{1-loop}(a)$. For the $SU(2)$ gauge group the space of conjugacy classes of Hermitian matrices is one dimensional and we parameterize it by $a$. As a result the measure is
$\dd a\, e^{-4\pi \Im \tau a^{2}}a^{2}Z_\text{1-loop}(a)$.
The derivative w.r.t both $\tau$ and $\bar{\tau}$ brings down a factor of $a^{2}$. Hence, the functions $f_{k}$ and $g_{l}$ are simply
$f_{k}(a) = a^{2k}$ and $g_{l}(a) = a^{2l}$.
From 
\begin{align}
\det_{kl}a_{k}^{2l}\det_{mn}a_{m}^{2n} = \left(\det_{kl}a_{k}^{2l} \right)^{2} = 
\prod_{k<l}(a^{2}_{k} - a^{2}_{l})^{2},
\end{align}
we have 
\begin{align}
 \det \mathcal{M}_{(n)} &= 
  \frac{1}{n!}\int_{0}^{\infty}\prod_{j=0}^{n-1}\dd x_{j}\,\sqrt{x_{j}}e^{-4\pi \Im \tau x_{j}} 
  Z_{1\mbox{\scriptsize -loop}}(\sqrt{x}_{j})\prod_{k<l}(x_{k} - x_{l})^{2},
 \end{align}
 where in the last step we changed variables of integration to $x_{j} = a_{j}^{2}$. Due to the presence of Vandermonde determinant $\prod_{k<l}(a_{k}^{2} - a_{l}^{2})^{2}$ the above expression can be recognized as a matrix integral. However in this case the eigenvalues are $a_{j}^{2}$ (i.e. $x_{j}$) and not $a_{j}$. As a result this expression doesn't come from an integral over Hermitian matrices but rather over positive matrices $W$, i.e. a 
 Wishart-Laguerre matrix model \cite{Akemann:2011csh}.
From
\begin{align}
\prod_{j=0}^{n-1}\sqrt{x_{j}} = \exp(\frac{1}{2}\sum_{j=0}^{n-1}\log x_{j}) \to \exp (\frac{1}{2}\tr \log W),
\end{align}  
we see that we are to compute the {\bf dual} matrix model partition function
\begin{align}
\label{3.8}
\int [\dd W]\, e^{-V(W)} ,\qquad V(W) =  4\pi \Im \tau \tr W - \frac{1}{2}\tr \log W - \tr \log Z_\text{1-loop}(\sqrt W).
\end{align} 
For this treatment to valid the 1-loop partition function must depend only on the conjugacy classes of $W$, i.e.
\begin{align}
\la{3.9}
Z_\text{1-loop}(W)  = Z_\text{1-loop}(\tr W , \tr W^{2} , \cdots).
\end{align}
This statement is trivially true for $SU(2)$ theories but as we shall see this will pose novel problems in the higher rank case.

\paragraph{Double scaling limit of the dual matrix model}
To evaluate the scaling function $F^{(2)}(\kappa; N)$ 
we need $\mathcal{M}_{(n)}$ in the limit (\ref{1.1}). The potential in (\ref{3.8}) is then 
\begin{align}
V(W) = \frac{n}{4\,\pi^{2}\,\kappa}\tr W - \frac{1}{2}\tr\log(W) - \tr \log Z_{\mbox{\scriptsize 1-loop}}(\sqrt{W}).
\end{align} 
In this expression the first factor depends on $n$ while the other two factors are single trace deformations that contribute at a sub-leading order in $n$. As explained in \cite{Grassi:2019txd}, in the double scaling limit the typical eigenvalue of $W$ is $\sim\kappa$. This combined with the trace structure of $V$ allows  for a $\frac{1}{n}$ expansion of $\log \mathcal{M}_{(n)}$ i.e.
\be
    \log \mathcal{M}_{(n)} = \sum_{k = 0}^{\infty}n^{2-k}C_{k}(\kappa). \label{3.11}
\ee
We can now compute $\mathcal{M}_{(n)}$ by treating the factor $Z_{\mbox{\scriptsize 1-loop}}$ as a perturbation around the 
Gaussian matrix model,  i.e. around the $\mathcal{N} = 4$ theory,
$\det \mathcal{M}_{(n)} = \ev{Z_{\mbox{\scriptsize 1-loop}}(W)}$.
For a single trace operator $\mathcal{O}$,  we can replace 
$\ev{e^{\mathcal{O}}} \to  e^{\ev{\mathcal{O}}}$ up to terms that are subleading at large $n$.
Since $Z_{\mbox{\scriptsize 1-loop}}$ is single trace, we have simply
\begin{align}
\log\det \mathcal{M}_{(n)} &= \log\ev{Z_{\mbox{\scriptsize 1-loop}}(W)}  \stackrel{n\to \infty}{=} 
\ev{\log Z_{\mbox{\scriptsize 1-loop}}(W)} + O\left(1\right).
\end{align}
The expectation value in the r.h.s. can be evaluated by integrating $Z_{\mbox{\scriptsize 1-loop}}$ weighted by the joint eigenvalue distribution function for positive matrices. The eigenvalue distribution is governed by 
Mar\v{c}enko-Pastur law  \cite{marvcenko1967distribution}. In the large $n$ limit the result is then 
\begin{align}
\la{3.13}
    \log\det \mathcal{M}_{(n)} \stackrel{n\to \infty}{=} n\int_{0}^{4}\dd x\, \rho(x)Z_{\mbox{\scriptsize 1-loop}}( 4\,\pi^{2}\,\kappa x)+\mc O(1),\qquad \text{with}\qquad 
    \rho(x) = \frac{1}{2\pi}\sqrt{\frac{4}{x}-1}.
\end{align}
Using this expression we compute the log of \eqref{3.3} while keeping in mind that $n$ and $\kappa$ have to be varied together. Subtracting the $\mathcal{N}=4$ contributions the final result is:
\begin{align}
\la{3.14}
    F^{(2)}(\kappa; 2) = \int_{0}^{4}\rho(x)\Big(\log Z_{\mbox{\scriptsize 1-loop}}(4\,\pi^{2}\,\kappa x) + \kappa \partial_{\kappa} Z_{\mbox{\scriptsize 1-loop}}(4\,\pi^{2}\,\kappa x)\Big).
\end{align}
Finally,
by using the series expansion for $\log Z_{\mbox{\scriptsize 1-loop}}$, it is possible to resum (\ref{3.14}) in the form  \eqref{1.8}.

\subsection{Higher rank extension for $SU(N)$ theories}

Now we turn to a generalization of the GKT approach which  enables us to compute both $F^{(2)}(\kappa,N)$ and $F^{(3)}(\kappa,N)$
for any $N$. We begin with $F^{(2)}(\kappa,N)$, while
the extension to $F^{(3)}(\kappa,N)$ will be obvious once we are done.  Again we start by writing  $\det \mc M_{(n)}$ as an integral of a determinant
\be
\det\mathcal{M}_{(n)} = \frac{1}{n!}\int\prod_{i=0}^{n-1}[\dd a_{i}]\, e^{-4\pi \Im \tau \tr a_{i}^{2}}Z_\text{1-loop}(a_{i})\prod_{j < i}(\tr a_{i}^{2} - \tr a_{j}^{2})^{2}.
\ee
We can see from this expression that the positive matrix ensemble emerges once again. The eigenvalues of this matrix are $\tr(a_{i}^{2})$. But unlike the rank one case there are additional variables since an $SU(N)$ matrix has $N-1$ independent eigenvalues. To make progress, we need to separate 
$\tr(a_{i}^{2})$ out of  the rest of these variables. To start, we consider the $\mathcal{N}=4$ theory by setting $Z_\text{1-loop}\to 1$. In this case,
$\tr a_{i}^{2}$ is already separated. We go from Cartesian coordinates for eigenvalues to polar coordinates after which $\tr a^{2}$ becomes the radial coordinate. Hence,
\begin{align}
\la{3.16}  
\det \mathcal{M}_{(n)} = \frac{C^{n}_{N}}{n!}\int_{0}^{\infty}\prod_{j=0}^{n-1}\dd x_{j}\,\sqrt{x_{j}^{N^{2}-3}}e^{-4\pi \Im \tau x_{j}} \prod_{k<l}(x_{k} - x_{l})^{2}, \end{align}  
where $x_{j} = \tr a_{j}^{2}$ and $C_{N}$ is an integral over the $(N-1)$-sphere, 
$C_{N} = \frac{1}{2}\int_{S^{N-1}} \dd\Omega\, D(\Omega)$, 
where $D(\Omega)$ is (implicitly) determined by polar decomposition  
\begin{align}
\delta(\tr a)\prod_{\nu<\mu}(a_{\mu} - a_{\nu})^{2} = \Big(\sum_{\mu=1}^{N}a_{\mu}^{2}\Big)^{\frac{N^{2}-N-1}{2}}D(\Omega).
\end{align}
As a result, the previous treatment based on the Wishart-Laguerre type matrix model generalizes 
straightforwardly to $SU(N)$ $\mathcal{N}=4$ theory
\begin{align}
    \det \mathcal{M}_{(n)} &= \frac{C_{N}^{n}}{n!}\int[\dd W]\,e^{-V(W)} ,\qquad
    V(W) =  4\pi \Im \tau \tr W - \frac{N^{2}-3}{2}\tr \log W .
\end{align}
This fails to be the case when we consider $\mathcal{N}=2$ theories because $Z_{\mbox{\scriptsize 1-loop}}$ is not a function of just $\tr(a^{2})$ but rather depends on 
(products of ) $\tr a^{k}$ with $2 \le k \le N$. This means that in \eqref{3.16}, $Z_{\mbox{\scriptsize 1-loop}}$ is a function not only of radial variable $x_{i}$ but also of angular variables $\Omega_{i}$. At this stage $Z_{\mbox{\scriptsize 1-loop}}$ is not an observable in the matrix model, but it becomes such after 
integrating out the angular variables. Hence, we define the quantity $\mathcal{Z}_{\mbox{\scriptsize 1-loop}}$  by
\begin{align}
\mathcal{Z}_{\mbox{\scriptsize 1-loop}} = \int_{S^{N-1}}\,\prod_{i=0}^{n-1}\dd\Omega_{i}\, D(\Omega_{i}) \, Z_{\mbox{\scriptsize 1-loop}}(x_{i},\Omega_{i}).
\end{align}
The important point is that this is  a class invariant function in the matrix model, cf. (\ref{3.9}),
\be
\mathcal{Z}_{\mbox{\scriptsize 1-loop}} = \mathcal{Z}_{\mbox{\scriptsize 1-loop}}(\tr W , \tr W^{2} , \cdots),
\ee
because  $\mathcal{Z}_{\mbox{\scriptsize 1-loop}}$ is a symmetric function of $x_{i}$ and any symmetric function of $x_{i}$ 
can be converted into function of traces of powers of $W$. Now, for a general function $K(x,\Omega)$ of the form 
\begin{align}
K(x,\Omega) = \exp\left(\sum_{k}f_{k}(\Omega)x^{2k} \right),
\end{align}
we can write 
\begin{align}
\la{3.22}
\mathcal{K}(W) &= 
 \frac{1}{C_{N}^{n}}\int_{S^{N-1}}\,\prod_{i=0}^{n-1}\dd\Omega_{i}\, D(\Omega_{i}) \, K(x_{i},\Omega_{i})
= \exp\left(\sum_{\vec{k}} \frac{\mathfrak{f}_{\vec{k}}}{n^{\#\vec{k}-1}S(\vec{k})}\prod_{k\in \vec{k}}\tr(W^{k}) \right),
\end{align} 
where
%
$\# \vec{k}$ is the number of non-zero entries of $\vec{k}$ and $S(\vec{k})$ is a symmetry factor which takes into account the degeneracies of entries of $k$. Both it and $n^{\#\vec{k}-1}$ have been included for the later convenience. 

Defining the angular expectation value $\llangle f \rrangle$ of $f(\Omega)$ to be,
\begin{align}
  \llangle f \rrangle &= \frac{1}{C_{N}}\int_{S^{N-1}}\,\dd\Omega\, D(\Omega) \, f(\Omega),
\end{align}
it is possible  to show that in the large $n$ limit \footnote{In other words, $\mathcal{K}(W)$ in the large $n$ limit is analogous to the effective action resulting from a path integral with tadpoles.}:
\begin{align}
\la{3.24}
 \mathfrak{f}_{\vec{k}} &= \left\{\begin{array}{cccc} \llangle f_{k} \rrangle &  & \mbox{if } \vec{k} = (k) \\
                    \llangle \prod_{k \in \vec{k}} (f_{k} - \llangle f_{k} \rrangle) \rrangle & &  \mbox{if } \#\vec{k} > 1
 \end{array}\right. 
 \end{align}
  The explicit calculation of the relevant angular integrals is explained in App.~\ref{app:angular}.
  
  We can now treat the double scaling limit perturbatively. In this limit the typical eigenvalue of the matrix $W$ is of the order of coupling $\kappa$, as a result $\tr(W^{k})$ contributes on the order of $n\kappa^{k}$. Hence, any operator with $\#\vec{k}$-traces contributes as $n^{\#\vec{k}}\kappa^{\sum_{k\in \vec{k}}k}$. It is clear from \eqref{3.24} that $\mathfrak{f}_{\vec{k}}$ is independent of $n$. This, combined with the explicit factor of $\frac{1}{n^{\#\vec{k}-1}}$ in \eqref{3.22}, means that higher trace operators are suppressed by just the right power of $n$ in $\mathcal{K}(W)$ and they contribute to the same order as single trace operators. Thus, this large $n$ limit receives corrections from non-planar diagrams even at leading order. 
Moreover to the leading order in $\frac{1}{n}$ we can again replace  
$\ev{e^{\mathcal{K}(W)}} \to e^{\ev{\mathcal{K}(W)}}$. Setting $\mathcal{K}$ to be $\mathcal{Z}_{\mbox{\scriptsize 1-loop}}$ we see that 
\begin{align}
\log\det \mathcal{M}_{(n)} = \log\ev{\mathcal{Z}_{\mbox{\scriptsize 1-loop}}} \stackrel{n\to \infty}{=}
 \ev{\log(\mathcal{Z}_{\mbox{\scriptsize 1-loop}})} +  O(1).
\end{align}
Using the large $n$ limit of Mar\v{c}enko-Pastur law, cf. (\ref{3.13}),
\begin{align}
\la{3.26}
\ev{\log(\mathcal{Z}_{\mbox{\scriptsize 1-loop}})} = \int_{0}^{4}\dd x\, \rho(x)\eval{\mathcal{Z}_{\mbox{\scriptsize 1-loop}}}_{W \to 4\,\pi^{2}\,\kappa x},\qquad \text{with}\qquad 
\rho(x) = \frac{1}{2\pi}\sqrt{\frac{4}{x}-1},
\end{align}
we can obtain  $F^{(2)}(\kappa,N)$ in the same fashion as in the $SU(2)$ case
\begin{align}
\log F^{(2)}(\kappa,N) &= \int_{0}^{4}\rho(x)\Big(\log\mathcal{Z}^{\prime}_{\mbox{\scriptsize 1-loop}}(4\,\pi^{2}\,\kappa x) + 
\kappa\,\partial_{\kappa} \mathcal{Z}^{\prime}_{\mbox{\scriptsize 1-loop}}(4\,\pi^{2}\,\kappa x)\Big),\notag \\
\mathcal{Z}^{\prime}_{\mbox{\scriptsize 1-loop}}(x) &= \sum_{i} c_{i} x^{i},\qquad 
c_{i} = \sum_{\vec{k},\sum_{k\in \vec{k}}k=i}\frac{\mathfrak{f}_{\vec{k}}}{S(\vec{k})}.
\end{align}
Hence, our final formula reads 
\begin{align}
\la{3.28}
\log F^{(2)}(\kappa;N) &= 
 \sum_{j = 1}^{\infty}c_{j}\frac{(j+1)2^{2j}\Gamma(j + \frac{1}{2})}{\sqrt\pi\,\Gamma(j+2)}(4\,\pi^{2}\,\kappa)^{j},
\end{align}
where the various $\Gamma$-functions come from  elementary integrals of the Mar\v{c}enko-Pastur distribution.

\subsection{Application to the five $\mc N=2$ superconformal $SU(N)$ gauge theories}

Let us summarize and illustrate in detail how (\ref{3.28}) may be applied to 
the specific $\cN=2$ theories in Sec.~(\ref{sec:models}) in order to obtain the  scaling functions $F^{(\Delta)}(\kappa, N)$. The relevant steps are :
\begin{enumerate}
\item Take the  {\em interacting action} $S_{\rm int}(a)$, cf. \eqref{2.3},
and 
%
convert $\tr_{\mc R}(\circ)$ into  traces in the fundamental
representation using the general relations derived in \cite{Billo:2019fbi}. This allows to  write 
 \be
 \la{3.29}
 S_{\rm int} = -\sum_{n=2}^{\infty} \sigma_{n}(a), 
 \ee
 where $\sigma_{n}(a)$ is a homogeneous polynomial in the traces $\tr(a^{k})$ evaluated in the fundamental representation.
 
 \item Compute the coefficients $\{c_{n}\}$ defined by 
 \be
 \la{3.30}
\widetilde{f}^{(2)}(\kappa; N) =  \sum_{n=2}^{\infty}c_{n}\,\kappa^{n} = \log\big\llangle 
 \exp(\sum_{n=2}^{\infty}\sigma_{n}(a)\,(4\pi^{2}\kappa)^{n})
 \big\rrangle,
\ee
where the angular bracket denotes angular integration and can be computed as in App.~\ref{app:angular}.

\item The scaling function for the $(\tr \varphi^{2})^{n}$ tower is obtained from, cf. (\ref{3.28}), 
 \be
 \la{3.31}
  \log F^{(2)}(\kappa, N) = \sum_{n=2}^{\infty}\frac{(n+1)\,2^{2n}\,\Gamma(n+\frac{1}{2})}{\sqrt\pi\,\Gamma(n+2)}\,c_{n}\,(4\pi^{2}\kappa)^{n}.
  \ee

\item The scaling function for the $\tr\varphi^{3}\,(\tr\varphi^{2})^{n}$ tower is similarly obtained as,
 \be
  \log F^{(3)}(\kappa, N) = \sum_{n=2}^{\infty}\frac{(n+1)\,2^{2n}\,\Gamma(n+\frac{1}{2})}{\sqrt\pi\,\Gamma(n+2)}\,d_{n}\,(4\pi^{2}\kappa)^{n},
  \ee
  where now (the denominator may be found in (\ref{B.4}) ) 
   \be
   \la{3.33}
\widetilde{f}^{(3)}(\kappa; N) =  \sum_{n=2}^{\infty}d_{n}\,\kappa^{n} = \log \frac{\left\llangle [\tr(a^{3})]^{2}\ 
 \exp(\sum_{n=2}^{\infty}\sigma_{n}(a)\,(4\pi^{2}\kappa)^{n})
 \right\rrangle}{\left\llangle [\tr(a^{3})]^{2} 
 \right\rrangle}.
\ee
 \end{enumerate}
 
\subsubsection{Explicit expansions}

Let us give explicit expansions of $\log F^{(\Delta)}(\kappa; N)$, $\Delta=2,3$, 
valid for generic models and rank.
For any of the five models their structure is 
\begin{align}
\la{3.34}
& \log F^{(2)}(\kappa; N) = 
\rf^{(2)}_{3}\,\zeta(3)\, \kappa ^2 
+\rf^{(2)}_{5} \,\zeta(5)\,\kappa ^3
+\bigg[\rf^{(2)}_{3^{2}}\,\zeta(3)^{2}+\rf^{(2)}_{7}\,\zeta(7)\bigg]\,\kappa ^4
+\bigg[\rf^{(2)}_{3,5}\,\zeta(3)\,\zeta(5)+\rf^{(2)}_{9}\,\zeta(9)\bigg]\,\kappa ^5 \spek
+ \bigg[\rf^{(2)}_{3^{3}}\,\zeta(3)^{3}+\rf^{(2)}_{5^{2}}\,\zeta(5)^{2}+\rf^{(2)}_{3,7}\,\zeta(3)\,\zeta(7)+\rf^{(2)}_{11}\,\zeta(11)\bigg]\,\,\kappa ^6\spek
+ \bigg[\rf^{(2)}_{3^{2},5}\,\zeta(3)^{2}\,\zeta(5)+\rf^{(2)}_{5,7}\,\zeta(5)\,\zeta(7)+\rf^{(2)}_{3,9}\,\zeta(3)\,\zeta(9)+\rf^{(2)}_{13}\,\zeta(13)\bigg]\,\,\kappa ^7\spek
+ \bigg[\rf^{(2)}_{3^{4}}\,\zeta(3)^{4}+\rf^{(2)}_{3,5^{2}}\,\zeta(3)\,\zeta(5)^{2}+\rf^{(2)}_{3^{2},7}\,\zeta(3)^{2}\,\zeta(7)+\rf^{(2)}_{7^{2}}\,\zeta(7)^{2}+\rf^{(2)}_{5,9}\,\zeta(5)\,\zeta(9)\spek
+\rf^{(2)}_{3,11}\,\zeta(3)\,\zeta(11)+\rf^{(2)}_{15}\,\zeta(15)
\bigg]\,\kappa ^8+\cdots
\end{align}
where $\rf^{(2)}_{\circ} = \rf^{(2)}_{\circ}(N)$. 
Notice  that (i) the first $\zeta(3)$ term is absent in the $\mathbf E$ model, and (ii) all terms involving powers of $\zeta(3)$ or 
products of $\zeta(3)$ with other $\zeta$ functions are absent in both the $\mathbf A$ and $\mathbf E$ models. 
The same structure of the expansion and special vanishing
properties hold for the second tower, i.e. for $\log F^{(3)}(\kappa; N)$. 
In this case we shall denote the coefficients as $\rf^{(3)}_{\circ}$. The explicit results for each model 
are collected in App.~\ref{app:results}. Up to the $\kappa^{4}$ term ($\kappa^{5}$ in the  $\mathbf E$ model) they 
read
\begin{align}
\la{3.35}
\log\, & F^{\mathbf A\,(2)}(\kappa; N) = 
-\frac{9\,\zeta (3)}{2}\,\kappa^{2}+\frac{25  (2 N^2-1) \,\zeta 
(5)}{N (N^2+3)}\,\kappa ^3-\frac{1225  (8 N^6+4 N^4-3 N^2+3) \,\zeta (7)}{16 N^2 (N^2+1) (N^2+3) (N^2+5)}\,\kappa^{4}+\cdots,
\notag \\
\log\, & F^{\mathbf B\,(2)}(\kappa; N) = 
-\frac{9 (N-3) (N-2) (N+1) \,\zeta (3)}{4 N (N^2+1)}\,\kappa^{2}+\frac{25 (N-2) (2 N^4-6 N^3-15 N^2+15) \,\zeta(5)}{2 N^2 (N^2+1) (N^2+3)}\,\kappa^{3}\spek
+\bigg[\frac{315 (N-3) (N-2) (N+2) 
(N+3) \,\zeta (3)^2}{4 (N^2+1)^2 (N^2+3) (N^2+5)}\spek
-\frac{245 (N-2) (40 
N^6-172 N^5-564 N^4-120 N^3+1185 N^2+480 N-945) \,\zeta (7)}{32 N^3 
(N^2+1) (N^2+3) (N^2+5)}\bigg]\,\kappa^{4}+\cdots,
\notag \\
\log\, & F^{\mathbf C\,(2)}(\kappa; N) = 
-\frac{9 (N-1) (N+2) (N+3) \,\zeta (3)}{4 N 
(N^2+1)}\,\kappa^{2}+\frac{25 (N+2) (2 N^4+6 N^3-15 N^2+15)\, \zeta 
(5)}{2 N^2 (N^2+1) (N^2+3)}\,\kappa^{3}\spek
+\kappa ^4 \bigg[\frac{315 (N-3) (N-2) (N+2) 
(N+3) \zeta (3)^2}{4 (N^2+1)^2 (N^2+3) (N^2+5)}\spek
-\frac{245 (N+2) (40 
N^6+172 N^5-564 N^4+120 N^3+1185 N^2-480 N-945) \zeta (7)}{32 N^3 
(N^2+1) (N^2+3) (N^2+5)}\bigg]\,\kappa^{4}+\cdots,
\notag \\
\log\, & F^{\mathbf D\,(2)}(\kappa; N) = 
-\frac{9  (2 N^2-3)\, \zeta (3)}{N (N^2+1)}\,\kappa^{2}
+\frac{50  (5 N^4-2 N^3-15 N^2+8 N+15)\, \zeta (5)}{N^2 (N^2+1) (N^2+3)}\,\kappa^{3}\spek
+\bigg[\frac{315 (N-3) (N-2) (N+2) (N+3) \zeta (3)^2}{(N^2+1)^2 (N^2+3) 
(N^2+5)}\spek
-\frac{735 (42 N^6-40 N^5-168 N^4+240 N^3+315 N^2-320 N-315) 
\zeta (7)}{8 N^3 (N^2+1) (N^2+3) (N^2+5)}\bigg]\,\kappa^{4}+\cdots,
\notag \\
\log\, & F^{\mathbf E\,(2)}(\kappa; N) = 
0\cdot\zeta(3)\,\kappa^{2}
-\frac{100 (N-2) (N+2)\, \zeta (5)}{N (N^2+1) 
(N^2+3)}\,\kappa^{3}+\frac{3675  (N-2) (N+2) (N^2-2) \,\zeta (7)}{N^2 
(N^2+1) (N^2+3) (N^2+5)}\,\kappa^{4}\spek
-\frac{15876  (N-2) (N+2) (7 N^4-25 N^2+36)\, \zeta (9)}{N^3 
(N^2+1) (N^2+3) (N^2+5) (N^2+7)}\,\kappa^{5}
+\cdots,
\end{align} 
and, for the $\tr(a^{3})$ tower, 
{\small
\begin{align}
\la{3.36}
\log\, & F^{\mathbf A\,(3)}(\kappa; N) = 
-\frac{9 \, \zeta (3)}{2}\,\kappa^{2}+\frac{25 (N-1) (N+1) (2 
N^4+45 N^2+105) \,\zeta (5)}{N (N^2+5) (N^2+7) (N^2+9)}\,\kappa^{3}\spek
-\frac{1225  (8 N^8+260 N^6+281 N^4-378 N^2+693) \,\zeta (7)}{16 N^2 
(N^2+5) (N^2+7) (N^2+9) (N^2+11)}\,\kappa^{4}+\cdots,
\notag \\
\log\, & F^{\mathbf B\,(3)}(\kappa; N) = 
-\frac{9 (N-3) (N^4-N^3+9 N^2-35 N-70)\, \zeta (3)}{4 N 
(N^2+5) (N^2+7)}\,\kappa^{2}\spek
+\frac{25  (2 N^7-10 N^6+31 N^5-320 N^4-168 
N^3+1800 N^2+1575 N-3150)\, \zeta (5)}{2 N^2 (N^2+5) (N^2+7) 
(N^2+9)}\,\kappa^{3}\spek
+\bigg[\frac{315 (N-3) (N+3) (N^6+90 N^4-471 N^2-8260) 
\,\zeta (3)^2}{4 (N^2+5)^2 (N^2+7)^2 (N^2+9) (N^2+11)}\spek
-\frac{245 (40 
N^9-252 N^8+580 N^7-13860 N^6-19475 N^5+87822 N^4+204030 N^3-238140 
N^2-440055 N+436590)\, \zeta (7)}{32 N^3 (N^2+5) (N^2+7) (N^2+9) 
(N^2+11)}\bigg]\,\kappa^{4}+\cdots,
\notag \\
\log\, & F^{\mathbf C\,(3)}(\kappa; N) = 
-\frac{9 (N+3) (N^4+N^3+9 N^2+35 N-70)\, \zeta (3))}{4 N 
(N^2+5) (N^2+7)}\kappa^{2}\spek
+\frac{25  (2 N^7+10 N^6+31 N^5+320 N^4-168 
N^3-1800 N^2+1575 N+3150)\, \zeta (5)}{2 N^2 (N^2+5) (N^2+7) 
(N^2+9)}\,\kappa^{3}\spek
+\bigg[\frac{315 (N-3) (N+3) (N^6+90 N^4-471 N^2-8260) \,
\zeta (3)^2}{4 (N^2+5)^2 (N^2+7)^2 (N^2+9) (N^2+11)}\spek
-\frac{245 (40 
N^9+252 N^8+580 N^7+13860 N^6-19475 N^5-87822 N^4+204030 N^3+238140 
N^2-440055 N-436590)\, \zeta (7)}{32 N^3 (N^2+5) (N^2+7) (N^2+9) 
(N^2+11)}\bigg]\,\kappa^{4}+\cdots,
\notag \\
\log\, & F^{\mathbf D\,(3)}(\kappa; N) = 
-\frac{9 (2 N^4+31 N^2-105) \, \zeta (3)}{N (N^2+5) 
(N^2+7)}\,\kappa^{2}+\frac{50 (5 N^6-6 N^5+160 N^4-114 N^3-900 N^2+840 
N+1575) \,\zeta (5)}{N^2 (N^2+5) (N^2+7) (N^2+9)}\,\kappa^{3}\spek
+\bigg[\frac{315 
(N-3) (N+3) (N^6+90 N^4-471 N^2-8260)\, \zeta (3)^2}{(N^2+5)^2 
(N^2+7)^2 (N^2+9) (N^2+11)}\spek
-\frac{2205 (14 N^8-40 N^7+770 N^6-1160 
N^5-4879 N^4+11440 N^3+13230 N^2-24640 N-24255)\, \zeta (7)}{8 N^3 
(N^2+5) (N^2+7) (N^2+9) (N^2+11)}\bigg]\,\kappa^{4}+\cdots,
\notag \\
\log\, & F^{\mathbf E\,(3)}(\kappa; N) = 
0\cdot\zeta(3)\,\kappa^{2}
-\frac{300 (N^4+19 N^2-140) \zeta (5)}{N (N^2+5) (N^2+7) 
(N^2+9)}\,\kappa^{3}+\frac{11025  (N-2) (N+2) (N^4+33 N^2-154) \,\zeta 
(7)}{N^2 (N^2+5) (N^2+7) (N^2+9) (N^2+11)}\,\kappa^{4}\spek
-\frac{7938 (41 
N^8+1745 N^6-19474 N^4+73200 N^2-123552) \zeta (9)}{N^3 (N^2+5) 
(N^2+7) (N^2+9) (N^2+11) (N^2+13)}\,\kappa^{5}+\cdots.
\end{align} 
}
Of course, specialization of (\ref{3.35}) to $SU(2)$ is in full agreement with (\ref{1.7}).
Also, specialization of (\ref{3.35}) and (\ref{3.36}) to 
$SU(3)$ agrees with (\ref{1.9}) and (\ref{1.10}). The $SU(3)$ and $SU(4)$ expansions at order $\mc O(\kappa^{10})$
are collected in App.~\ref{app:specialization}.

\paragraph{Remark 1:} There is a simple formal duality between $\mathbf{B}$ and $\mathbf{C}$ models expressed by the relations
\be
\log F^{\mathbf B\,(\Delta)}(\kappa; N)  = \log F^{\mathbf C\,(\Delta)}(-\kappa; -N),\qquad \Delta=2,3,
\ee
that are consequence of the specific matter content in Tab.~\ref{tab:scft}.

\paragraph{Remark 2:} The expansions (\ref{3.35}) and (\ref{3.36}) show that the two-point scaling functions do not exponentiate in the simple
way as in the $SU(2)$ theory, i.e. $\log F^{(\Delta)}$ is not a simple series linear in the $\zeta$-numbers. 
This makes any attempt to a full resummation little promising. Nevertheless, our approach makes it easy to resum special
contributions. The example of the first non-trivial terms, i.e. those proportional to simple powers of $\zeta(5)$,  is treated
in App.~\ref{app:zeta5}. 

\section{Three loop diagram analysis in $\mc N=1$ superspace}
\la{sec:diagrams}

As first mentioned in \cite{Bourget:2018obm}, there is a special interest in understanding the topology of Feynman diagrams in the large charge limit of chiral correlators in $\cN=2$ theories with $SU(N)$ gauge group. Here we will show that the diagrams contributing to the 
double scaling limit are specific maximally non-planar diagrams.

We consider a four dimensional Euclidean spacetime and follow the $\mathcal{N}=1$ superspace formalism as well as the diagrammatic difference between $\mathcal{N}=2$  and $\mathcal{N}=4$. Indeed the scaling functions in (\ref{1.2})
precisely account for the matter content of the difference theory. 
We refer to App.~\ref{app:rules} for the complete expression of the Lagrangian and Feynman rules (see \cite{Billo:2019fbi} for a more detailed description of the tools). We limit our analysis to the diagrams contributing the maximal transcendentality at each perturbative order.

\subsection{Tree level}

Our previous discussion has concerned correlation functions for a specific class of chiral operators that we can 
generically write as $\cO_{\Delta,n}(x) = \Phi_\Delta \;(\tr\varphi^2)^n (x)$, where $\Phi_\Delta = \tr \varphi^\Delta$. Such operators have scaling dimension $\Delta+2n$ and can be written as
\begin{align}\label{4.1}
\cO_{\Delta,n}= R^{(\cO)}_{a_1\dots a_{\Delta+2n}} \varphi^{a_1}\dots\varphi^{a_{\Delta+2n}},
\end{align}
where $R^{(\cO)}$ is a totally symmetric tensor, whose expression is encoded in the trace structure\footnote{Note the difference between $R^{(\cO)}$ defining the full operator $\cO_{\Delta,n}$ and $R^{(\Delta)}$ which defines $\Phi_\Delta$, and so specifies the tower}. 

We study the flat space correlation function between a chiral and an antichiral operator. According to (\ref{2.4}), we can write
\begin{align}\label{4.2}
\langle\cO_{\Delta,n}(x)\overline\cO_{\Delta,n}(0)\rangle=  \frac{G_{\mathcal{O} \bar{\mathcal{O}}}(g,n,N)}{(4\pi^2x^2)^{\Delta+2n}}~,
\end{align}
where the 2-pt coefficient $G_{\mathcal{O} \bar{\mathcal{O}}}$ is captured by the matrix model. Our aim is to provide a 
direct field theory analysis that identifies all the Feynman diagrams contributing to the correlator \eqref{4.2}
and surviving the double scaling limit (\ref{1.1}).

\begin{figure}[ht]
\begin{center}
\includegraphics[scale=0.7]{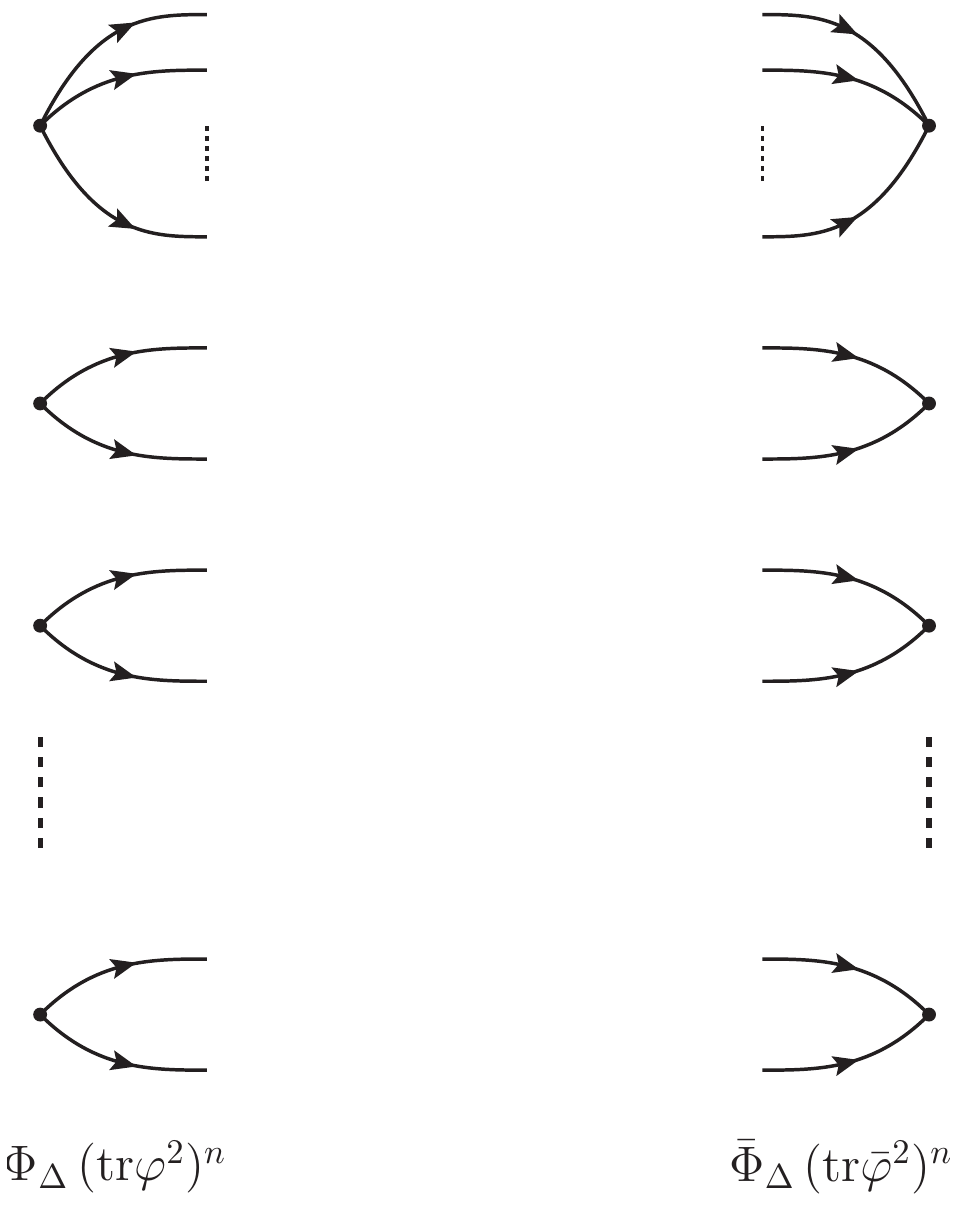}
\end{center}
\caption{The $\cN=4$ result is the full tree level contraction of this diagram. The chiral operator is placed in $x$, the anti-chiral in $0$.}
\label{Fig::N4}
\end{figure}

We start with the $\cN=4$ result for the correlator $\langle\cO_{\Delta,n}(x)\overline\cO_{\Delta,n}(0)\rangle_{\cN=4}$, 
which corresponds the  denominator of \eqref{1.2}. 
In this case the correlator is not only of the form \eqref{4.2}, but also is closed with tree level propagators only
\begin{equation}
\Vev{\varphi^a(x)\bar\varphi^b(0)} = \frac{\delta^{ab}}{4\pi^2x^2},
\end{equation}
so that it reads
\begin{equation}
\langle\cO_{\Delta,n}(x)\overline\cO_{\Delta,n}(0)\rangle_{\cN=4} =\,\left(\frac{g^2}{8\pi^2}\right)^{2n+\Delta}\!\!\! (2n+\Delta)!\; R^{(\cO)}\cdot R^{(\cO)} \frac{1}{(4\pi^2x^2)^{\Delta+2n}}~,
\end{equation}
namely it corresponds to the full contraction of the $R^{(O)}$ tensors, as reported in Figure \ref{Fig::N4}. 

Even though the Feynman diagram analysis can be pursued for any $\Phi_{\Delta}$, 
in this paper we will write explicit results for the towers $\Phi_2$ and $\Phi_3$. 
Note that for $\Delta=2$ we simply reabsorb $\Phi_2 = \tr \varphi^2$ inside $(\tr\phi^2)^n$ 
in order to simplify the notation. Thus the operators we focus on are 
\begin{align}
\cO_{2,n}= (\tr\varphi^2)^n~, \hspace{1cm} \cO_{3,n}= \tr\varphi^3(\tr\varphi^2)^n~.
\end{align}
Their tree level contraction, dropping the space-time dependence, are (see \cite{Bourget:2018obm, Beccaria:2018xxl})
\begin{align}\label{4.6}
G^{(2)}_{\cN=4}(g,n,N)&= \,\left(\frac{g^2}{8\pi^2}\right)^{2n} n!\;\frac{\Gamma \left(\frac{N^2-1}{2}+n\right)}{\Gamma \left(\frac{N^2-1}{2}\right)}~,\\
G^{(3)}_{\cN=4}(g,n,N)&=\,\left(\frac{g^2}{8\pi^2}\right)^{2n+3} \frac{1}{4}\, d_{a_1a_2a_3}\;\frac{1}{4}\,d^{a_1a_2a_3}~ n!\,\frac{\Gamma \left(\frac{N^2-1}{2}+n+3\right)}{\Gamma \left(\frac{N^2-1}{2}+3\right)},
\end{align}
where $\frac{1}{4}d_{abc} := R^{(3)}$ is the totally symmetric 3-indices tensor defining $\tr\varphi^3$ (see \eqref{dabc}).

The generalization for any $\cO_{\Delta,n}$ easily follows. This operator is specified by a certain $\Phi_\Delta$, thus by a totally symmetrized tensor $R^{(\Delta)}$. Its tree level contraction turns out to be:
\begin{align}\label{4.7}
G^{(\Delta)}_{\cN=4}(g,n,N)&=\,\left(\frac{g^2}{8\pi^2}\right)^{2n+\Delta} R^{(\Delta)}\cdot R^{(\Delta)} ~ n!\;\frac{\Gamma \left(\frac{N^2-1}{2}+n+\Delta\right)}{\Gamma \left(\frac{N^2-1}{2}+\Delta\right)}.
\end{align}

\subsection{$\cN=2$ corrections and maximally non-planar diagrams}

\begin{figure}[htb]
\hspace{2.5cm}
\includegraphics[scale=0.5]{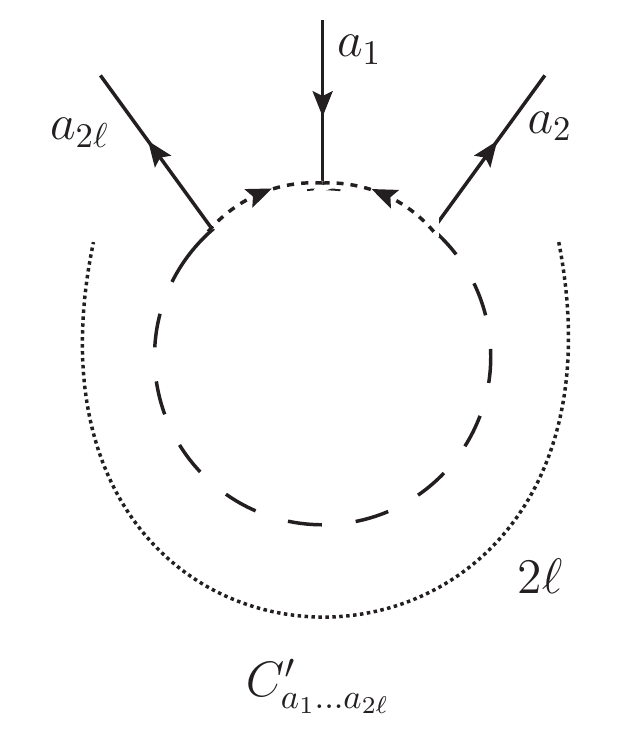}
\hspace{1.5cm}
\includegraphics[scale=0.5]{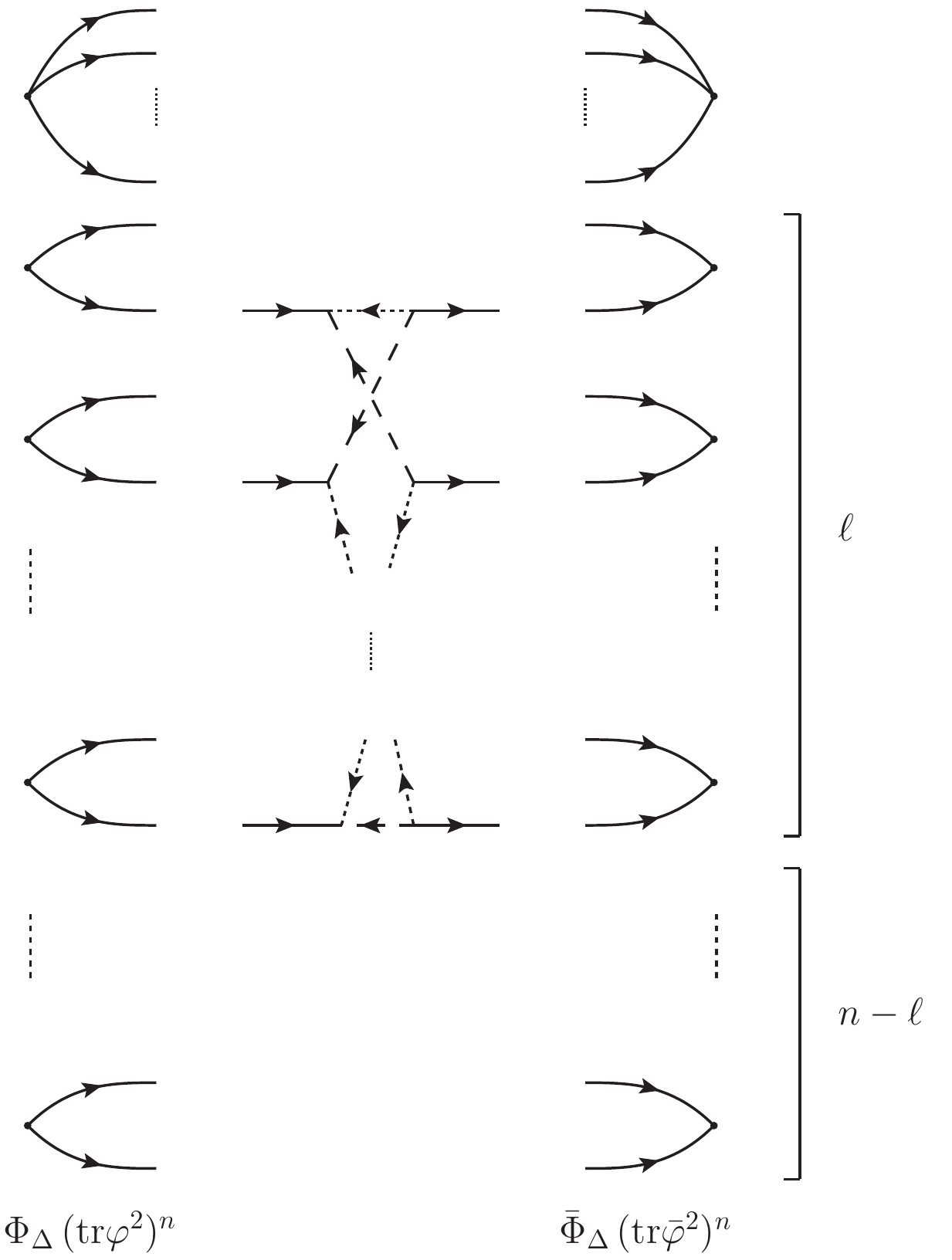}
\caption{Generic $2\ell$-legs diagram with its color factor in the difference theory. The straight lines represent $\varphi,\bar{\varphi}$ fields, the dashed line generically represents the hypermultiplet loop. On the right we see how to insert it in the maximal non planar way, which gives the leading order in the double scaling limit}
\label{Fig::lloops}
\end{figure}

Our goal here is to identify the class of diagrams contributes to the leading order in $n$ providing the double scaling limit for each perturbative order. We \emph{claim} a general behavior for any operator $\cO_{\Delta,n}$ and for any transcendentality $\zeta(2\ell-1)$ contributing to $g^{2\ell}$ order, following a very simple reasoning. If we want to reproduce the leading terms $g^{2\ell} n^\ell$, at $g^{2\ell}$ order there is a unique way to obtain a $n^\ell$ term to achieve the correct double scaling limit, that is a diagram with a hypermultiplet loop with $2\ell$ adjoint chiral legs. It is built up with $\ell~$ $\widetilde{Q}\Phi Q$  and $\ell~$ $ Q^\dagger\Phi^\dagger \widetilde{Q}^\dagger$ vertices, represented in Figure \ref{fig:Feynmatter}. Each vertex brings a $g$ factor. Then, the only way to get a $n^\ell$ scaling is to insert this diagram inside  $\ell$ out of $n$ pairs of traces, see Figure \ref{Fig::lloops}. Hence, the only contribution in the double scaling limit comes from this $2\ell$-leg diagram inserted in a \emph{ maximally non-planar way}.

We motivate this statement and we provide a formal computation for the generic $\ell$-loops contribution and for a general $\cO_{\Delta,n}$ tower. In the next subsections we prove it for the two loops ($\ell=2$) and three loops ($\ell=3$) cases, specifically for the $\cO_{2,n}$ and $\cO_{3,n}$ towers and making a direct comparison with the matrix model computations.

The diagram in Figure \ref{Fig::lloops} can be factorized into three contributions: the 
Feynman loop integral $W_{2\ell}(g,x)$, a symmetry factor $S(\ell,n)$ and the color factor $K^{\Phi}(N,\ell)$. 
We discuss separately each of them.

\paragraph{Loop integral}
We have a factor of $(\pm i \sqrt{2} g)$ for each vertex, while the superspace integral can be mapped to the evaluation of the L-loop contribution of ladder diagrams to the four-point function in $\phi^3$-theory, which was computed in \cite{Usyukina:1993ch}. This analogy was exploited in Appendix B of \cite{Billo:2017glv} for the $\ell=2,3$ cases, in general this integral is always finite and yields
\begin{align}
W_{2\ell}(g,x)=2^{\ell}g^{2\ell}\frac{(2\ell)!}{(\ell!)^2}\zeta(2\ell-1)\frac{(-1)^\ell}{(16\pi^2)^\ell}\frac{1}{(4\pi^2x^2)^{2\ell}} = (-1)^\ell \frac{(2\ell)!}{(\ell!)^2} \zeta(2\ell-1)\left(\frac{g^2}{8\pi^2}\right)^\ell\frac{1}{(4\pi^2x^2)^{2\ell}}~,
\end{align}
where $\zeta(2\ell-1)$ is the Riemann zeta function, which counts the transcendentality order of the perturbative expansion.
Note that the insertion of these diagrams preserves the spacetime structure of the propagator, so that the structure of the correlator \eqref{4.2} is correctly preserved. Therefore, from now on, we simply drop the spacetime dependence.

\paragraph{Symmetry factor}
The important contribution is $\ell!(\ell-1)!$, due to the number of independent hypermultiplet loops.
Then, the only way to obtain a leading $n^\ell$ contribution is to insert the $2\ell$-leg diagram inside the maximal number of $\tr\varphi^2 \tr\bar{\varphi}^2$ pairs. So we have
\begin{align}
S(\ell,n)= \ell !\; (\ell-1)!\;\binom{n}{\ell}^2\ .
\end{align}
\paragraph{Color factor}
The color factor is the more involved part, since we are considering the maximal non-planar diagram. We provide a recipe to capture the leading order in $n$ and we test it for the first non trivial orders.\\
The color factor from the open $2\ell$-legs diagram in the difference theory precisely reproduces the trace combination $C'_{a_1\dots a_{2\ell}}$ that we already found in the matrix model expansion \eqref{defC}. (see App.~\ref{app:rules} for the explanation of its diagrammatic origin).
After the non-planar insertion of this diagram like in Figure \ref{Fig::lloops}, the leading order will be the contraction of the $C'_{a_1\dots a_{2\ell}}$ color factor with the $\Phi_\Delta \bar{\Phi}_\Delta$ part of the correlator, defined by the tensor $R^{(\Delta)}$. We clarify this statement with the two specific examples.\\
The $\Phi^{(2)}$ result is particularly easy, $C'_{a_1\dots a_{2\ell}}$ can be contracted only with color delta functions. We obtain a totally contracted, fully symmetrized tensor
\begin{align}\label{C2l}
C'_{(a_1a_1\dots a_{\ell}a_{\ell})}:=C'_{(2\ell)}~.
\end{align}
The $\Phi^{(3)}$ result is more involved, since we need to contract $C'_{a_1\dots a_{2\ell}}$ with the two $R^{(3)}$ tensors defining the operators. We obtain a tensor that can be formally written as
\begin{align}\label{R3C2lR3}
R^{(3)}\cdot C'_{2\ell} \cdot R^{(3)}~.
\end{align}
In the next section we compute this tensor in the $\ell=2$ and $\ell=3$ cases.
After this contraction we are left with $n-\ell$ pairs of untouched traces that will be contracted analogously to the $\cN=4$ case.
After the ratio with the $\cN=4$ contribution \eqref{4.6}, we can write the explicit results for the $\Phi^{(2)}$ and $\Phi^{(3)}$ towers
\begin{align}
K^{(2)}(N,\ell) &= \frac{(n-\ell)!}{n!}\;\frac{\Gamma\left(\frac{N^2-1}{2}\right)}{\Gamma\left(\frac{N^2-1}{2}+\ell\right)} C'_{2\ell}, \notag \\
K^{(3)}(N,\ell) &= \frac{(n-\ell)!}{n!}\;\frac{\Gamma\left(\frac{N^2-1}{2}+3\right)}{\Gamma\left(\frac{N^2-1}{2}+\ell+3\right)}\frac{R^{(3)}\cdot C'_{2\ell} \cdot R^{(3)}}{R^{(3)}\cdot R^{(3)}}~.
\end{align}
The generalization for a generic tower $\Phi^{(\Delta)}$ immediately follows
\begin{align}
K^{\Delta}(N,\ell)= \frac{(n-\ell)!}{n!}\;\frac{\Gamma\left(\frac{N^2-1}{2}+\Delta\right)}{\Gamma\left(\frac{N^2-1}{2}+\ell+\Delta\right)}\frac{R^{(\Delta)}\cdot C'_{2\ell} \cdot R^{(\Delta)}}{R^{(\Delta)}\cdot R^{(\Delta)}} ~.
\end{align}

\paragraph{Total result}
In total we get a very compact expression for the generic $\ell$-loops result with transcendentality $\zeta(2\ell-1)$ of the correlator \eqref{1.2} in the double scaling limit
\begin{align}\label{4.14}
F^{(2)}\big|_{\zeta(2\ell-1)} &=  (-1)^\ell \frac{(2\ell)!}{(\ell!)^2}\frac{\zeta(2\ell-1)}{2^\ell \, \ell}\, \kappa^\ell \;\frac{\Gamma\left(\frac{N^2-1}{2}\right)}{\Gamma\left(\frac{N^2-1}{2}+\ell\right)}  C'_{2\ell}, \notag \\
F^{(3)}\big|_{\zeta(2\ell-1)} &=  (-1)^\ell \frac{(2\ell)!}{(\ell!)^2}\frac{\zeta(2\ell-1)}{2^\ell \, \ell}\, \kappa^\ell \;\frac{\Gamma\left(\frac{N^2-1}{2}+3\right)}{\Gamma\left(\frac{N^2-1}{2}+\ell+3\right)}  \frac{R^{(3)}\cdot C'_{2\ell} \cdot R^{(3)}}{R^{(3)}\cdot R^{(3)}}~.
\end{align}
The generalization for a generic tower $\Phi^{(\Delta)}$ is 
\begin{align}\label{4.15}
F^{(\Delta)}\big|_{\zeta(2\ell-1)}=  (-1)^\ell \frac{(2\ell)!}{(\ell!)^2}\frac{\zeta(2\ell-1)}{2^\ell \, \ell}\, \kappa^\ell \;\frac{\Gamma\left(\frac{N^2-1}{2}+\Delta\right)}{\Gamma\left(\frac{N^2-1}{2}+\ell+\Delta\right)}\frac{R^{(\Delta)}\cdot C'_{2\ell} \cdot R^{(\Delta)}}{R^{(\Delta)}\cdot R^{(\Delta)}}~.
\end{align}
Now we can enforce this statement providing an explicit computation at two and three loops order for the $\Phi^{(2)}$ and $\Phi^{(3)}$ towers. In particular, we will see that the color factor worked out in \eqref{4.14} precisely reproduces the matrix model results.

\subsection{Two loop diagrams: $\left. F^{(\Delta)}\right|_{\zeta(3)}$}

\begin{figure}[htb]
\begin{center}
\includegraphics[scale=0.6]{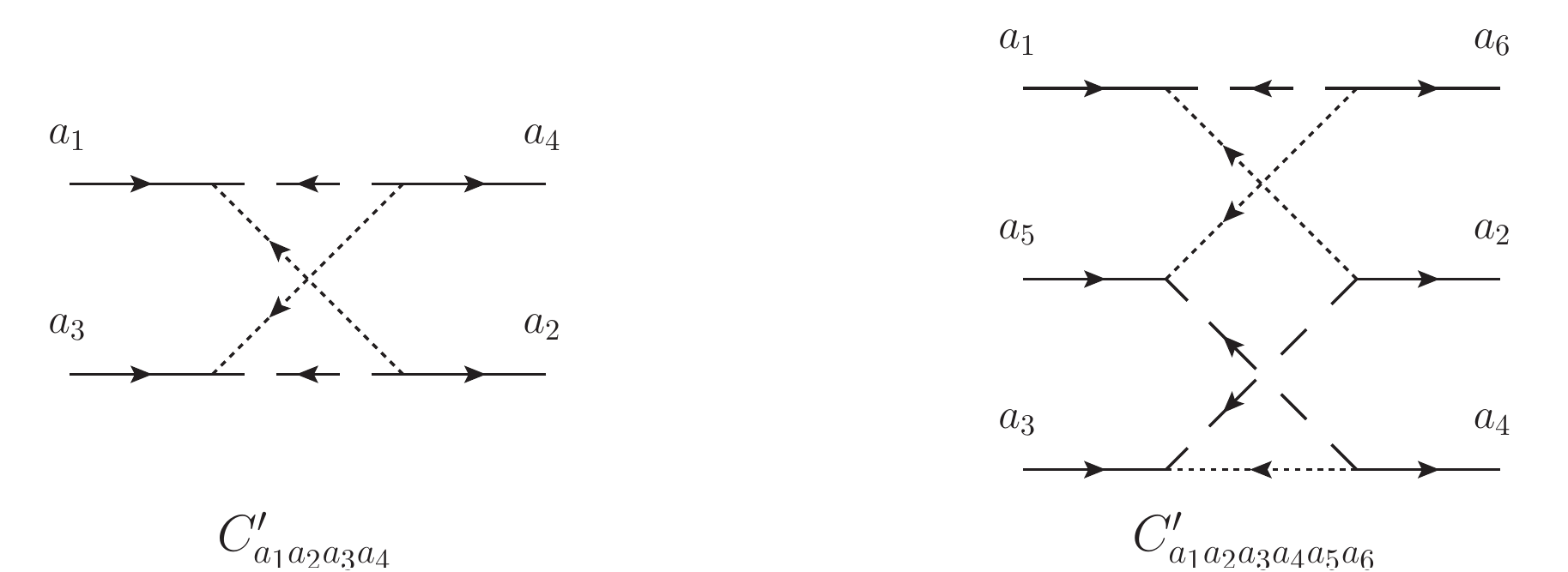}
\end{center}
\caption{Box and exagon diagram contributing to 2 and 3 loops order, with their color factors}
\label{Fig::23loops}
\end{figure}

As explained before, the unique contribution at $g^4$ order in the double scaling limit is represented by the first diagram of Figure \ref{Fig::23loops}.
In the $\Phi^{(2)}$ tower the color factor of this diagram must be totally self-contracted, generating a totally symmetrized expression (following \eqref{C2l}):
\begin{equation}\label{C4}
C'_{(a,a,b,b)}=C'_{(4)}
\end{equation}
The $\Phi^{(3)}$ tower is defined by the tensor $R^{a_1a_2a_3}=\frac{1}{4}d^{a_1a_2a_3}$. The total color factor will be a sum over all the possible way of contracting $C'$ with two $R^{(3)}$ tensors
\begin{align}\label{R3C4R3}
R^{(3)}\cdot C'_{4}\cdot R^{(3)} &= \frac{1}{4}d_{a_1a_2a_3} \frac{1}{4}d^{a_1a_2a_3}C'_{(b,b,c,c)} + \frac{1}{4}d_{a_1a_2b_1} \frac{1}{4}d^{a_1a_2b_2}C'_{(b_1,b_2,c,c)}\notag \\
&+ \frac{1}{4}d_{a_1b_1c_1} \frac{1}{4}d^{a_1b_2c_2}C'_{(b_1,b_2,c_1,c_2)}.
\end{align}

 \begin{table}[H]
\begin{center}
\be
\def\arraystretch{1.5}
\begin{array}{ccc}
\toprule
\textsc{theory}\phantom{\bigg|} & C'_{(4)} & \dfrac{R^{(3)}\cdot C'_{4}\cdot R^{(3)}}{R^{(3)}\cdot R^{(3)}} \\
\midrule
\mathbf{A}\phantom{\bigg|} & -\dfrac{3}{2}(N^4-1) & -\dfrac{3}{2} \left(N^2+5\right) \left(N^2+7\right) \\
\mathbf{B}\phantom{\bigg|} & -\dfrac{3(N^2-1)(N+1)(N-2)(N-3)}{4N}~~ & ~
-\dfrac{3 (N-3) \left(N^4-N^3+9 N^2-35 N-70\right)}{4 N} \\
\mathbf{C}\phantom{\bigg|} & -\dfrac{3(N^2-1)(N-1)(N+2)(N+3)}{4N}~~ & ~
-\dfrac{3 (N+3) \left(N^4+N^3+9 N^2+35 N-70\right)}{4 N} \\
\mathbf{D}\phantom{\bigg|} & -\dfrac{3(2N^2-3)(N^2-1)}{N} & -\dfrac{3 \left(2 N^4+31 N^2-105\right)}{N} \\
\!\mathbf{E} & 0 & 0 \\
\bottomrule
\end{array}\notag 
\ee
\end{center}
\caption{Theory dependent coefficients determining the two-loop $\zeta(3)$ contribution to the  
scaling functions $F^{(\Delta)}$ for the two towers with $\Delta=2,3$.}
\label{tab:RC4R}
\end{table}

To evaluate \eqref{C4} and \eqref{R3C4R3} we follow the procedure of App. \ref{app:rules}, using \eqref{colfact2} and \eqref{fussion}. The final result in terms of rational functions in $N$ is obtained using FormTracer \cite{Cyrol:2016zqb}.\\
Substituting $\ell=2$ inside \eqref{4.15}, the two loops results for the two towers are
\begin{align}
F^{(2)}\big|_{\zeta(3)} &=  \frac{3\,\zeta(3)\, \kappa^2}{N^4-1} C'_4, \notag \\
F^{(3)}\big|_{\zeta(3)} &= \frac{3\, \zeta(3)\,\kappa^2}{(N^2+5)(N^2+7)} \frac{R^{(3)}\cdot C'_{4}\cdot R^{(3)}}{R^{(3)}\cdot R^{(3)}},
\end{align}
where the color factors for all the $SU(N)$ conformal theories are reported in Table \ref{tab:RC4R}.
We see a perfect match with the $\zeta(3)$ terms of the matrix model results in (\ref{3.35}).

\subsection{Three loop diagrams: $\left. F^{(\Delta)}\right|_{\zeta(5)}$}

The three loops case is technically more involved, but conceptually 
it is all encoded inside the generalized \eqref{4.15} formula. 
Now the diagram to be inserted has an exagon shape, see Figure \ref{Fig::23loops} inserted in the maximally non-planar way.  
Substituting $\ell=3$ inside \eqref{4.15}, the three-loops 
results for the two towers are
\begin{align}
F^{(2)}\big|_{\zeta(5)} &= -\frac{20}{3}\zeta(5)\kappa^3 \frac{1}{(N^4-1)(N^2+3)} C'_6 , \notag \\
F^{(3)}\big|_{\zeta(5)} &=  -\frac{20}{3}\zeta(5)\kappa^3\frac{1}{(N^2+5)(N^2+7)(N^2+9)} 
\frac{R^{(3)}\cdot C'_{6}\cdot R^{(3)}}{R^{(3)}\cdot R^{(3)}} ,
\end{align}
where again the color factors are explicitly computed for all the $SU(N)$ conformal theories, using the same procedure as before, and are reported in Table \ref{tab:RC6R} for both the towers.

\begin{table}[htb]
\begin{center}
\be
\def\arraystretch{1.5}
\begin{array}{ccc}
\toprule
\textsc{theory} &  C'_{6}  & \dfrac{R^{(3)}\cdot C'_{6}\cdot R^{(3)}}{R^{(3)}\cdot R^{(3)}} \\
\midrule
\mathbf{A} & -\frac{15(N^4-1)(2N^2-1)}{4 N} &
-\frac{15 (N-1) (N+1) \left(2 N^4+45 N^2+105\right)}{4 N} \\
\mathbf{B} & -\frac{15 (N-2)(N^2-1) (2N^4-6N^3-15N^2+15)}{8 N^2} & 
-\frac{15 \left(2 N^7-10 N^6+31 N^5-320 N^4-168 N^3+1800 N^2+1575 N-3150\right)}{8 N^2} \\
\mathbf{C} & -\frac{15 (N+2)(N^2-1) (2N^4+6N^3-15N^2+15)}{8 N^2} & 
-\frac{15 \left(2 N^7+10 N^6+31 N^5+320 N^4-168 N^3-1800 N^2+1575 N+3150\right)}{8 N^2} \\
\mathbf{D} & -\frac{15(N^2-1)(5N^4-2N^3-15N^2+8N+15)}{2 N^2} & 
-\frac{15 \left(5 N^6-6 N^5+160 N^4-114 N^3-900 N^2+840 N+1575\right)}{2 N^2}\\
\mathbf{E} & \frac{15 (N^2-1)(N^2-4)}{N} & \frac{45 \left(N^4+19 N^2-140\right)}{N} \\
\bottomrule
\end{array}\notag
\ee
\end{center}
\caption{Theory dependent coefficients determining the three loop  $\zeta(5)$ contribution to the  
scaling functions $F^{(\Delta)}$ for the two towers with $\Delta=2,3$.}
\label{tab:RC6R}
\end{table}

\noindent Again we find a perfect match with the $\zeta(5)$ coefficients of the matrix model expressions \eqref{3.35}

\subsection{Summary of the diagrammatical analysis}

In summary, we have confirmed our previous claim
by explicit calculations and comparison with the matrix model results (\ref{3.35}). The 
$\zeta(2\ell-1)\,g^{2\ell}$ contributions to the scaling function $F^{(\Delta)}(\kappa; N)$ 
for $\ell=2,3$ and $\Delta=2,3$
come indeed from a diagram with a hypermultiplet loop with $2\ell$ adjoint chiral legs 
that is inserted into the tree diagram, see Figure \ref{Fig::lloops}, in 
a maximally non-planar way. The pattern is reasonably preserved at higher perturbative orders, since this is the only way to produce the necessary power of $n$ needed to 
survive the double scaling limit. This analysis provides a intriguing evidence of the duality between the rank of the gauge group $N$ and the number of the operator insertions $n$, as suggested in \cite{Grassi:2019txd}.

\section{One-point  Wilson functions: collecting more data for $SU(3)$ and $SU(4)$ theories}
\la{sec:wilson-exp}

The second class of observables that we are going to discuss are one-point Wilson functions for which we 
want to analyze the double scaling limit. 
As we pointed out in the Introduction, the available data for the one-point Wilson scaling function
is limited to the $SU(2)$ case and the  $\mathbf A$ model with $SU(3)$ gauge group. 
In this section, we exploit localization to collect additional explicit 
data for all models in the $SU(3)$ and $SU(4)$ theories and for both the $(\tr \varphi^{2})^{n}$ 
and $\tr \varphi^{3}\,(\tr \varphi^{2})^{n}$
large R-charge chiral primaries. This work will be useful to formulate some conjectures 
that we shall prove by using the higher rank  dual matrix model.

\subsection{One-point Wilson scaling functions for the two $SU(3)$ theories}

Let us recall that in the $SU(3)$  $\mathbf A$ model one obtains at finite $n$ \cite{Beccaria:2018owt} ($\zeta_{n}\equiv \zeta(n)$) \footnote{
We \underline{only} write the contributions that are going to 
survive in the scaling limit. These are the transcendentality structures that appear in the expansion 
of $\exp(-S_{\rm int})$. That this happens in general will be proved later.}
\begin{align}
\la{5.1}
F_{\mc W, n}^{\mathbf{A}\,(2)}(g; 3) &= 
1-\frac{9\,n\,(n+4)\,\zeta_{3}}{32\,\pi^{4}}\,g^{4}
+\frac{25\,n\,(254+927\,n+850\,n^{2}+375\,n^{3}+42\,n^{4})\,\zeta_{5}}{2304\,(2+3\,n+3\,n^{2})\,\pi^{6}}
\,g^{6}+\cdots,
\end{align}
and then, taking the limit (\ref{1.4}), one has 
\begin{align}
\la{5.2}
F^{\mathbf{A}\,(2)}_{\mc W}(\kappa; 3)  &= 1-\frac{9\,\zeta_{3}}{2}\,\kappa^{2}
+\frac{175\, \zeta_{5}}{18}\,\kappa^{3}+
\bigg(\frac{81\, \zeta_{3}^2}{8}-\frac{12005\, \zeta_{7}}{576}\bigg)\,\kappa^{4}\notag \\
& +\bigg(\frac{1491\, \zeta_{9}}{32}-\frac{175\, \zeta_{3}\,\zeta_{5}}{4}\bigg)\,\kappa^{5}+
 \bigg(-\frac{243\, \zeta_{3}^3}{16}+\frac{30625\, 
\zeta_{5}^2}{648}+\frac{12005\, \zeta_{3}\,\zeta_{7}}{128}-\frac{2247091 
\zeta_{11}}{20736}\bigg)\,\kappa ^6+\cdots,
\end{align}
that may be exponentiated in the simple form (\ref{1.11}).
Repeating the calculations in \cite{Beccaria:2018owt} for the  $\mathbf{B}$ model we find 
\footnote{We checked until $n=14$ that is dimension 28 where mixing is rather hard. This is possible because $N$ is fixed
as explained in \cite{Beccaria:2018owt}.}
\begin{align}
\la{5.3}
& F^{\mathbf{B}\,(2)}_{\mc W, n}(g; 3) = 
1-\frac{25\, n\, (12 n^4+3 n^3+104 n^2-117 n+142)\, \zeta_{5}}{1152 \,
\pi ^6 (3 n^2+3 n+2)}\,g^{6}\notag \\
& +\frac{1225\, n\, (12 n^5+27 n^4+110 n^3+109 
n^2-74 n+296) \,\zeta_{7}}{36864\, \pi ^8\, (3 n^2+3 n+2)}\,g^{8}\notag \\
& -\frac{147 \, n\, (108 n^6+459 n^5+1443 n^4+3135 n^3+1577 n^2+2646 n+5032) \, \zeta_{9}}
{32768 \, \pi ^{10} \,(3 n^2+3 n+2)}\,g^{10}+\cdots
\end{align}
In the limit (\ref{1.4}) we get 
\begin{align}
\la{5.4}
F^{\mathbf{B}}_{\mc W}(\kappa; 3) &= 1-\frac{50 \, \zeta_{5}}{9}\,\kappa^{3}+\frac{1225 \,\zeta_{7}}{36}
\,\kappa^{4}-\frac{1323\,\zeta_{9}}{8}\,\kappa^{5}+\bigg(\frac{1250 \,\zeta_{5}^2}{81}
+\frac{1960805\,\zeta_{11}}{2592}\bigg)\,\kappa^{6}\notag \\
&+\bigg(-\frac{30625\, \zeta_{5}\,\zeta_{7}}{162}-\frac{17688385 \zeta_{13}}{5184}\bigg)\,\kappa^{7}+\cdots,
\end{align}
and, remarkably, (\ref{5.4}) can once again be written in exponential form in terms of simple $\zeta$-numbers
\begin{align}
\la{5.5}
F^{\mathbf{B}}_{\mc W}(\kappa; 3)   &= \exp\bigg(
-\frac{50  \zeta_{5}}{9}\,\kappa^{3}+\frac{1225 \, \zeta_{7}}{36}\,\kappa^{4}-\frac{1323\,\zeta_{9}}{8}
\,\kappa^{5}+\frac{1960805\,\zeta_{11}}{2592}\,\kappa^{6}
-\frac{17688385 \,\zeta_{13}}{5184}\,\kappa^{7}+\cdots
\bigg).
\end{align}
Comparing with (\ref{1.11}), we also remark that we have model dependence, as it would be natural to expect.

Similar calculations may be done by considering the other large R-charge tower
$\tr\varphi^{3}\,(\tr\varphi^{2})^{n}$. Now, in the $\mathbf{A}$ model we find
\begin{align}
\la{5.6}
F_{\mc W, n}^{\mathbf{A}\,(3)}(g; 3) &= 
1-\frac{9\,(n+1)\,(n+6)\,\zeta_{3}}{32\,\pi^{4}}\,g^{4}\notag \\
& +\frac{5\,(15060+33926 n+26460 n^{2}+9559 n^{3}+1725 n^{4}+105 n^{5})\,
\zeta_{5}}{1152\,(20+12n+3n^{2})\,\pi^{6}}
\,g^{6}+\cdots,
\end{align}
and then, taking the limit (\ref{1.4}),
\begin{align}
\la{5.7}
F_{\mc W}^{\mathbf{A}\,(3)}(\kappa; 3) &= 1-\frac{9\,\zeta_{3}}{2}\,\kappa^{2}
+\frac{175\, \zeta_{5}}{18}\,\kappa^{3}+
\bigg(\frac{81\, \zeta_{3}^2}{8}-\frac{12005\, \zeta_{7}}{576}\bigg)\,\kappa^{4}\notag \\
& +\bigg(\frac{1491\, \zeta_{9}}{32}-\frac{175\, \zeta_{3}\,\zeta_{5}}{4}\bigg)\,\kappa^{5}+
 \bigg(-\frac{243\, \zeta_{3}^3}{16}+\frac{30625\, 
\zeta_{5}^2}{648}+\frac{12005\, \zeta_{3}\,\zeta_{7}}{128}-\frac{2247091 
\zeta_{11}}{20736}\bigg)\,\kappa ^6\notag\\
& +\bigg(
\frac{1575\, \zeta_{3}^2\, \zeta_{5}}{16}-\frac{2100875\, \zeta_{5}\, \zeta_{7}}
{10368}-\frac{13419\, \zeta_{3}\, \zeta_{9}}{64}+\frac{5400395\, \zeta_{13}}{20736}
\bigg)\,\kappa^{7}+\cdots\ ,
\end{align}
that can be written in exponentiated form as 
\begin{align}
\la{5.8}
F_{\mc W}^{\mathbf{A}\,(3)}(\kappa; 3)  &= \exp\bigg(
-\frac{9\,\zeta_{3}}{2}\,\kappa^{2}+\frac{175\,\zeta_{5}}{18}\,\kappa^{3}
-\frac{12005\,\zeta_{7}}{576}\,\kappa^{4}+\frac{1491\,\zeta_{9}}{32}\,\kappa^{5}
-\frac{2247091\, \zeta_{11}}{20736}\,\kappa^{6}
+\frac{5400395\,\zeta_{13}}{20736}\,\kappa^{7}+\cdots
\bigg).
\end{align}

\paragraph{Remark} At finite $n$, the expansions  (\ref{5.1}) and (\ref{5.6}) are similar but non-trivially related. As expected
there is no obvious transformation of $n$ relating the two towers. Nevertheless, in the scaling limit, the
expression (\ref{5.7}) is equal to (\ref{5.2}) (and of course the same holds for the exponentiated form), i.e. 
\be
F^{\mathbf{A}\,(2)}_{\mc W}(\kappa; 3) = F^{\mathbf{A}\,(3)}_{\mc W}(\kappa; 3).
\ee
which will turn out to be a special case of the universality relation (\ref{1.13}), to be proved later.

To confirm (\ref{1.13}), we also evaluate the scaling function $F_{\mc W}^{(3)}$ in  the $\mathbf{B}$ model. 
In this case, we find
\begin{align}
\la{5.10}
F_{\mc W, n}^{\textbf{B}\,(3)}(g; 3) &= 
1-\frac{5\,  (4440 + 9944 n + 5400 n^2 + 2581 n^3 + 465 n^4 + 60 n^5)\, \zeta_{5}}{1152 \,
\pi ^6 (3 n^2+12 n+20)}\,g^{6}\notag \\
& + \frac{175\, (18480 + 50414 n + 37867 n^2 + 19531 n^3 + 5819 n^4 + 945 n^5 + 84 n^6)
 \,\zeta_{7}}{36864\, \pi ^8\, (3 n^2+12 n+20)}\,g^{8}\notag \\
&  -\frac{21 \, (520800 + 1609864 n + 1578564 n^2 + 915854 n^3 + 351435 n^4 + 
 80976 n^5 + 11151 n^6 + 756 n^7) \, \zeta_{9}}
{32768 \, \pi ^{10} \,(3 n^2+12 n+20)}\,g^{10}+\cdots
\end{align}
The scaling limit gives again the same result as for the $(\tr \varphi^{2})^{n}$ tower of operators, i.e.
\be
F^{\mathbf{B}\,(2)}_{\mc W}(\kappa; 3) = F^{\mathbf{B}\,(3)}_{\mc W}(\kappa; 3),
\ee
supporting the claim that (\ref{1.13}) has a chance to hold in any model.

\subsection{Scaling functions for the $SU(4)$ theories}

We have also analyzed the five $\mathbf{ABCDE}$ theories for $SU(4)$ gauge group. In this case they are all distinct. 
The analysis is computationally rather demanding and we did not collect long expansions. 
Nevertheless, we checked exponentiation in all cases, at least up to terms $\sim \zeta(9)$, as well as the validity of the 
tower-independence (\ref{1.13}) -- hence from now on we shall drop the tower label.
The first 4 terms of the five scaling functions are 
\begin{align}
\la{5.12}
\log F_{\mc W}^{\mathbf A}(\kappa; 4) &= -\frac{9\,\zeta(3)}{2}\,\kappa^{2}+\frac{325\,\zeta(5)}{36}\,\kappa^{3}-\frac{41405\,\zeta(7)}{2304}\,\kappa^{4}+\frac{9429\,\zeta(9)}{256}\,\kappa^{5}+\cdots, \notag  \\
\log F_{\mc W}^{\mathbf B}(\kappa; 4) &= \frac{3\,\zeta(3)}{8}\,\kappa^{2}-\frac{1675\,\zeta(5)}{144}\,\kappa^{3}+\frac{70315\,\zeta(7)}{1024}\,\kappa^{4}-\frac{1055873\,\zeta(9)}{3072}\,\kappa^{5}+\cdots, \notag  \\
\log F_{\mc W}^{\mathbf C}(\kappa; 4) &=  -\frac{39\,\zeta(3)}{8}\,\kappa^{2}+\frac{1375\,\zeta(5)}{144}\,\kappa^{3}-\frac{57085\,\zeta(7)}{3072}\,\kappa^{4}+\frac{115325\,\zeta(9)}{3072}\,\kappa^{5}+\cdots, \notag  \\
\log F_{\mc W}^{\mathbf D}(\kappa; 4) &= -\frac{21\,\zeta(3)}{4}\,\kappa^{2}+\frac{725\,\zeta(5)}{72}\,\kappa^{3}-\frac{88445\,\zeta(7)}{4608}\,\kappa^{4}+\frac{58751\,\zeta(9)}{1536}\,\kappa^{5}+\cdots, \notag  \\
\log F_{\mc W}^{\mathbf E}(\kappa; 4) &= 0\cdot\zeta(3)\,\kappa^{2}-\frac{100\,\zeta(5)}{9}\,\kappa^{3}+\frac{1225\,\zeta(7)}{18}\,\kappa^{4}
-343\,\zeta(9)\,\kappa^{5}+\cdots.
\end{align}
Notice that the expansions obey the  following relations to be proved and generalized in the next section
\begin{align}
\la{5.13}
& \log F_{\mc W}^{\mathbf A}(\kappa; 4)-2\,\log F_{\mc W}^{\mathbf C}(\kappa; 4)+\log F_{\mc W}^{\mathbf D}(\kappa; 4)=0,\notag \\
& \log F_{\mc W}^{\mathbf B}(\kappa; 4)-\log F_{\mc W}^{\mathbf C}(\kappa; 4)+\log F_{\mc W}^{\mathbf D}(\kappa; 4)-\log F_{\mc W}^{\mathbf E}(\kappa; 4) = 0.
\end{align}

\paragraph{Summary of the extended (higher rank) explicit results}

In summary, by considering the $SU(3)$ and $SU(4)$ theories, we have collected strong evidence that the one-loop Wilson scaling functions
are (i) independent on which tower is used and (ii) exponentiate in a sum of simple $\zeta$-numbers. This last feature is very promising
and hints for a simple relation with the interacting action of the model. Also, it seems a good starting point to attempt to derive all-order 
resummations. In the next section, we shall prove these claims.

\section{One-point  Wilson functions  from the dual matrix model}
\la{sec:wilson-th}


Our main tool will again be a dual matrix model of Wishart-Laguerre type. 
The dual matrix model 
serves the same purpose as the in the two-point functions we dealt with earlier: 
it takes care of mixing induced by localization. 
However, the emergence of matrix model is more subtle in the case of the one-point Wilson functions. 
Instead of being an exact solution to the mixing problem, it is an asymptotic solution in the large R-charge limit.
To set the stage for the general treatment we turn back to  $SU(2)$ result
\eqref{1.6} and give a proof that will be the basis for its $SU(N)$ generalization.

    The main idea behind our proof is to define a truncated version  
    $E_{n}(a)$ of the Wilson loop matrix model operator $\mc W(a)$ defined in \eqref{2.8} . 
    This amounts to the splitting  
    \begin{align}
        \mc W(a) = E_{n}(a) + \Delta_{n}(a), \qquad
        E_{n}(a) = \sum_{k=0}^{n}\frac{(2\pi)^{2k}}{(2k!)}a^{2k}, \qquad 
        \Delta_{n}(a) = \sum_{k=n+1}^{\infty}\frac{(2\pi)^{2k}}{(2k!)}a^{2k},      
    \end{align}    
    where, in this $SU(2)$ case $a$ is the first of two eigenvalues of the traceless Hermitian matrix.  
    Since $E_{n}(a)$ only has terms up to degree $n$ we can write it as 
    \begin{align}
        E_{n}(a) = \frac{(2\pi)^{2n}}{(2n!)}\phi_{n} + \sum_{k=0}^{n-1}c_{k}\phi_{k}.
    \end{align}

    Here $\phi_{n}=:(\tr a^{2})^{n}:$ is the operator (including mixing) that corresponds to $(\tr \varphi^{2})^{n}$ in the matrix model. 
    As a result of mixing described by \eqref{2.7}, it is a polynomials in $a^{2}$ with the leading term $a^{2n}$.

     By definition of $\phi_{n}$, we can exploit orthogonality to lower dimensional operators and write
     \begin{align}
    \la{6.3}
        \ev{\phi_{n}\,E_{n}} &= \frac{(2\pi)^{2n}}{(2n!)}\ev{\phi_{n} \, \phi_{n}}.
    \end{align}
    The prefactor on the r.h.s. of (\ref{6.3}) is the same for both $\mathcal{N}=4$ and $\mathcal{N}=2$ theories 
    and cancels in the ratio
    defining the scaling functions. In light of this 
    (\ref{1.6}) is equivalent to the statement that in the large $n$ limit we can approximate 
    $\ev{\phi_{n}\,\mc W }$ by the first non-zero term in its series expansion, i.e. $\Delta_{n}$ 
    contributes to $F^{(2)}_{\mathcal{W}}(\kappa,2)$ at a subleading order in $n$. 
    Appendix \ref{simple_observable_appendix} sets out a sufficient condition for this to hold. 
    Applied to this case it reads
    \begin{align}
        \lim_{n \to \infty}\frac{1}{n^{2}} \frac{\ev{ a^{2n+2k+2}\, \phi_{n}}}{\ev{ a^{2n+2k}\,\phi_{n}}} &= 0.
    \end{align}
    This condition is indeed satisfied in the double scaling limit, but we leave the demonstration of this fact 
    to Sec.~(\ref{subsec:SU2_bound}).

    \subsection{Generalization to $SU(N)$ theories}
        Another way of framing the proof in last section is that instead of directly dealing with 
        one-point Wilson functions,
         we can also consider a sequence of two-point functions that converges to it in the large $n$ limit. 
         Furthermore, we can expect (as shown later in the double scaling limit)
         that the large $n$ limit is again determined by the contribution of the first term in the 
         Wilson loop's expansion that has a non-zero two-point function with $\phi_{n}$. 
         For $SU(N)$ this term is the one proportional to $\tr a^{2n}$. As a result we would like to prove that
        \begin{align}
            \label{6.5}
            \lim_{n \to \infty}\ev{\phi_{n}\, \mc W} &=\frac{1}{N} \lim_{n \to \infty}\ev{\phi_{n}\,\tr a^{2n}}.
        \end{align}
        Using the results in App.~\ref{simple_observable_appendix} we see that a sufficient condition 
        for this to be the true is that
        \begin{align}
            \lim_{n \to \infty}\frac{1}{n^2}\frac{\ev{\phi_{n}\,\tr a^{2n+2k+2}}}{\ev{\phi_{n}\, \tr a^{2n+2k}}} \to 0.
        \end{align}
        We leave the verification that this is indeed the case to Sec.~(\ref{subsec:SUN_bound}).

        At a first glance the situation is markedly different from the previous study of extremal  two-point functions. 
        Because $\tr a^{2n}$ can't be reduced to a function of $\tr a^{2}$ for $N>2$, 
        we can't deal with mixing by simply writing 
        the two-point function in \eqref{6.5} as a determinant.   
        Another way of stating this is to recall the change to polar variables 
        from $a_{\mu}$ and note that unlike $\phi_{n}$, $\tr a^{n}$ 
        is a non-trivial function of angular variables and this function 
        is strongly dependent of $n$. Remarkably, as we shall 
        see shortly, it is this strong dependence of angular part on $n$ that ensures that for large $n$ the mixing problem 
        can be solved by an ``effective'' matrix integral. 

        \paragraph{Large $n$ limit of the angular integrals}
        
            \label{subsec:saddle_point}
            Changing to polar coordinates, we have
            \begin{align}
                \ev{\phi_{n}\,\tr a^{2n}} &= \frac{1}{Z_{S^{4}}}\int r^{n-1}\dd r \,\dd \Omega \, r^{N^2 -3} 
                D(\Omega)\, \phi_{n}(r) r^{2n}A_{n}(\Omega)\exp(-4\pi \Im\tau\, r^2) 
                Z_{\mbox{\scriptsize 1-loop}}(r,\Omega).
            \end{align}
            Here, $A_{n}(\Omega)$ is determined by restricting $\tr a^{2n}$ to the sphere  $\tr a^2 = 1$. 
            The idea now is to use it in the large $n$ limit to do the angular integration first.
            To illustrate the idea in a clear fashion we first consider the $U(N)$ case. Since
$\tr a^{2n} = \sum_{\mu =1}^{N} a_{\mu}^{2n}$, 
            when $a$ is a point on the unit sphere, we have $a_{\mu} \le 1$ and  so, in the large $n$ limit, 
            $\tr a^{2n}$ vanishes almost 
            everywhere except around the $2N$ points where one of the coordinates is $\pm 1$ and all others are $0$. 
            Moreover, it goes to zero extremely quickly around these points. 
            As a result we can treat the angular integral by saddle point approximation around these points. 
            Besides, each of these $2N$ points gives the same result.

            Although for $SU(N)$ the situation is somewhat more complicated due to the $\tr a = 0$ constraint, 
            the angular integral is still well approximated by a saddle saddle point approximation around the points 
            that maximize $\tr a^{2n}$. 

            The constrained extrema of $\tr a^{2n}$ are studied in App.~\ref{appendix_extrema}. 
            Here, we just state the relevant results. 
            The set of point that maximize $\tr a^{2n}$ is the same for all $n > 2$. There are $2N$ such points, one being:
            \begin{align}
            \la{6.8}
                a_{0} = \left(\frac{1}{\sqrt{N\,(N-1)}}, \frac{1}{\sqrt{N\,(N-1)}} , \cdots , 
                \frac{1}{\sqrt{N\,(N-1)}} , -\sqrt{\frac{N-1}{N}}\right).
            \end{align}
            The other ones are related by a permutation of coordinates to either $a_{0}$ or $-a_{0}$. Since,
            \begin{align}
                 \label{6.9}
                 \tr a_{0}^{n} = [N(N-1)]^{-\frac{n}{2}}((N-1) + (1-N)^n),
             \end{align} 
            any even symmetric function of $a_{\mu}$ (i.e. any function of traces of $a$ invariant under $a \to -a$) 
            takes the same value on any of these points. As a result, to the leading order in $n$, we have 
            \begin{align}
                \label{6.10}
                \ev{\phi_{n}\,\tr a^{2n}} &= \frac{c_{n}}{Z_{S^4}}\int \dd r \, r^{N^2 -2}\, \phi_{n}(r) 
                r^{2n}\,\exp(-4\pi \Im\tau\, r^2) Z_{\mbox{\scriptsize 1-loop}}(ra_{0}) + \cdots , 
            \end{align}
            where $c_{n}$ is a constant that is the same for both $\mathcal{N}=2$ and $\mathcal{N}=4$ theories. 
            It can be determined straightforwardly from saddle point approximation but it irrelevant to our results 
            so we shall not compute it.
        
        \paragraph{The large $n$ effective matrix model}
            
            Equation \eqref{6.10} gives us an effective partition function for the large $n$ limit, which given by
            \begin{align}
                \label{6.11}
                Z_{\rm eff} &= \int \dd r \, r^{N^2 -2} \,\exp(-4\pi \Im\tau\, r^2) Z_{\mbox{\scriptsize 1-loop}}(ra_{0}).   
            \end{align}    
            This leads us to a much simpler ``$SU(2)$ like'' matrix model where we have a much better hope of 
            solving the mixing problem. In fact the salient details are exactly the same:
            \begin{itemize}
                \item
                    There a single variable $r$.
                \item
                    $\phi_{n}(r)$ has the leading term $r^{2n}$ and the subleading terms are determined by the 
                    condition that $\ev{\phi_{n}\, r^{2k}} = 0$ for $k < n$.
                \item
                    The theory has a single parameter $\tau$ and a derivative of $Z_{\rm eff}$ 
                    with respect to $\tau$ brings down a factor of $r^2$.
            \end{itemize}
            As a result we can write down the determinant formula:
            \begin{align}
                 \ev{\phi_{n}\,\tr a^{2n}} &= \frac{c_{n} \det \mathcal{M}_{(n+1)}}{c_{n-1} \det\mathcal{M}_{(n)}},
                  \qquad \text{with}\quad 
                 \mathcal{M}_{kl} = \frac{1}{Z_{\rm eff}}\frac{\partial^{k+l} Z_{\rm eff}}{\partial{\tau}^{k}
                 \partial{\bar{\tau}^{l}}}.
            \end{align} 
            The only difference from the GKT result for $SU(2)$ is the presence of $c_{n}$ in the above expression. 
            But $c_{n}$ gets no contribution from $Z_{\mbox{\scriptsize 1-loop}}$ to the leading order in $n$. 
            As a result they are same both for $\mathcal{N}=4$ and $\mathcal{N}=2$ theories and disappear when 
            taking the ratio of the correlation function for the two theories.

            At this stage, using the dual matrix model and following the same step as for $SU(2)$, 
            we can straightaway write the result for $F^{(2)}_{\mathcal{W}}(\kappa,N)$. It is
            \begin{align}
                \label{6.13}
                F^{(2)}_{\mathcal{W}}(\kappa,N) = \int_{0}^{4}dx\,\rho(x)
                \Big(\log Z_{\mbox{\scriptsize 1-loop}}(4\,\pi^{2}\,\kappa\, x\, a_{0}) + 
                \kappa\, \partial_{\kappa} Z_{\mbox{\scriptsize 1-loop}}(4\,\pi^{2}\,\kappa\, x\, a_{0})\Big),
            \end{align}
            where $a_{0}$ is in (\ref{6.8}).
        
        \subsection{Universality of large $n$ limit}
            We point out another feature of the result obtained above, tying up a loose end in the previous discussion.
            The factor of $r^{N^2 -2}$ in the $SU(2)$ like action in \eqref{6.11} which is a remnant of the $SU(N)$ 
            theory we started with doesn't play any part in it \eqref{6.13}. 
            This factor contributes to the $\log W$ term of the potential for effective matrix model and has 
            two related effects:
            \begin{itemize}
                \item 
                    It changes the $\mathcal{N}=4$ results.
                \item
                    It changes the eigenvalue distribution of the matrix $W$ we are integrating over. 
                    But this change doesn't affect the large $n$ result and 
                    changes only the subleading correction of order $\frac{1}{n}$ in 
                    $F^{(2)}_{\mathcal{W}}(\kappa,N)$. 
            \end{itemize}
            This remains true if we insert any function $\mathcal{O}(a)$ in the partition function 
            $Z_{\rm eff}$ or equivalently change 
            $\phi_{n}(a)$ to $\mathcal{O}(a)\phi_{n}(a)$. The only effect will be to change $c_{n}$ and contribute to 
            $\log W$ term in the dual matrix model with a coefficient proportional to $R$-charge of $\mathcal{O}$, i.e.
            \begin{align}
            \la{6.14}
                \frac{\langle \mathcal{O}\, \phi_{n}\, \mc W\rangle^{\cN=2}}
                {\langle \mathcal{O}\,\phi_{n}\,\mc W\rangle^{\cN=4}} &= 
                F^{(2)}_{\mathcal{W}}(\kappa,N),
            \end{align}
            as long as R-charge of the terms in $\mathcal{O}(a)$ is bounded. 
 The previous relation (\ref{1.13}) is nothing but a direct consequence of (\ref{6.14}) and is thus proved.
 The tower-independent scaling function $F_{\mc W}$ is provided by the r.h.s. of (\ref{6.13}).

\subsection{Application to the five $\mc N=2$ superconformal $SU(N)$ gauge theories}

We now apply the master formula (\ref{6.13}) to the $\mathbf{ABCDE}$ models. We straightaway obtain
{\small
\begin{align}
\la{6.15}
& \log  F_{\mc W}^{\mathbf{A}}(\kappa; N) = 
-\frac{9  \zeta (3)}{2}\,\kappa^{2}+\frac{25  (N^2-N+1) \zeta 
(5)}{3 (N-1) N}\,\kappa^{3}-\frac{245  (N^2-N+1)^2 \zeta (7)}{16 
(N-1)^2 N^2}\,\kappa^{4}\spek
+\frac{189  (3 N^6-9 N^5+19 N^4-23 N^3+19 
N^2-9 N+3) \zeta (9)}{20 (N-1)^3 N^3}\,\kappa ^5 \spek
-\frac{847  (N^2-N+1) 
(N^6-3 N^5+7 N^4-9 N^3+7 N^2-3 N+1) \zeta (11)}{16 (N-1)^4 N^4}\,\kappa ^6+\cdots, \notag \\
& \log  F_{\mc W}^{\mathbf{B}}(\kappa; N) = 
\frac{9  (N-3) (N-2) \zeta (3)}{4 (N-1) N}\,\kappa ^2-\frac{25 
(N-2) (3 N^3-10 N^2+10 N-5) \zeta (5)}{2 (N-1)^2 
N^2}\,\kappa ^3 \spek
+\frac{245 (N-2) (33 N^5-169 N^4+355 N^3-395 N^2+221 
N-63) \zeta (7)}{32 (N-1)^3 N^3}\,\kappa ^4 \spek
-\frac{189 (N-2) (331 
N^7-2332 N^6+7252 N^5-12950 N^4+14294 N^3-9828 N^2+3828 N-765) \zeta 
(9)}{40 (N-1)^4 N^4}\,\kappa^{5}\spek
+\frac{2541  (N-2) (123 N^9-1109 
N^8+4547 N^7-11126 N^6+17906 N^5-19698 N^4+14838 N^3-7419 N^2+2217 
N-341) \zeta (11)}{32 (N-1)^5 N^5}\,\kappa ^6+\cdots, \notag \\
& \log  F_{\mc W}^{\mathbf{C}}(\kappa; N) = 
-\frac{9 (3 N^2-7 N+6) \zeta (3)}{4 (N-1) N}\,\kappa^{2}+\frac{25 
 (N^4-4 N^3+10 N^2-15 N+10) \zeta (5)}{2 (N-1)^2 
N^2}\,\kappa^{3}\spek
-\frac{245 (3 N^6-17 N^5+63 N^4-155 N^3+249 N^2-251 
N+126) \zeta (7)}{32 (N-1)^3 N^3}\,\kappa^{4}\spek
+\frac{189  (9 N^8-66 
N^7+324 N^6-1106 N^5+2646 N^4-4424 N^3+5076 N^2-3819 N+1530) \zeta 
(9)}{40 (N-1)^4 N^4}\,\kappa^{5}\spek
-\frac{2541 (N^{10}-9 N^9+55 N^8-240 
N^7+762 N^6-1778 N^5+3054 N^4-3825 N^3+3405 N^2-2045 N+682) \zeta 
(11)}{32 (N-1)^5 N^5}\,\kappa^{6}+\cdots, \notag \\
& \log  F_{\mc W}^{\mathbf{D}}(\kappa; N) = 
-\frac{9 (N^2-3 N+3) \zeta (3)}{(N-1) N}\,\kappa^{2}+\frac{50 
 (N^4-5 N^3+14 N^2-22 N+15) \zeta (5)}{3 (N-1)^2 N^2}\,\kappa^{3}\spek
 -\frac{245  (N^6-7 N^5+29 N^4-75 N^3+123 N^2-125 N+63) \zeta (7)}{8 
(N-1)^3 N^3}\,\kappa^{4}\spek
+\frac{189  (3 N^8-27 N^7+148 N^6-532 N^5+1302 
N^4-2198 N^3+2532 N^2-1908 N+765) \zeta (9)}{10 (N-1)^4 
N^4}\,\kappa^{5}\spek
-\frac{847 (N^{10}-11 N^9+75 N^8-345 N^7+1122 N^6-2646 
N^5+4566 N^4-5730 N^3+5105 N^2-3067 N+1023) \zeta (11)}{8 (N-1)^5 
N^5}\,\kappa^{6}+\cdots\notag \\
& \log  F_{\mc W}^{\mathbf{E}}(\kappa; N) = 
-\frac{100  (N-2)^2 \zeta (5)}{3 (N-1) N}\,\kappa^{3}+\frac{245 
 (N-2)^2 (N^2-2 N+2) \zeta (7)}{(N-1)^2 N^2}\,\kappa^{4}\spek
 -\frac{189 (N-2)^2 (41 N^4-164 N^3+308 N^2-288 N+144) \zeta (9)}{5 
(N-1)^3 N^3}\,\kappa^{5}\spek
+\frac{847 (N-2)^2 (N^2-2 N+2) (23 N^4-92 
N^3+156 N^2-128 N+64) \zeta (11)}{2 (N-1)^4 N^4}\,\kappa^{6}+\cdots.
\end{align}
}

\noindent
Of course, specialization of (\ref{6.15}) to $SU(2)$, $SU(3)$,  and $SU(4)$ reproduces perfectly the 
previous partial results (\ref{1.7}), (\ref{1.11}), (\ref{5.5}), and (\ref{5.12}).

\paragraph{Remark:} We now easily understand the reason behind the  two constraints (\ref{5.13}). 
To this aim, we remark that the main formula (\ref{6.13}) shows that  $\log F_{\mc W}$ is linear in the 
interacting action. This allows to prove that for any $N$ we have exactly 
\begin{align}
\la{6.16}
& \log F_{\mc W}^{\mathbf A}(\kappa; N)-2\,\log F_{\mc W}^{\mathbf C}(\kappa; N)+\log F_{\mc W}^{\mathbf D}(\kappa; N)=0,\notag \\
& \log F_{\mc W}^{\mathbf B}(\kappa; N)-\log F_{\mc W}^{\mathbf C}(\kappa; N)+\log F_{\mc W}^{\mathbf D}(\kappa; N)-\log F_{\mc W}^{\mathbf E}(\kappa; N) = 0.
\end{align}
Indeed, the interacting action is linear in the number of fundamental, symmetric and antisymmetric representations and these numbers obey 
the above relations. There are also constant terms appearing in the rewriting of traces in terms of traces in the fundamental, but these terms drop
since the sum of coefficients in (\ref{6.16}) is zero.

\subsubsection{All-order resummation of the one-loop Wilson scaling functions}

As a further application of the formula (\ref{6.13}) we can extend fixed $N$ expansions like (\ref{5.3}) and (\ref{5.5}) as far as needed.
In particular, for those two $SU(3)$ models one finds the long expansions
\begin{align}
\la{6.17}
\log F_{\mc W}^{\mathbf{A}}(\kappa; 3) &= -\frac{9  \zeta (3)}{2}\,\kappa^{2}+\frac{175  \zeta 
(5)}{18}\,\kappa^{3}-\frac{12005 \zeta (7)}{576}\,\kappa^{4}+\frac{1491  
\zeta (9)}{32}\,\kappa^{5}-\frac{2247091  \zeta 
(11)}{20736}\,\kappa^{6}+\frac{5400395 \zeta 
(13)}{20736}\,\kappa^{7} \spek 
-\frac{568668815  \zeta 
(15)}{884736}\,\kappa^{8}+\frac{261350914825  \zeta 
(17)}{161243136}\,\kappa^{9}-\frac{8943246419107  \zeta 
(19)}{2149908480}\,\kappa^{10}+\frac{1552522828675  \zeta 
(21)}{143327232}\,\kappa^{11} \spek
-\frac{10606464907364417  \zeta 
(23)}{371504185344}\,\kappa^{12}+\frac{14121732251822125  \zeta 
(25)}{185752092672}\,\kappa^{13}-\frac{8742217069824025 \zeta 
(27)}{42807066624}\,\kappa^{14}\spek
+\frac{2464280244310231795 \zeta 
(29)}{4458050224128}\,\kappa^{15}-\frac{3437685880746945869965  \zeta 
(31)}{2282521714753536}\,\kappa ^{16}\spek
+\frac{392539284372606415825 
\zeta (33)}{95105071448064}\,\kappa^{17}+\cdots,\notag \\
\log F_{\mc W}^{\mathbf{B}}(\kappa; 3) &= -\frac{50  \zeta (5)}{9}\,\kappa^{3}+\frac{1225  \zeta 
(7)}{36}\,\kappa^{4}-\frac{1323 \kappa ^5 \zeta (9)}{8}+\frac{1960805 \kappa ^6 
\zeta (11)}{2592}-\frac{17688385  \zeta 
(13)}{5184}\,\kappa ^7 \spek
+\frac{142167025  \zeta 
(15)}{9216}\,\kappa ^8-\frac{2834936114725  \zeta 
(17)}{40310784}\,\kappa ^9+\frac{17410184710919  \zeta 
(19)}{53747712}\,\kappa ^{10}\spek
-\frac{18006386209175  \zeta 
(21)}{11943936}\,\kappa ^{11}+\frac{164517685436679575  \zeta 
(23)}{23219011584}\,\kappa ^{12}-\frac{1560309607284420125  \zeta 
(25)}{46438023168}\,\kappa ^{13}\spek
+\frac{4547264436973375  \zeta 
(27)}{28311552}\,\kappa ^{14}-\frac{861837268172768598385  \zeta 
(29)}{1114512556032}\,\kappa ^{15}\spek
+\frac{133626553771108660672025  
\zeta (31)}{35664401793024}\,\kappa ^{16} -\frac{144697635847665710153575 
 \zeta (33)}{7925422620672}\,\kappa^{17}+\cdots.
\end{align}
We remind that in the $SU(2)$ theory,  the analogous expansion is  (\ref{1.7}), cf. also (\ref{1.6}), and one 
has the all-order series coefficients 
\be
\la{6.18}
\log F_{\mc W}^{\mathbf A}(\kappa; 2) = \frac{8}{\sqrt\pi}\,\sum_{n=1}^{\infty}(-1)^{n}\,
\frac{4^{n}-1}{(n+1)^{2}\,n!}\,\Gamma\left(n+\frac{3}{2}\right)\,\zeta(2n+1)\,\kappa^{n+1},
\ee
leading to the following integral representation \footnote{
Notice that the successive derivation of (\ref{6.19}) in \cite{Grassi:2019txd} was done independently and with a different method 
strongly suggesting that there are no non-perturbative ambiguities in the reconstruction from the weak-coupling expansion, at
least in the half-plane $\text{Re}(\kappa)>0$.
}
\be
\la{6.19}
\log F_{\mc W}^{\mathbf A}(\kappa; 2) = 4\,\int_{0}^{\infty}\frac{dt\,e^{t}}{t\,(e^{t}-1)^{2}}\,\left[
-3+4\,J_{0}\left(t\,\sqrt{\kappa}\right)-J_{0}\left(2\,t\,\sqrt{\kappa}\right)
\right] .
\ee
For the $SU(3)$ expansions in (\ref{6.17}), guided by (\ref{6.18}), we easily find
\begin{align}
\la{6.20}
\log F_{\mc W}^{\mathbf{A}}(\kappa; 3) &=  
\frac{4}{\sqrt\pi}\,\sum_{n=1}^{\infty}(-1)^{n}\,\frac{3^{-n}\,(-1-2^{1+2n}+3^{1+2n})}{(n+1)^{2}\,n!}\,\Gamma\left(n+\frac{3}{2}\right)\,\zeta(2n+1)\,\kappa^{n+1}, \\
\log F_{\mc W}^{\mathbf{B}}(\kappa; 3) &= 
\frac{4}{\sqrt\pi}\,\sum_{n=1}^{\infty}(-1)^{n}\,\frac{3^{-1-n}\,(-1+3^{2+2n}-4^{1+n}-4^{1+2n})}{(n+1)^{2}\,n!}\,\Gamma\left(n+\frac{3}{2}\right)\,\zeta(2n+1)\,\kappa^{n+1},\notag
\end{align}
as can be checked by reproducing (\ref{6.17}). \footnote{We checked agreement with many more terms, 
a task that is possible due to  (\ref{6.13}).}
The sums in (\ref{6.20}) can be written in integral form by using the identity
\be
\int_{0}^{\infty}dt\,\frac{t^{p}\,e^{t}}{(e^{t}-1)^{2}} = p!\,\zeta(p),\qquad p>1,
\ee
and we obtain 
\begin{align}
\la{6.22}
\log F_{\mc W}^{\mathbf{A}}(\kappa; 3) &= 4\,\int_{0}^{\infty}\frac{dt\,e^{t}}{t\,(e^{t}-1)^{2}}\,\left[
-7+6\,J_{0}\left(t\,\sqrt{\tfrac{\kappa}{3}}\right)+3\,J_{0}\left(2\,t\,\sqrt{\tfrac{\kappa}{3}}\right)-2\,J_{0}\left(t\,\sqrt{3\,\kappa}\right)
\right] , \\
\log F_{\mc W}^{\mathbf{B}}(\kappa; 3) &= 2\,\int_{0}^{\infty}\frac{dt\,e^{t}}{t\,(e^{t}-1)^{2}}\,\left[
-5+4\,J_{0}\left(t\,\sqrt{\tfrac{\kappa}{3}}\right)+4\,J_{0}\left(2\,t\,\sqrt{\tfrac{\kappa}{3}}\right)
-4\,J_{0}\left(t\,\sqrt{3\,\kappa}\right)+J_{0}\left(4\,t\,\sqrt{\tfrac{\kappa}{3}}\right)
\right] , \notag
\end{align}
with a structure close to the $SU(2)$ expression 
(\ref{6.19}). 

\medskip
It is now a straightforward exercise to repeat the same analysis for a general 
$SU(N)$ gauge group. The final result is remarkably neat. Let us introduce the notation 
\be
\la{6.23}
\widetilde J_{0}(x) = J_{0}(x)-1.
\ee
Then, for the five models we obtain (of course, only 3 expressions are independent thanks to (\ref{6.16}))

\paragraph{$\mathbf A$ model}
\begin{align}
\la{6.24}
\log & F_{\mc W}^{\mathbf{A}}(\kappa; N) =4\,\int_{0}^{\infty}\frac{dt\,e^{t}}{t\,(e^{t}-1)^{2}}\, 
\bigg[
N\,\widetilde J_{0}\bigg(t\,\sqrt\frac{2\,(N-1)\,\kappa}{N}\bigg)+N\,(N-1)\,\widetilde J_{0}\bigg(t\,\sqrt\frac{2\,\kappa}{N\,(N-1)}\bigg)\notag \\
& -(N-1)\,\widetilde J_{0}\bigg(t\,\sqrt\frac{2\,N\,\kappa}{N-1}\bigg)
\bigg].
\end{align}

\paragraph{$\mathbf B$ model}
\begin{align}
\la{6.25}
\log & F_{\mc W}^{\mathbf{B}}(\kappa; N) =\int_{0}^{\infty}\frac{dt\,e^{t}}{t\,(e^{t}-1)^{2}}\,\bigg[
2\,(N-1)(N-2)\,\widetilde J_{0}\bigg(t\,\sqrt\frac{2\,\kappa}{N\,(N-1)}\bigg)+N\,(N-1)\,\widetilde J_{0}\bigg(t\,\sqrt\frac{8\,\kappa}{N\,(N-1)}\bigg)\notag \\
& +2\,(N-1)\,\widetilde J_{0}\bigg(t\,(N-2)\,\sqrt\frac{2\,\kappa}{N\,(N-1)}\bigg)+2\,(N-2)\,\widetilde J_{0}\bigg(t\,\sqrt\frac{2\,(N-1)\,\kappa}{N}\bigg)\notag \\
&-4\,(N-1)\,\widetilde J_{0}\bigg(t\,\sqrt\frac{2\,N\,\kappa}{N-1}\bigg)+2\,\widetilde J_{0}\bigg(t\,\sqrt\frac{8\,(N-1)\,\kappa}{N}\bigg)
\bigg].
\end{align}

\paragraph{$\mathbf C$ model}
\begin{align}
\la{6.26}
\log & F_{\mc W}^{\mathbf{C}}(\kappa; N) =\int_{0}^{\infty}\frac{dt\,e^{t}}{t\,(e^{t}-1)^{2}}\,\bigg[
2\,(N-1)(N+2)\,\widetilde J_{0}\bigg(t\,\sqrt\frac{2\,\kappa}{N\,(N-1)}\bigg)
+(N-2)\,(N-1)\,\widetilde J_{0}\bigg(t\,\sqrt\frac{8\,\kappa}{N\,(N-1)}\bigg)\notag \\
& +2\,(N-1)\,\widetilde J_{0}\bigg(t\,(N-2)\,\sqrt\frac{2\,\kappa}{N\,(N-1)}\bigg)
+2\,(N+2)\,\widetilde J_{0}\bigg(t\,\sqrt\frac{2\,(N-1)\,\kappa}{N}\bigg)-4\,(N-1)\,\widetilde J_{0}\bigg(t\,\sqrt\frac{2\,N\,\kappa}{N-1}\bigg)
\bigg].
\end{align}

\paragraph{$\mathbf D$ model}
\begin{align}
\la{6.27}
\log & F_{\mc W}^{\mathbf{D}}(\kappa; N) =2\,\int_{0}^{\infty}\frac{dt\,e^{t}}{t\,(e^{t}-1)^{2}}\,\bigg[
4\,(N-1)\,\widetilde J_{0}\bigg(t\,\sqrt\frac{2\,\kappa}{N\,(N-1)}\bigg)
+(N-2)\,(N-1)\,\widetilde J_{0}\bigg(t\,\sqrt\frac{8\,\kappa}{N\,(N-1)}\bigg)\notag \\
& +2\,(N-1)\,\widetilde J_{0}\bigg(t\,(N-2)\,\sqrt\frac{2\,\kappa}{N\,(N-1)}\bigg)
+4\,\widetilde J_{0}\bigg(t\,\sqrt\frac{2\,(N-1)\,\kappa}{N}\bigg)-2\,(N-1)\,\widetilde J_{0}\bigg(t\,\sqrt\frac{2\,N\,\kappa}{N-1}\bigg)
\bigg].
\end{align}

\paragraph{$\mathbf E$ model}
\begin{align}
\la{6.28}
\log & F_{\mc W}^{\mathbf{E}}(\kappa; N) =2\,\int_{0}^{\infty}\frac{dt\,e^{t}}{t\,(e^{t}-1)^{2}}\,\bigg[
(N-1)^{2}\,\widetilde J_{0}\bigg(t\,\sqrt\frac{8\,\kappa}{N\,(N-1)}\bigg)
+2\,(N-1)\,\widetilde J_{0}\bigg(t\,(N-2)\,\sqrt\frac{2\,\kappa}{N\,(N-1)}\bigg)\notag \\
& -2\,(N-1)\,\widetilde J_{0}\bigg(t\,\sqrt\frac{2\,N\,\kappa}{N-1}\bigg)
+\widetilde J_{0}\bigg(t\,\sqrt\frac{8\,(N-1)\,\kappa}{N}\bigg)
\bigg]
\end{align}
As a final remark, it may be interesting to stress that, 
the function $\log F_{\mc W}(\kappa; N)$ admits a finite non-trivial
limit when $N\to \infty$. This can be verified using \eqref{6.9}, which shows that in this limit the traces at the saddle point are simply $\tr a_{0}^{n} = \pm 1$\footnote{Here we point out that this is the case for $U(N)$ for any $N$. As $N \to \infty$, the $SU(N)$ saddle point moves closer and closer to the $U(N)$ saddle point, as should be the case since both $U(N)$ and $SU(N)$ theories have the same large $N$ limit.}. This can also be seen from the explicit expansion in \eqref{6.15} Taking this limit in the above expressions and defining, cf. (\ref{6.23}),
\be
\widetilde J_{1}(x) = x\,J_{1}(x)-\frac{x^{2}}{4},
\ee
we can write following representations for the $N\to \infty$ limit of the scaling functions 
\begin{align}
\log F_{\mc W}^{\mathbf{A}}(\kappa; \infty) &= 4\,\int_{0}^{\infty}\frac{dt\,e^{t}}{t\,(e^{t}-1)^{2}}\,
\bigg[
\widetilde{J}_0(\sqrt{2\,\kappa }\ t)+\widetilde{J}_1(\sqrt{2\,\kappa }\ t)
\bigg], \notag \\
\log F_{\mc W}^{\mathbf{B}}(\kappa; \infty) &= -2\,\int_{0}^{\infty}\frac{dt\,e^{t}}{t\,(e^{t}-1)^{2}}\,
\bigg[
\widetilde{J}_0(\sqrt{2\,\kappa }\ t)-\widetilde{J}_0(2 \, \sqrt{2\,\kappa }\ t)-3 \,\widetilde{J}_1(\sqrt{2\,\kappa }\ t)
\bigg], \notag \\
\log F_{\mc W}^{\mathbf{C}}(\kappa; \infty) &= 6\,\int_{0}^{\infty}\frac{dt\,e^{t}}{t\,(e^{t}-1)^{2}}\,
\bigg[
\widetilde{J}_0( \sqrt{2\,\kappa }\ t)+\widetilde{J}_1(\sqrt{2\,\kappa }\ t)
\bigg], \notag \\
\log F_{\mc W}^{\mathbf{D}}(\kappa; \infty) &= 8\,\int_{0}^{\infty}\frac{dt\,e^{t}}{t\,(e^{t}-1)^{2}}\,
\bigg[
\widetilde{J}_0( \sqrt{2\,\kappa }\ t)+\widetilde{J}_1( \sqrt{2\,\kappa }\ t)
\bigg], \notag \\
\log F_{\mc W}^{\mathbf{E}}(\kappa; \infty) &= 2\,\int_{0}^{\infty}\frac{dt\,e^{t}}{t\,(e^{t}-1)^{2}}\,
\bigg[
\widetilde{J}_0(2 \,\sqrt{2\,\kappa }\ t)+4\, \widetilde{J}_1( \sqrt{2\,\kappa }\ t)
\bigg].
\end{align}
Such large charge and large $N$ simultaneous limit, with $N\ll n$, has been recently considered also in 
$O(N)$ invariant scalar theories \cite{Alvarez-Gaume:2019biu}.

\section{The heavy BPS regime of  one-point Wilson functions}
\la{sec:resum}

As we  remarked at the end of the Introduction, the large $\kappa$ expansion of the expressions 
(\ref{6.24}-\ref{6.28})  is  potentially rather interesting since non-perturbative corrections of the form 
$\sim \exp(-c\,\sqrt\kappa)$
are expected to be present and associated with 
heavy electric BPS states (matter hypermultiplets and reduced vector multiplet) with masses $\sim \sqrt\kappa$ in the double scaling limit.
Hence, the large $\kappa$ limit probes the weak coupling BPS states in the moduli space point 
selected by the relevant saddle point associated with the large R-charge insertion. In this section, we 
present the tools that are needed to compute the $\kappa\gg 1$ expansion of (\ref{6.24}-\ref{6.28}) and 
discuss the  detailed matching with the  mass spectrum of heavy BPS states.

\subsection{Large $\kappa$ expansion and non-perturbative corrections}

The example of $SU(2)$ has been discussed in \cite{Grassi:2019txd}. Here, we want to present 
some general expressions that may be used for all other cases. To this aim, it will be enough to 
revisit the $SU(2)$ case and work out the $SU(3)$ $\mathbf{A}$ and $\mathbf{B}$ models. All other cases
may be treated by the same formulas. It is convenient to write the resummed scaling function
(\ref{6.24}) for $N=2,3$ and (\ref{6.25}) for $N=3$ in the form 
\begin{align}
\la{7.1}
\log F_{\mc W}^{\mathbf{A}}(\kappa; 2) &= 4\,\bigg[4\,\mc B(\sqrt\kappa)-\mc B(2\sqrt\kappa)\bigg],\notag \\
\log F_{\mc W}^{\mathbf{A}}(\kappa; 3) &= 4\,\bigg[6\,\mc B(\tfrac{1}{\sqrt 3}\,\sqrt\kappa)
+3\,\mc B(\tfrac{2}{\sqrt 3}\,\sqrt\kappa)-2\,\mc B(\sqrt 3\,\sqrt\kappa)\bigg],\notag \\
\log F_{\mc W}^{\mathbf{B}}(\kappa; 3) &= 2\,\bigg[4\,\mc B(\tfrac{1}{\sqrt 3}\,\sqrt\kappa)
+4\,\mc B(\tfrac{2}{\sqrt 3}\,\sqrt\kappa)-4\,\mc B(\sqrt 3\,\sqrt\kappa)+\mc B(\tfrac{4}{\sqrt 3}\,\sqrt\kappa)\bigg],
\end{align}
where the regulated $\mc B$ function is 
\be
\mc B_{\eta}(x) = \int_{0}^{\infty}\frac{dt\,e^{t}}{t\,(e^{t}-1)^{2}}\,t^{\eta}\,[J_{0}(t\,x)-1],\qquad \eta>0,
\ee
and the limit $\eta\to 0$ is taken in (\ref{7.1}). \footnote{The $\eta\to 0$ limit is finite since the integrand of the combinations in (\ref{7.1})
have no singularities at $t=0$.}
The large $x$ expansion of this function has a perturbative part $\mc B_{\rm P}$ 
plus a non-perturbative contribution $\mc B_{\rm NP}$ that is 
exponentially suppressed at large $x$. The perturbative part can be computed easily by Mellin transform methods 
and amounts to 
\be
\la{7.3}
\mc B_{\rm P, \eta}(x) = \frac{x^{2}}{4}\,(\log x-\log 2 +\gamma_{\rm E}-\eta^{-1})+\frac{1}{12}\,(\log x+12\,\log(\text{A})-\log 2-1),
\ee
where $\text{A}$ is Glaisher's constant ($\log{\rm A} = \frac{1}{12}-\zeta'(-1)$). Notice that 
the singular term $\sim x^{2}\eta^{-1}$ always correctly cancels in the combinations appearing
in (\ref{6.24}-\ref{6.28}). \footnote{
Just to give an example, for the  $\mathbf{A}$ model one has, cf. (\ref{6.24}),
\be\notag
N\,\left[\sqrt\frac{2\,(N-1)\,\kappa}{N}\,\right]^{2}+N\,(N-1)\,\left[\sqrt\frac{2\,\kappa}{N\,(N-1)}
\,\right]^{2}-(N-1)\,\left[\sqrt\frac{2\,N\,\kappa}{N-1}\,\right]^{2}=0.
\ee
}
Remarkably, the terms in (\ref{7.3}) exhaust all contributions that are not exponentially suppressed as $\kappa\to \infty$, i.e. there are
no algebraically decaying inverse powers of $\kappa$.
 
\medskip
The non-perturbative part is regular for $\eta\to 0$. To determine it we can write 
\be
(\frac{1}{x}\mc B'(x))' = \sum_{p=1}^{\infty}p\,\int_{0}^{\infty}dt\,e^{-p\,t}\frac{t}{x}\,J_{2}(tx) = 
\frac{1}{x^{3}}\sum_{p=1}^{\infty}\left[2p-\frac{p^{2}(2p^{2}+3x^{2})}{(p^{2}+x^{2})^{3/2}}\right].
\ee
where we applied a simple differential operator to get a convergent sum. In particular, this expression can be evaluated at $\eta=0$.
To extract the non-perturbative part of the infinite sum, we convert it into a contour integral 
using the standard kernel $\pi\,\cot(\pi\,p)$ and deforming the $p$ integration contour over the semi-infinite line $[i\,x, +i\,\infty)$,
see e.g. \cite{SchaferNameki:2006gk}. This gives the representation 
\begin{align}
\la{7.5}
\mc B_{\rm NP}(x) &= \sum_{m=1}^{\infty} e^{-2\,\pi\,m\,x}\,\bigg(
\frac{\sqrt{x}}{2 \pi\, m^{3/2}}+\frac{11}{32 \pi 
^2 \,m^{5/2}}\frac{1}{\sqrt{x}}-\frac{31 }{1024 \pi ^3 \, m^{7/2}}\frac{1}{x^{3/2}}
+\frac{177 }{16384 \pi ^4 \, m^{9/2}}\frac{1}{x^{5/2}} \spek
-\frac{7125 }{1048576 \pi ^5 
m^{11/2}}\,\frac{1}{x^{7/2}}+\frac{102165 }{16777216 \pi ^6 \,
m^{13/2}}\,\frac{1}{x^{9/2}}+\cdots
\bigg).
\end{align}
Applying (\ref{7.3}) and (\ref{7.5}) to the specific cases in (\ref{7.1}) we then obtain 
\begin{align}
\la{7.6}
\log F_{\mc W}^{\mathbf{A}}(\kappa; 2) &= -4\,\log (2)\,\kappa+\frac{1}{2}\,\log\kappa+12\,\log{\rm A}-\frac{4}{3}\,\log 2-1 \notag \\
&+\frac{8\,\kappa^{1/4}}{\pi}\,e^{-2\,\pi\,\sqrt\kappa}\,\left(1+\frac{11}{16\,\pi}\,\frac{1}{\sqrt\kappa}-\frac{31}{512\,\pi^{2}}\,\frac{1}{\kappa}+\cdots\right) + \cdots, 
\notag \\
\log F_{\mc W}^{\mathbf{A}}(\kappa; 3) &= -2\,(3\log 3-2\log 2)\,\kappa+\frac{7}{6}\,\log\kappa+28\,\log{\rm A}-\frac{4}{3}\,\log 2-\frac{11}{6}\,\log 3-\frac{7}{3}\notag \\
&+\frac{4\,(27\kappa)^{1/4}}{\pi}\,e^{-2\,\pi\,\sqrt\frac{\kappa}{3}}\,\left(1+\frac{11}{16\,\pi}\,\sqrt\frac{3}{\kappa}-\frac{93}{512\,\pi^{2}}\,\frac{1}{\kappa}+\cdots\right) + \cdots, 
\notag \\
\log F_{\mc W}^{\mathbf{B}}(\kappa; 3) &= -2\,(3\log 3-4\log 2)\,\kappa+\frac{5}{12}\,\log\kappa+10\,\log{\rm A}+\frac{1}{6}\,\log 2-\frac{13}{12}\,\log 3-\frac{5}{6}+\notag \\
&+\frac{4\,(27\kappa)^{1/4}}{3\,\pi}\,e^{-2\,\pi\,\sqrt\frac{\kappa}{3}}\,\left(1+\frac{11}{16\,\pi}\,\sqrt\frac{3}{\kappa}-\frac{93}{512\,\pi^{2}}\,\frac{1}{\kappa}+\cdots\right) + \cdots, 
\end{align}
where we have written the perturbative part plus the first terms of the leading non-perturbative correction. The subleading non-perturbative corrections are rather different in the two $SU(3)$
models and can be studied from the higher order terms with $m\ge 2$ in (\ref{7.5}). Of course the first of (\ref{7.6})
agrees with  GKT result, see their Eq.~(4.21).  Notice that, as remarked in \cite{Grassi:2019txd}, 
the term $\sim (e^{-2\pi\sqrt\kappa})^{2}$ 
cancels in $\log F_{\mc W}(\kappa; 2)$. Actually, one can check that all even powers of 
$\sim e^{-2\pi\sqrt\kappa}$ cancel, but that this does not happen for higher rank gauge groups, 
even considering only the  $\mathbf{A}$ model.

\subsection{Identification of the relevant BPS spectrum at large $\kappa$}

To conclude this section, we give a quantitative explanation of the various terms 
appearing in the resummation formulae for the 
scaling functions, Eqs. (\ref{6.24})--(\ref{6.28}). To this aim one can consider
the $\kappa\gg 1$ limit and, in particular, the non-perturbative
corrections. From the expansion (\ref{7.5}), we can identify the $N$-dependent
coefficient of $t$ in the $\widetilde{J}_{0}$ functions 
with the exponent in the exponentially suppressed terms. This is in turn 
proportional to the mass of degenerate heavy states. Their multiplicity is proportional to the $N$-dependent
prefactors of the $\widetilde{J}_{0}$ functions. The peculiar algebraic dependence on $N$ allows to identify the origin of the 
various terms in the resummed scaling functions.

The $\widetilde{J}_{0}$ functions in (\ref{6.24})--(\ref{6.28}) appear always as a group with positive (integer) coefficients
and argument $\sim \sqrt\kappa$ proportional to 
\be
\la{7.7}
\frac{1}{\sqrt{N(N-1)}},\qquad \text{or}\qquad \sqrt\frac{N-1}{N}.
\ee
Besides, there is a single negative term common to all models and reading
\be
\la{7.8}
-4\,(N-1)\,\widetilde J_{0}\bigg(t\,\sqrt\frac{2\,N\,\kappa}{N-1}\bigg).
\ee
The quantities in (\ref{7.7})  are the components of $a_{0}$ in (\ref{6.8}). 
This is not surprising because 
(\ref{6.13}) shows that $a_{0}$ is indeed the relevant point on the sphere $\tr a^{2}=1$
governing the large $n$
contributions to the Wilson scaling function. As a consequence, we can read the mass spectrum by 
expanding $\Phi$ around $\sqrt \kappa\, a_{0}$. 

\paragraph{Hypermultiplets}

Hypermultiplets get mass from the Yukawa-type coupling and the associated heavy states 
turn out to be in correspondence with the positive contributions with Bessel function arguments proportional to 
(\ref{7.7}). Let us look in detail to the  $\mathbf A$ model case. 
From the term $\sim \widetilde Q\,\Phi\,Q$ and replacing $\Phi\to \sqrt\kappa\,a_{0}$ we get a mass
spectrum with $2N\times (N-1)$ masses $\sim \sqrt\frac{\kappa}{N(N-1)}$ and $2N\times 1$ masses $\sim\sqrt\frac{\kappa\,(N-1)}{N}$,
cf. (\ref{6.8}), in agreement with the positive contributions in (\ref{6.24}).  The same exercise should be repeated for the other 
models taking into account the representation content. 
As a consistency check, we can verify that 
in all $\mathbf{ABCDE}$ models  the ratio between the 
sum of the prefactors of positive terms and the sum of dimensions of matter representations is constant, i.e. independent on $N$. 
For instance, in the 
$\mathbf A$ model we have 
\be
4\,N+4\,N(N-1) = 4\,N^{2} = 2\times \left[2\,N\,\dim \Yfund\right]~,
\ee
and, similary, in  the  other models  we have, cf. Tab.~\ref{tab:scft}, \footnote{$\dim\Ysymm = N(N+1)/2$, $\dim \Yasymm = N(N-1)/2$.}
\begin{align}
& \mathbf{B}:  2\,(N-1)(N-2)+N(N-1)+2\,(N-1)+2\,(N-2)+2  = 3\,N\,(N-1) \notag \\
&  \qquad\qquad = 2\times \left[
(N-2)\,\dim \Yfund+\dim \Ysymm\right], \notag \\
& \mathbf{C}:  2\,(N-1)(N+2)+(N-2)(N-1) +2\,(N-1)+2\,(N+2) = 3\,N\,(N+1) \notag \\
& \qquad\qquad = 2\times \left[
(N+2)\,\dim \Yfund+\dim \Yasymm\right], \notag \\
& \mathbf{D}:  8\,(N-1)+2\,(N-2)(N-1)+4\,(N-1)+8 =  2\,N\,(N+3) \notag \\
& \qquad\qquad = 2\times \left[
4\,\dim \Yfund+2\,\dim \Yasymm\right], \notag \\
& \mathbf{E}: 2(N-1)^{2}+4(N-1)+2  = 2\,N^{2} = 
2\times \left[
\dim \Ysymm+\dim \Yasymm\right].
\end{align}

\paragraph{$W$-multiplet}

The common term (\ref{7.8}) is instead due to the heavy states in the model-independent gauge sector.
In this case the mass spectrum can be computed  by considering the quartic coupling 
$\sim \tr[W, \Phi]^2$ where $W = W^a T^a$ are the $SU(N)$ gauge fields. 
The $U(N-1)$ unbroken gauge symmetry at $ \Phi = \sqrt\kappa\, a_{0}$ predicts 
$(N-1)^2$ massless gauge bosons. The remaining $N^2-1-(N-1)^2 = 2\,(N-1)$ fields are massive $W$-bosons in the effective 
large $n$ limit. They are associated with the hermitian traceless $W$ matrices  $W_{ij}^{(\ell)} = \frac{1}{\sqrt 2}
(\delta_{i, \ell}\delta_{j,N}+ \delta_{i, N}\delta_{j,\ell})$ and $\widetilde W_{ij}^{(\ell)} = \frac{i}{\sqrt 2}
(\delta_{i, \ell}\delta_{j,N}- \delta_{i, N}\delta_{j,\ell})$,for $\ell=1, \dots, N-1$.
The common mass is obtained evaluating the commutator with $a_{0}$. This gives a factor $N$ times the repeated component of 
$a_{0}$, i.e. $N\times \frac{1}{\sqrt{N(N-1)}} = \sqrt\frac{N}{N-1}$, in agreement with (\ref{7.8}), including 
multiplicity.

\section*{Acknowledgements}

We thank C. Angelantonj, L. Bianchi, M. Bill\'o,  D. Orlando,  L. Tizzano, and  P. West for several useful discussions. 
MB and AH were supported by  the INFN grant GSS (Gauge Theories, Strings and Supergravity).
FG was supported by  the INFN grant ST\&FI (String Theory and Fundamental Interactions). 
We also thank the organizers of the XV Avogadro Meeting on Strings, Supergravity, and Gauge Theories, during which part of the work presented in this paper was done.

\appendix

\section{Field theory action and Feynman rules}
\label{app:rules}

We work in $\cN=1$ superspace formalism and we consider 
the diagrammatic difference of the $\cN=2$ SYM theory with respect to the 
$\cN=4$ theory. We schematically review these techniques and our conventions.

The $\cN=2 $ theory contains both gauge fields, organized in an $\cN=2$ vector multiplet, and matter fields, organized in hypermultiplets. In terms of $\cN=1$ superfields
\begin{align}\label{A.1}
\mathrm{Vector}_{(\cN=2)} &= \big(V, \Phi \big) ~~\mathrm{adjoint~of~}SU(N) \notag \\
\mathrm{Hyper}_{(\cN=2)} &= \big(Q, \widetilde{Q} \big)~~\mathrm{representations~} \cR, \bar{\cR}  ~\mathrm{of~}SU(N)~,
\end{align}
where $V$ is a $\cN=1$ vector superfield, $\Phi, Q, \widetilde{Q}$ are $\cN=1$ chiral superfields. 

In the Fermi-Feynman gauge we separate the part of the action which only involves the adjoint fields
\begin{align}
	\label{A.2}
		S_{\mathrm{gauge}}&=\!\int\!d^4x\,d^2\theta\,d^2\bar{\theta}\,\Big(
		- V^a\square V^a+\Phi^{\dagger a}\Phi^a
		+\frac{i}{4}g f^{abc}\,\big[\bar{D}^2(D^{\alpha} V^a)\big]\, V^b\, (D_{\alpha} V^c) 
		\nonumber\\
		& \qquad\qquad+ 2\,i g f^{abc}\,\Phi^{\dagger a}V^b\Phi^c
		+\cdots\Big),
\end{align}
where the dots stand for higher order vertices and $f^{abc}$ are the structure constants of $SU(N)$.

\begin{figure}[H]
\begin{center}
	\includegraphics[scale=0.6]{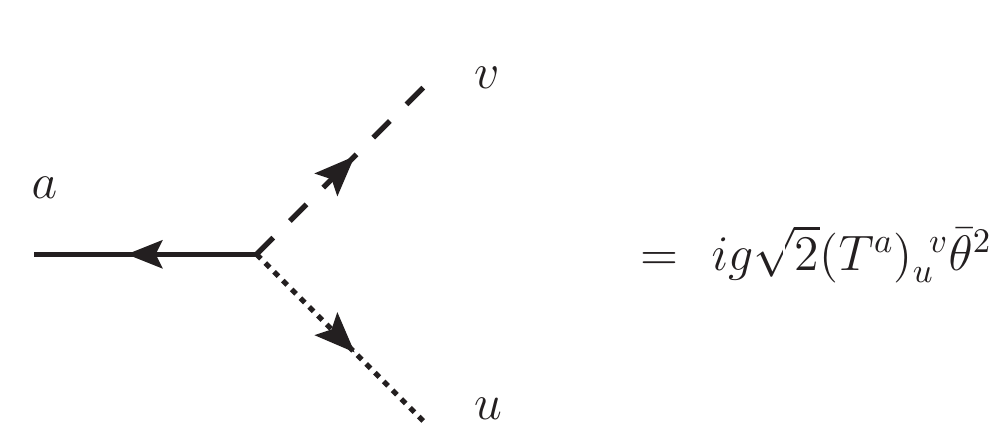}\hspace{1.5cm}
	\includegraphics[scale=0.6]{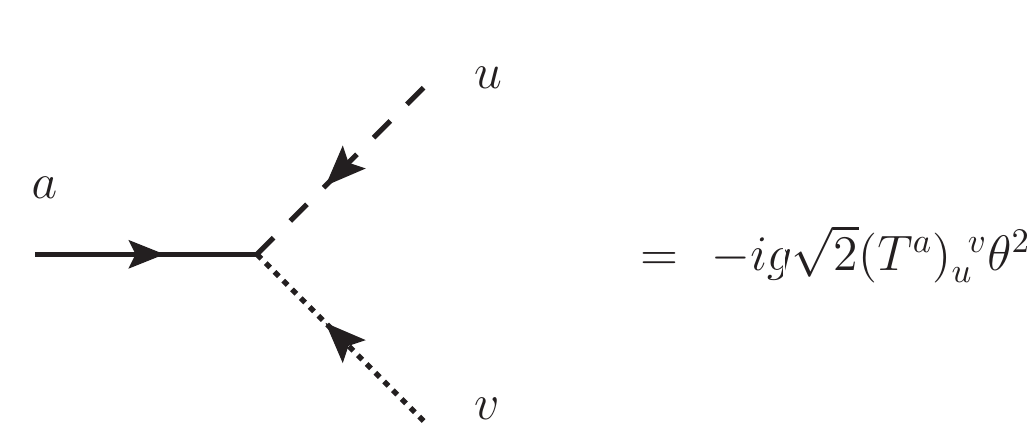} 
	
	\caption{Feynman rules involving $\Phi$, $Q$ and $\tilde{Q}$ chiral fields, with solid, dashed and dotted lines respectively}
	\label{fig:Feynmatter}
\end{center}
\end{figure}

The action for the matter part, again in the Fermi-Feynman gauge, is
\begin{align}
	\label{A.3}
		S_{\mathrm{matter}}& =
		\!\int\!d^4x\,d^2\theta\,d^2\bar{\theta}\,\Big(Q^{\dagger\,u} Q_{u}
		+ 2g\,Q^{\dagger\,u} V^a (T^a)_{u}^{\,v}\,Q_{v}
		+ \widetilde{Q}^{u}\,\widetilde{Q}^\dagger_{u} 
		- 2g\,\widetilde{Q}^{u} \,V^a (T^a)_{u}^{\,v}\,\widetilde{Q}^\dagger_{v}+\cdots\notag\\
		&\qquad\qquad
		+ i\sqrt{2}g\,\widetilde{Q}^{u}\Phi^a (T^a)_{u}^{\, v} Q_{v}\,\bar{\theta}^2
		- i\sqrt{2}g\,Q^{\dagger\,u}\Phi^{\dagger\,a}(T^a)_{u}^{\,v} 
		\widetilde{Q}^\dagger_{v }\,\theta^2
		\Big),
\end{align}
where by $T^a$ we denote the $SU(N)$ generators in the representation $\cR$, and $u,v=1,\ldots \mathrm{dim}_\cR$ includes the cases in which $\cR$ is reducible, namely it contains several copies of a given irreducible representation.\\
In Figure \ref{fig:Feynmatter} we draw the Feynman rules that we need in the present paper.

\noindent The total action for the $\cN=2$ theory is simply
\begin{equation}
	\label{A.4} 
		 S_{\cN=2} = S_{\mathrm{gauge}} +S_{\mathrm{matter}}~.
\end{equation}
The $\cN=4$ SYM theory can be seen as a particular $\cN=2$ theory containing a vector multiplet and an hypermultiplet, both in the adjoint representation of the gauge group. 
So the field content is:
\begin{align}\label{N4fields}
\mathrm{Vector}_{(\cN=4)} = \begin{cases}
\mathrm{Vector}_{(\cN=2)} &= \big(V, \Phi \big) ~~\mathrm{adj~of~}SU(N) \notag \\
\mathrm{Hyper}_{(\cN=2)} &= \big(H, \widetilde{H} \big)~~\mathrm{adjoint~of~}SU(N)~,
\end{cases}
\end{align}
Thus we can write
\begin{equation}
	\label{A.5}
		S_{\cN=4}= S_{\mathrm{gauge}} + S_H~,
\end{equation}
where $S_H$ has the same structure as $S_{\mathrm{matter}}$  with $Q_u$, $\widetilde Q^u$ replaced by $H_a$,$\widetilde H_a$ and the generator components $(T_a)_u^{\, v}$ by the structure constants $i f_{abc}$.

\begin{figure}[H]
	\begin{center}
		\includegraphics[scale=0.5]{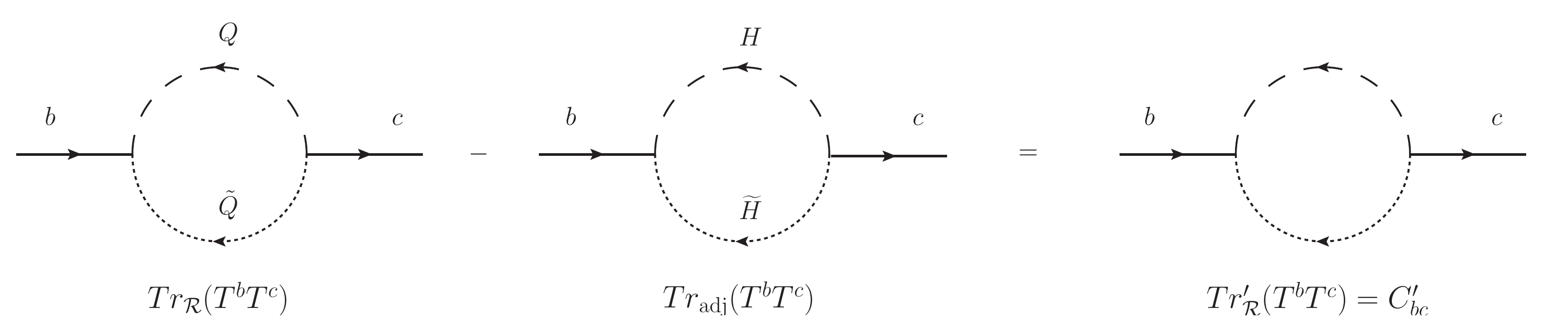}
	\end{center}
	\caption{One-loop correction to $\Phi$ propagator in the difference theory. The color factor is proportional to the $\beta_0$ coefficient, so it vanishes for conformal theories.}
	\label{fig:1loop}
\end{figure}

{From} (\ref{A.4}) and (\ref{A.5}) it is easy to realize that the total action of our $\cN=2$ theory can be written as
\begin{equation}
S_{\cN=2} = S_{\cN=4} - S_H + S_{\mathrm{matter}}~.
\end{equation}
Given any observable $\cA$ of the $\cN=2$ theory, which also exists in the
$\cN=4$ theory, we can write
\begin{equation}
\label{A.7}
\Delta\cA= \mathcal{A}_{\cN=2} - \mathcal{A}_{\cN=4} = \mathcal{A}_{\mathrm{matter}} - \mathcal{A}_H~.
\end{equation}
Thus, if we compute the difference with respect to the $\cN=4$ result, we have to consider only 
diagrams where the hypermultiplet fields, either of the $Q$, $\widetilde Q$ type or of
the $H$, $\widetilde{H}$ type, propagate in the internal lines, and then take the difference
between the $(Q,\widetilde{Q})$ and the $(H,\widetilde{H})$ diagrams.
This procedure reduces in a significant way
the number of diagrams to be computed. 
The first simple example is the 1-loop correction to the chiral $\Phi$ propagator.
The two diagrams involving $Q$s and $H$s fields have the same Feynman rules and generate the same loop integral, but differ in their color structure. The color combination precisely accounts for the $C^\prime$ tensor that we find in the matrix model, see Figure~\ref{fig:1loop}.

We can generalize this fact for higher order corrections: the only contributions to the difference theory come from a series of building blocks, made of hypermultiplet loops with insertions of adjoint lines, coming from $\Phi$ or $V$ fields. The number of insertions of adjoint lines  counts the power of $g$ and specifies the rank of the color tensor, which is always of the form $C'$, which we found inside the perturbative expansion of the matrix model, see equation \ref{defC}.
\begin{figure}[H]
\includegraphics[scale=0.47]{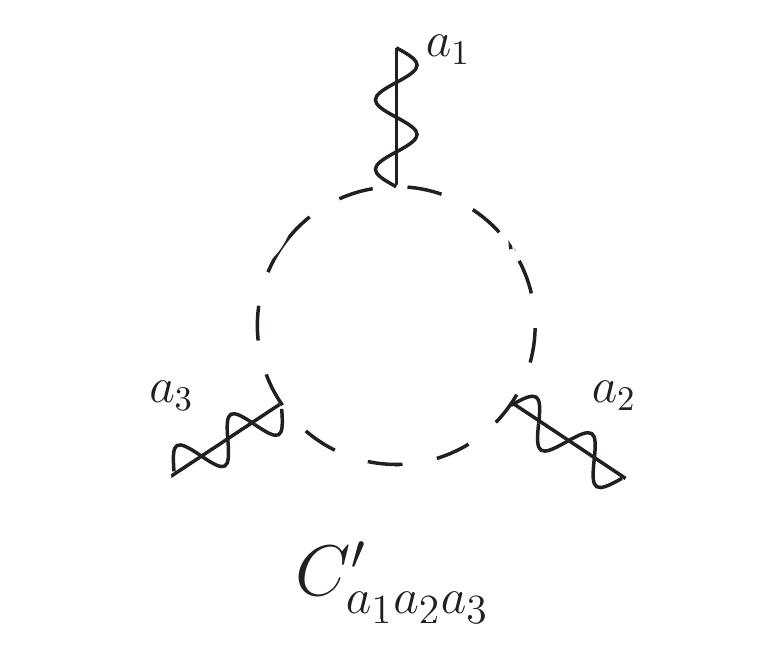}\hspace{0.2cm}
\includegraphics[scale=0.47]{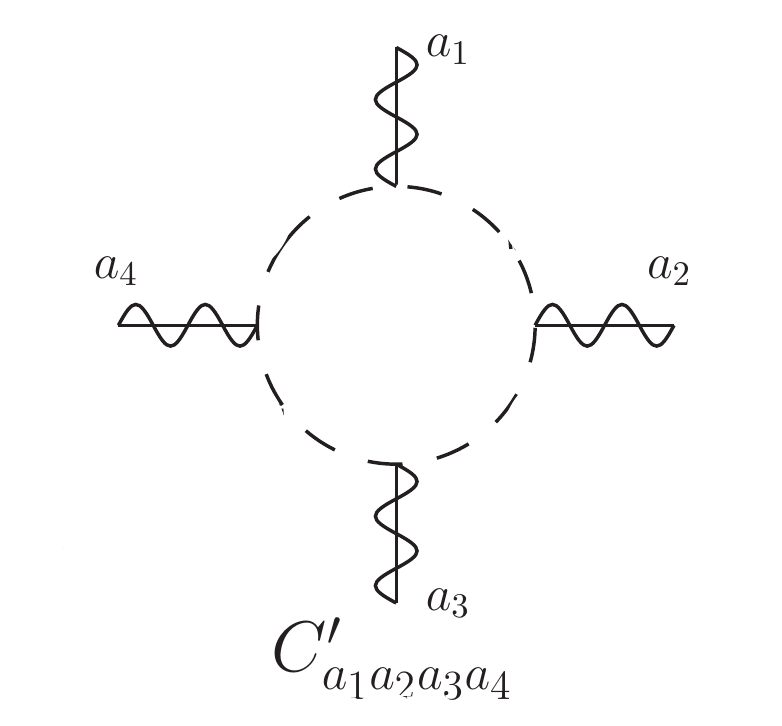}\hspace{.2cm}
\includegraphics[scale=0.47]{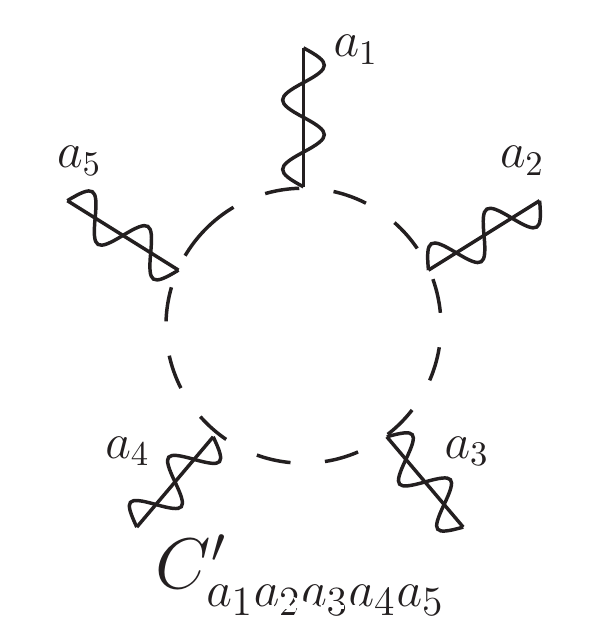}\hspace{.2cm}
\includegraphics[scale=0.47]{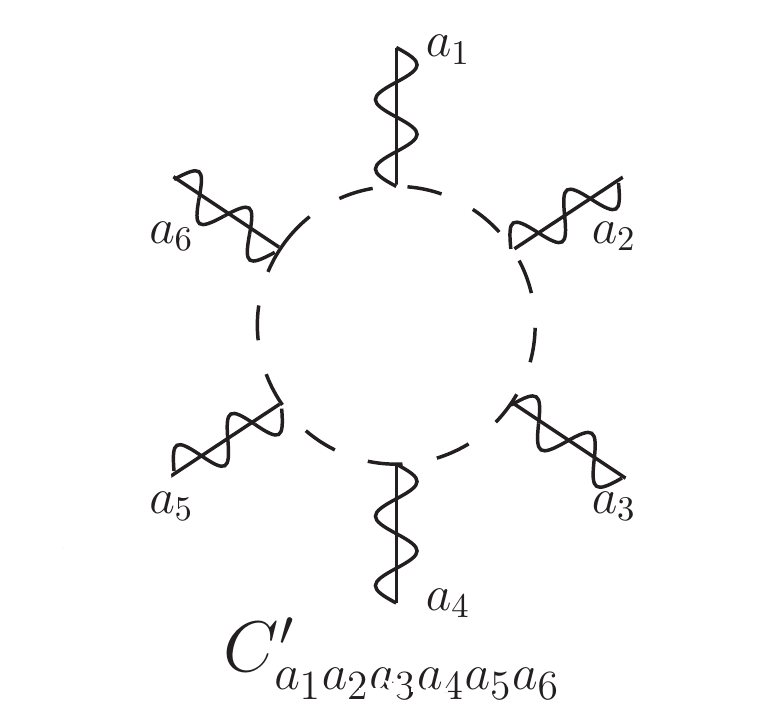}
\caption{Building blocks until $g^6$ order. The overlapped wavy/solid lines stand for generic adjoint fields ($\Phi$ or $V$).}
\label{Fig::BBlocks}
\end{figure}

\begin{figure}[H]
\begin{center}
\includegraphics[scale=0.5]{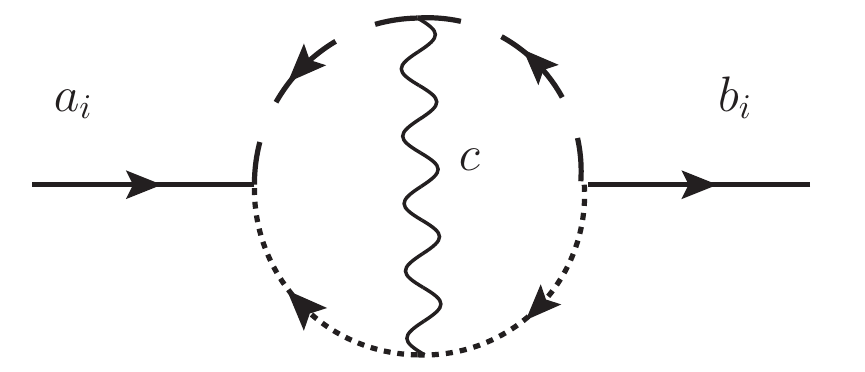}\hspace{1.5cm}
\includegraphics[scale=0.5]{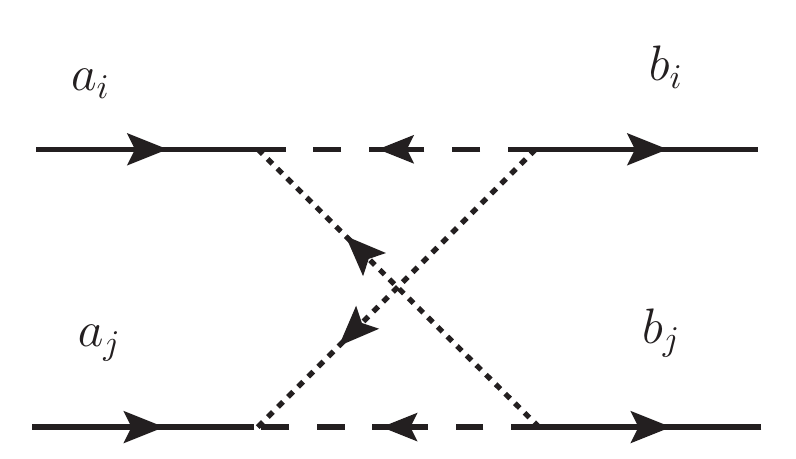}
\end{center}
\caption{The diagrams arising from the building block $C'_{(4)}$, with color factors $C'_{a_i,c,b_i,c}$ and $C'_{a_i,a_j,b_i,b_j}$ respectively. Only the box diagram on the right contributes to the leading order in the double scaling limit}
\label{Fig::2loops}
\end{figure}

Each Feynman diagram is built from these building blocks, after suitable contraction of the adjoint lines.
As an example we easily build all the diagrams coming at order $g^4$, contributing to chiral/anti-chiral correlators. Since all the diagrams built from $C'_{(2)}$ and $C'_{(3)}$ 
vanish due to conformal symmetry  \cite{Billo:2017glv} and since we have two ways to 
close the building block $C'_{(4)}$, there exist two possible diagrams at this order, 
see Figure \ref{Fig::2loops}.

The next orders will be more and more involved. Diagrams built from  $C'_{(4)}$, $C'_{(5)}$, $C'_{(6)}$ will appear at $g^6$ order (see \cite{Billo:2019fbi} for a $g^6$ analysis).

\subsection{Evaluation of the color factors}
The generators $T_a$ with $a=1,\ldots, N^2-1$ of the $\mathfrak{su}(N)$ Lie algebra satisfy the algebra
\begin{align}
	\label{sunalgebrar}
		\big[\,T_a\,,\,T_b\,\big] = i f_{abc}\, T_c~.
\end{align} 
Generators in the fundamental representation are indicated by $t_a$; they are Hermitean, traceless $N\times N$ matrices that we normalize by setting 
\begin{equation}
	\label{norm}
		\tr\,t_a t_b = \frac{1}{2}\,\delta_{ab}~.
\end{equation}
We introduce the totally symmetric tensor $d_{abc}$ as the symmetrized trace of 3 generators:
\begin{align}
	\label{dabc}
		\tr\left(\big\{\,t_a\,,\,t_b\big\}\,t_c\right) = \frac{1}{2}\,d_{abc}~,
\end{align}
Traces of a higher number of generators in the fundamental representation are determined by reducing contractions using 
the following fusion/fission identities:
\begin{align}
	\label{fussion}
		\tr\left(t_a M_1 t_a M_2\right) & = \frac{1}{2}\,\tr\, M_1\, \tr\,M_2
		-\frac{1}{2N}\,\tr\left(M_1 M_2\right)~,\\
		\tr\left(t_a M_1\right) ~\tr\left(t_a M_2\right) & = \frac{1}{2}\,\tr\left(M_1 M_2\right) -\frac{1}{2N}\,\tr\,M_1 \,\tr\,M_2~,
\end{align}
where $M_1$ and $M_2$ are arbitrary $(N\times N)$ matrices.

In a generic representation $\cR$ we have
\begin{equation}
	\label{indexR}
		\Tr_\cR T_a T_b = i_\cR\, \delta_{ab}~,
\end{equation}
where $i_\cR$ is the index of $\cR$.
Higher order traces define a set of cyclic tensors 
\begin{align}
	\label{defCapp}
		C_{a_1\ldots a_m} = \Tr_{\cR} T_{a_1}\ldots T_{a_m}~.
\end{align}
In our computations we encounter the particular combination of traces introduced in 
(\ref{defC}), namely
\begin{align}
	\label{defCagain}
		C^\prime_{a_1\ldots a_m} =\, 
		\Tr_{\cR} T_{a_1}\ldots T_{a_m}- 
		\Tr_{\mathrm{adj}} \,T_{a_1}\ldots T_{a_m}~,
\end{align}
and in particular:
\begin{equation}
	\label{C2is} 
	 C^\prime_{a_1 a_2} = \left(i_{\cR} - i_{\mathrm{adj}}\right) \delta_{a_1 a_2} = 
		\left(i_{\cR} - N\right) \delta_{a_1 a_2} = - \frac{\beta_0}{2}\, \delta_{a_1 a_2}
\end{equation}
where $\beta_0$ the one-loop coefficient of the $\beta$-function of the corresponding $\cN=2$ gauge theory. 
In superconformal models, one has $\beta_0=0$. 
If we consider a representation $\cR$ made of $N_F$ fundamental, $N_S$ symmetric and $N_A$ anti-symmetric representations, we have:
\begin{align}
	\label{RFSAb0}
		{\beta_0}=\big(N_F + N_S(N+2) + N_A(N-2) - 2N\big)=0
	\end{align}
Solutions of this equation for $N_F$, $N_S$ and $N_A$ determine the 5 superconformal theories for $SU(N)$ gauge group in Table \ref{tab:scft}.

Higher order $C^\prime$ tensors can be computed in terms of fundamental traces using the formula (see App. A of \cite{Billo:2019fbi} for more details):
\begin{align}
	\label{colfact2}
		C^\prime_{a_1\ldots a_m} & = \Big[(N_F + 2^{m-1} \big(N_S - N_A\big) 
		+ N \big(N_S + N_A -(1 + (-1)^m)\big) \Big]\, \tr T_{a_1}\ldots T_{a_m}
	\nonumber\\
	& ~~~+ \sum_{p=1}^{m-1} \binom{m}{p} \left(\frac{N_S + N_A}{2} - (-1)^{m-p}\right)\tr T_{a_1}\ldots T_{a_p}\, \tr T_{a_{p+1}}\ldots T_{a_m}~.		
\end{align}
which can be further reduced using \eqref{fussion}.

\section{Angular integration over $SU(N)$}
\la{app:angular}

An important step in the concrete application of (\ref{3.28}) is the calculation of the angular integration $\llangle \circ\rrangle$
over the sphere $S^{N-1}$  with traceless constraint.  It is closely related
to  the $\mc N=4$ expectation value with $SU(N)$ gauge group
\be
\langle \mc O(a) \rangle_{\mc N=4} = \mathscr N\,\int d^{N}a \, \prod_{\mu<\nu}^{N}(a_{\mu}-a_{\nu})^{2}\, \delta(\tr a)\, e^{-\tr (a^{2})} \, \mc O(a),
\ee
where $\mathscr N$ is taken such that $\langle 1 \rangle_{\mc N=4}=1$, and we have rescaled
the matrix $a$ in order to have Gaussian measure $\sim e^{-\tr a^{2}}$ as above.
To make the relation clear, let us consider a  homogeneous function $\mc O(\lambda \, a) = \lambda^{d_{\mc O}}\,\mc O(a)$. We can 
introduce polar coordinates and write \footnote{Notice that a factor $r^{-1}$ comes from the $\delta$ function.}
\be
\langle \mc O(a) \rangle_{\mc N=4} = C_{N}\,\llangle \mc O(a)\rrangle\,\int_{0}^{\infty}dr\,e^{-r^{2}}\,r^{N(N-1)+N-2+d_{\mc O}}.
\ee
Fixing $C_{N}$ by the requirement $\llangle 1 \rrangle = 1$ gives the explicit formula
\be
\la{B.3}
\llangle \mc O(a) \rrangle = \langle \mc O(a) \rangle_{\mc N=4}\,\frac{\Gamma\left(\frac{1}{2}(N^{2}-1)\right)}
{\Gamma\left(\frac{1}{2}(N^{2}-1+d_{\mathcal{O}})\right)}.
\ee
Using the results in \cite{Billo:2017glv} it is easy to compute this formula for operators $\mc O(a)$ with large dimension $d_{\mc O}$. Examples are 
\begin{align}
\la{B.4}
& \llangle \tr(a^{2})^{p}\,X\rrangle = \llangle X\rrangle, \qquad 
\llangle \tr(a^{4}) \rrangle = \frac{2 N^2-3}{N (N^2+1)}, \qquad 
\llangle \tr(a^{3})^{2}\rrangle = \frac{3 (N-2) (N+2)}{N (N^2+1) (N^2+3)}, \notag \\
& \llangle \tr(a^{6}) \rrangle = \frac{5 (N^4-3 N^2+3)}{N^2 (N^2+1) (N^2+3)}, \qquad 
\llangle \tr(a^{3})\,\tr(a^{5})\rrangle = \frac{15 (N-2) (N+2) (N^2-2)}{N^2 (N^2+1) (N^2+3) (N^2+5)},
\end{align}
and so on.

\section{Weak-coupling expansion of the scaling functions: Higher order terms}
\la{app:results}

In this Appendix, we give the terms in (\ref{3.34})
for the $\mathbf{A}$ and $\mathbf{E}$ models, 
keeping only the non-vanishing quantities. We avoid writing down similar expansions for the $\mathbf{BCD}$, 
however these results are available upon request.

\subsection{Scaling function $F^{(2)}(\kappa; N)$}
We define 
\begin{align}
P^{(2)}(k)= 2^k\frac{\Gamma\left(\frac{N^2+1}{2}+k\right)}{\Gamma\left(\frac{N^2+1}{2}\right)}~,
\end{align}
which will be useful in making all the formulas more compact.

\paragraph{Model A} 
\begin{align}
f^{(2)}_{3}&= -\frac{9}{2},  \notag \\
f^{(2)}_{5}&= \frac{25  \left(2 N^2-1\right)}{N \left(N^2+3\right)},  \notag \\
f^{(2)}_{7}&= -\frac{1225  \left(8 N^6+4 N^4-3 N^2+3\right)}{16 N^2 P^{(2)}(3)}, \notag \\
f^{(2)}_{9}&= \frac{1323\left(26 N^8+28 N^6-3 N^4+6 N^2-9\right) }{4 N^3 P^{(2)}(4)},  \notag \\
f^{(2)}_{11}&= -\frac{17787 \left(122 N^{10}+280 N^8+48 N^6-15 N^4+45\right)}{16 N^4 P^{(2)}(5)}, \notag \\
f^{(2)}_{5,5}&= \frac{5775 (N-2) (N+2) \left(N^6-N^4-43 N^2-37\right)}{2 \left(N^2+3\right) P^{(2)}(5)}, \notag \\
f^{(2)}_{13}&= \frac{552123\left(N^2+1\right) \left(34 N^{10}+110 N^8-29 N^6+20 N^4-15\right)}{8 N^5 P^{(2)}(6)}, \notag \\
f^{(2)}_{5,7}&= -\frac{15015 (N-2) (N+2) \left(22 N^8+11 N^6-1167 N^4-531 N^2+705\right)}{2 N \left(N^2+3\right) P^{(2)}(6)}, \notag \\
f^{(2)}_{15}&= -\frac{41409225 \left(540 N^{14}+3780 N^{12}+4676 N^{10}+440 N^8+329 N^6-735 N^4+735 N^2+315\right)}{512 N^6 P^{(2)}(7)}, \notag \\
f^{(2)}_{7,7}&=  \frac{1576575 (N^2-4)}{64 N^2 P^{(2)}(3)P^{(2)}(7)}\big(98 N^{14}+928 N^{12}-4151 N^{10}-44359 N^8-42036 N^6+26754 N^4 \notag \\&\qquad\qquad\qquad\qquad\qquad+14553 N^2-16299\big),  \notag \\
f^{(2)}_{5,9}&= \frac{675675 (N-2) (N+2) \left(23 N^{10}+70 N^8-1455 N^6-1335 N^4+192 N^2-855\right)}{4 N^2  \left(N^2+3\right) P^{(2)}(7)}.
\end{align}

\paragraph{Model E} 
\begin{align}
f^{(2)}_{5}&= -\frac{100 \left(N^2-4\right)}{N P^{(2)}(2)},  \notag \\
f^{(2)}_{7}&= \frac{3675 \left(N^4-6 N^2+8\right) \zeta (7)}{N^2 P^{(2)}(3)}, \notag \\
f^{(2)}_{9}&= -\frac{15876 \left(7 N^6-53 N^4+136 N^2-144\right) \zeta (9)}{N^3 P^{(2)}(4)}, \notag \\
f^{(2)}_{11}&= \frac{1067220 \left(N^2-4\right) \left(3 N^6-13 N^4+33 N^2-48\right)}{N^4 P^{(2)}(5)}, \notag \\
f^{(2)}_{5,5}&= \frac{23100 \left(N^2-4\right) \left(N^6+26 N^4-121 N^2-626\right)}{P^{(2)}(2)P^{(2)}(5)}, \notag \\
f^{(2)}_{13}&= -\frac{8281845 \left(N^2-4\right) \left(11 N^8-43 N^6+112 N^4-320 N^2+640\right)}{N^5 P^{(2)}(6)},  \notag \\
f^{(2)}_{5,7}&= -\frac{1801800 \left(N^2-4\right)^2 \left(N^6+43 N^4-121 N^2-883\right)}{N P^{(2)}(2)P^{(2)}(6)}, \notag \\
f^{(2)}_{15}&= \frac{289864575 (N-2) (N+2) \left(143 N^{10}-275 N^8-708 N^6+2880 N^4+4800 N^2-34560\right)}
{16 N^6 P^{(2)}(7)}, \notag \\
f^{(2)}_{7,7}&=  \frac{4729725 (N^2-4)}{4 N^2 P^{(2)}(3)P^{(2)}(7)}\big(31 N^{12}+1924 N^{10}-8334 N^8-68608 N^6+316415 N^4 \notag \\&\qquad\qquad\qquad\qquad\qquad+498684 N^2-3764112\big),  \notag \\
f^{(2)}_{5,9}&= \frac{2027025 (N^2-4) \left(27 N^{10}+1553 N^8-14171 N^6+11887 N^4+159104 N^2-521280\right)}{N^2 P^{(2)}(2)P^{(2)}(7)}.
\end{align}

\subsection{Scaling function $F^{(3)}(\kappa; N)$}
We define, similarly to what we did before: 
\begin{align}
P^{(3)}(k)= 2^k\frac{\Gamma\left(\frac{N^2+5}{2}+k\right)}{\Gamma\left(\frac{N^2+5}{2}\right)}.
\end{align}

\paragraph{Model A} 
\begin{align}
f^{(3)}_{3}&= -\frac{9}{2}, \notag \\
f^{(3)}_{5}&= \frac{25 \left(2 N^6+43 N^4+60 N^2-105\right) }{N P^{(3)}(3)}, \notag \\
f^{(3)}_{7}&=  -\frac{1225 \left(8 N^8+260 N^6+281 N^4-378 N^2+693\right) }{16 N^2P^{(3)}(4)},  \notag \\
f^{(3)}_{9}&= \frac{1323 \left(26 N^{10}+1154 N^8+2129 N^6-213 N^4+765 N^2-3861\right) }{4 N^3 P^{(3)}(5)}, \notag \\
f^{(3)}_{11}&= -\frac{17787 \left(122 N^{12}+6950 N^{10}+24848 N^8+8085 N^6-12645 N^4+15345 N^2+32175\right)}
{16 N^4 P^{(3)}(6)}, \notag \\
f^{(3)}_{5,5}&= \frac{5775 }{2 P^{(3)}(3)P^{(3)}(6)}\big(N^{14}+88 N^{12}+15 N^{10}-18088 N^8-39661 N^6+1053540 N^4\notag \\ &\qquad\qquad\qquad\qquad\qquad+4281405 N^2+4399500\big), \notag \\
f^{(3)}_{13}&=  \frac{552123 \left(N^2-1\right)}{8 N^5 P^{(3)}(7)}\big(34 N^{12}+2424 N^{10}+17285 N^8+29655 N^6+24450 N^4\notag \\ &\qquad\qquad\qquad\qquad\qquad+41145 N^2+16575\big), \notag \\
f^{(3)}_{5,7}&= -\frac{15015 \left(N^2-1\right) }{2 N P^{(3)}(3)P^{(3)}(7)}\big(22 N^{14}+2323 N^{12}+6951 N^{10}-473938 N^8-1641088 N^6\notag \\ &\qquad\qquad\qquad\qquad\qquad+30589515 N^4+136118115 N^2+144364500\big),\notag \\
f^{(3)}_{15}&= -\frac{41409225}{512 N^6 P^{(3)}(8)}\big(540 N^{16}+45624 N^{14}+445760 N^{12}+771476 N^{10}+62225 N^8\notag \\ &\qquad\qquad\qquad\qquad\qquad+212520 N^6-739410 N^4+1434300 N^2+508725\big), \notag \\
f^{(3)}_{7,7}&=  \frac{1576575 }{64 N^2 P^{(3)}(4)P^{(3)}(8)}\big(98 N^{20}+13168 N^{18}+195787 N^{16}-1872045 N^{14}-38053705 N^{12}\notag \\&\qquad\qquad\qquad\qquad\qquad+100653479 N^{10}+2925371701 N^8+6196842921 N^6\notag \\&\qquad\qquad\qquad\qquad\qquad-4983588981 N^4-7790105043 N^2+11951297820\big), \notag \\
f^{(3)}_{5,9}&= \frac{675675}{4 N^2 P^{(3)}(3)P^{(3)}(8)}\big(23 N^{18}+2775 N^{16}+14758 N^{14}-564239 N^{12}-2460606 N^{10}\notag \\&\qquad\qquad\qquad\qquad\qquad+41052509 N^8+170225310 N^6+132265575 N^4\notag \\&\qquad\qquad\qquad\qquad\qquad+70609635 N^2+273142260\big).
\end{align}

\paragraph{Theory E} 
\begin{align}
f^{(3)}_{5}&= -\frac{300 \left(N^4+19 N^2-140\right) }{N P^{(3)}(3)}, \notag \\
f^{(3)}_{7}&= \frac{11025 \left(N^6+29 N^4-286 N^2+616\right) \zeta (7)}{N^2 P^{(3)}(4)}, \notag \\
f^{(3)}_{9}&= \frac{7938 \left(41 N^8+1745 N^6-19474 N^4+73200 N^2-123552\right)}{N^3 P^{(3)}(5)}, \notag \\
f^{(3)}_{11}&= \frac{533610 \left(17 N^{10}+1022 N^8-11221 N^6+49278 N^4-150936 N^2+274560\right)}{N^4 P^{(3)}(6)}, \notag \\
f^{(3)}_{5,5}&= \frac{69300 \left(N^{12}+92 N^{10}+2936 N^8-6402 N^6-389297 N^4+198070 N^2+14709800\right)}
{P^{(3)}(3)P^{(3)}(6)}, \notag \\
f^{(3)}_{13}&= -\frac{552123}{N^5 P^{(3)}(7)}\big(451 N^{12}+37030 N^{10}-335741 N^8+1002420 N^6-2721600 N^4\notag \\ &\qquad\qquad\qquad\qquad\qquad+14421120 N^2-42432000\big),  \notag \\
f^{(3)}_{5,7}&= -\frac{5405400 }{N P^{(3)}(3)P^{(3)}(7)}\big(N^{14}+123 N^{12}+5062 N^{10}-24814 N^8-691079 N^6+3421571 N^4\notag \\ &\qquad\qquad\qquad\qquad\qquad+28187136 N^2-161534800\big), \notag \\
f^{(3)}_{15}&= \frac{289864575}{16 N^6 P^{(3)}(8)}\big(377 N^{14}+41078 N^{12}-212527 N^{10}-1183248 N^8\notag \\ &\qquad\qquad\qquad\qquad\qquad+12210960 N^6-39049920 N^4+8148480 N^2+223257600\big), \notag \\
f^{(3)}_{7,7}&=  \frac{14189175 }{4 N^2 P^{(3)}(4)P^{(3)}(8)}\big(31 N^{18}+5386 N^{16}+319785 N^{14}+1056052 N^{12}\notag \\ &\qquad\qquad\qquad\qquad\qquad-55231871 N^{10}-54950742 N^8+4657773463 N^6\notag \\ &\qquad\qquad\qquad\qquad\qquad-8789685416 N^4-165039789408 N^2+691827702720\big), \notag \\
f^{(3)}_{5,9}&= \frac{6081075 }{N^2 P^{(3)}(3)P^{(3)}(8)}\big(27 N^{16}+4355 N^{14}+232522 N^{12}\notag \\ &\qquad\qquad\qquad\qquad\qquad-1340066 N^{10}-29934729 N^8+223699871 N^6+827329700 N^4\notag \\ &\qquad\qquad\qquad\qquad\qquad-12426876800 N^2+38840613120\big).
\end{align}

    \section{Explicit expansions for the 2-point functions at low rank}
    \la{app:specialization}

    In this Appendix, we report the $\mc O(\kappa^{11})$ expansions of $\log F^{(\Delta)}(\kappa; N)$ for $\Delta=2,3$, $N=3,4$, i.e. the SU(3)
    and SU(4) theories, and for all the models $\mathbf{A}$ and $\mathbf{E}$. For $SU(3)$ model $\mathbf{E}$ is the same as model $\mathbf{B}$.
    There is of course agreement in the cases dealt with at lower order in (\ref{1.10}).

    \subsection{$SU(3)$}
    \paragraph{Model $\mathbf{A}$}
    \begin{align}
        \log & \, F^{\mathbf{A}\,(2)}(\kappa; 3) = -\frac{9 \kappa^2 \zeta (3)}{2}+\frac{425 \kappa^3 \zeta (5)}{36}-\frac{17885 \kappa^4 \zeta (7)}{576}+\frac{5565 \kappa^5 \zeta (9)}{64}+
        \kappa^6 \left(\frac{1925 \zeta (5)^2}{3456}-\frac{2668897 \zeta (11)}{10368}\right) \notag\notag \\
            &+\kappa^7 \left(\frac{32984237 \zeta (13)}{41472}-\frac{5005 \zeta (5) \zeta (7)}{864}\right)+\kappa^8 \left(\frac{35035 \zeta (7)^2}{2304}+\frac{146575 \zeta (5) \zeta (9)}{6144}-\frac{2245755655 \zeta (15)}{884736}\right) \notag \\
            &+\kappa^9 \left(-\frac{1519375 \zeta (5)^3}{4478976}-\frac{546184925 \zeta (11) \zeta (5)}{5971968}-\frac{3488485 \zeta (7) \zeta (9)}{27648}+\frac{669686057755 \zeta (17)}{80621568}\right)\notag \\
            &+\kappa^{10} \left(\frac{8083075 \zeta (7) \zeta (5)^2}{1492992}+\frac{4074100745 \zeta (13) \zeta (5)}{11943936}+\frac{77643709 \zeta (9)^2}{294912}+\frac{2905703801 \zeta (7) \zeta (11)}{5971968}\right. \notag\\
                &\phantom{aaaaaaaa}\left. -\frac{29805018472801 \zeta (19)}{1074954240}\right)+O\left(\kappa^{11}\right) \\
        \log & \, F^{\mathbf{A}\,(3)}(\kappa; 3) = -\frac{9 \kappa^2 \zeta (3)}{2}+\frac{100 \kappa^3 \zeta (5)}{9}-\frac{15925 \kappa^4 \zeta (7)}{576}+\frac{147 \kappa^5 \zeta (9)}{2}+\kappa^6 \left(\frac{1925 \zeta (5)^2}{3456}-\frac{8599591 \zeta (11)}{41472}\right) \notag\\
            &+\kappa^7 \left(\frac{3177031 \zeta (13)}{5184}-\frac{5005 \zeta (5) \zeta (7)}{864}\right)+\kappa^8 \left(\frac{35035 \zeta (7)^2}{2304}+\frac{146575 \zeta (5) \zeta (9)}{6144}-\frac{1660676875 \zeta (15)}{884736}\right) \notag \\
            &+\kappa^9 \left(-\frac{3488485 \zeta (7) \zeta (9)}{27648}-\frac{203900125 \zeta (5) \zeta (11)}{2239488}+\frac{29779203025 \zeta (17)}{5038848}\right) \notag \\
            &+\kappa^{10} \left(\frac{77643709 \zeta (9)^2}{294912}+\frac{1084748665 \zeta (7) \zeta (11)}{2239488}+\frac{12054174275 \zeta (5) \zeta (13)}{35831808}-\frac{81777816230539 \zeta (19)}{4299816960}\right)+O\left(\kappa^{11}\right)  
    \end{align}
    \paragraph{Model $\mathbf{B}$}
    \begin{align}
    \la{D.3}
         \log & \, F^{\mathbf{B}\,(2)}(\kappa; 3) = -\frac{25 \kappa^3 \zeta (5)}{18}+\frac{1225 \kappa^4 \zeta (7)}{144}-\frac{1323 \kappa^5 \zeta (9)}{32}+\kappa^6  \left(\frac{1925 \zeta (5)^2}{864}+\frac{1952335 \zeta (11)}{10368}\right)\notag \\
            &+\kappa^7 \left(-\frac{25025 \zeta (5) \zeta (7)}{864}-\frac{17402099 \zeta (13)}{20736}\right)+\kappa^8 \left(\frac{875875 \zeta (7)^2}{9216}+\frac{75075 \zeta (5) \zeta (9)}{512}+\frac{136961825 \zeta (15)}{36864}\right)\notag \\
            &+\kappa^9 \left(-\frac{1519375 \zeta (5)^3}{559872}-\frac{257382125 \zeta (11) \zeta (5)}{373248}-\frac{2977975 \zeta (7) \zeta (9)}{3072}-\frac{1325476118395 \zeta (17)}{80621568}\right)\notag \\
            &+\kappa^{10} \left(\frac{40415375 \zeta (7) \zeta (5)^2}{746496}+\frac{4722594305 \zeta (13) \zeta (5)}{1492992}+\frac{20369349 \zeta (9)^2}{8192}+\frac{6846364525 \zeta (7) \zeta (11)}{1492992}\right. \notag \\
                &\phantom{aaaaaaaa}\left. +\frac{1959930123437 \zeta (19)}{26873856}\right)+O\left(\kappa^{11}\right) \\
        \log & \, F^{\mathbf{B}\,(3)}(\kappa; 3) = -\frac{25 \kappa^3 \zeta (5)}{9}+\frac{1225 \kappa^4 \zeta (7)}{72}-\frac{1323 \kappa^5 \zeta (9)}{16}+\kappa^6 \left(\frac{1925 \zeta (5)^2}{864}+\frac{3908905 \zeta (11)}{10368}\right)\notag\\
            &+\kappa^7 \left(-\frac{25025 \zeta (5) \zeta (7)}{864}-\frac{34947341 \zeta (13)}{20736}\right)+\kappa^8 \left(\frac{875875 \zeta (7)^2}{9216}+\frac{75075 \zeta (5) \zeta (9)}{512}+\frac{138263125 \zeta (15)}{18432}\right)\notag \\
            &+\kappa^9 \left(-\frac{2977975 \zeta (7) \zeta (9)}{3072}-\frac{1547635375 \zeta (5) \zeta (11)}{2239488}-\frac{2696833678825 \zeta (17)}{80621568}\right)\notag \\
            &+\kappa^{10} \left(\frac{20369349 \zeta (9)^2}{8192}+\frac{41167100975 \zeta (7) \zeta (11)}{8957952}+\frac{28566741775 \zeta (5) \zeta (13)}{8957952}+\frac{32217517886983 \zeta (19)}{214990848}\right)+O\left(\kappa^{11}\right)
    \end{align}
    \subsection{$SU(4)$}
    \paragraph{Model $\mathbf{A}$}
    \begin{align}
        \log & \, F^{\mathbf{A}\,(2)}(\kappa; 4) =- \frac{9 \kappa^2 \zeta (3)}{2}+\frac{775 \kappa^3 \zeta (5)}{76}-\frac{1968575 \kappa^4 \zeta (7)}{82688}+\frac{4979583 \kappa^5 \zeta (9)}{82688}\notag\\ 
            &+\kappa^6 \left(\frac{205590 \zeta (5)^2}{141151}-\frac{24811876551 \zeta (11)}{152145920}\right) +\kappa^7 \left(\frac{438160723 \zeta (13)}{942080}-\frac{3750747 \zeta (5) \zeta (7)}{282302}\right)\notag \\ 
            &+\kappa^8 \left(\frac{547389173331 \zeta (7)^2}{17814385408}+\frac{110482515 \zeta (5) \zeta (9)}{2258416}-\frac{623128271382045 \zeta (15)}{451812524032}\right)\notag \\ 
            &+\kappa^9 \left(-\frac{263656250 \zeta (5)^3}{425470629}-\frac{978640180375 \zeta (11) \zeta (5)}{5732656896}-\frac{119950027197 \zeta (7) \zeta (9)}{523952512}+\frac{5207730821398105 \zeta (17)}{1235840139264}\right) \notag \\
            &+\kappa^{10} \left(\frac{7360327975 \zeta (7) \zeta (5)^2}{1015157992}+\frac{37291512115 \zeta (13) \zeta (5)}{63519744}+\frac{237530487051 \zeta (9)^2}{551528960}+\frac{2282495988773 \zeta (7) \zeta (11)}{2829914112}\right. \notag \\
                &\phantom{aaaaaaaa}\left.-\frac{68597690492349161 \zeta (19)}{5203537428480}\right)+O\left(\kappa^{11}\right) \\
        \log & \, F^{\mathbf{A}\,(3)}(\kappa; 4) = -\frac{9 \kappa^2 \zeta (3)}{2}+\frac{955 \kappa^3 \zeta (5)}{92}-\frac{429289 \kappa^4 \zeta (7)}{17664}+\frac{78057 \kappa^5 \zeta (9)}{1280} \notag \\
            &+\kappa^6 \left(\frac{293062 \zeta (5)^2}{475571}-\frac{68971014343 \zeta (11)}{423464960}\right)+\kappa^7 \left(\frac{387146868537 \zeta (13)}{846929920}-\frac{28131103 \zeta (5) \zeta (7)}{4755710}\right)\notag\\
            &+\kappa^8 \left(\frac{8779133363 \zeta (7)^2}{608730880}+\frac{7550829 \zeta (5) \zeta (9)}{330832}-\frac{57815474734017 \zeta (15)}{43362811904}\right)\notag\\
            &+\kappa^9 \left(-\frac{296509070 \zeta (5)^3}{3642398289}-\frac{372030392019 \zeta (11) \zeta (5)}{4504608512}-\frac{51442391 \zeta (7) \zeta (9)}{456320}+\frac{28934892829055213 \zeta (17)}{7219908182016}\right)\notag\\
            &+\kappa^{10} \left(\frac{95215713593 \zeta (7) \zeta (5)^2}{145695931560}+\frac{26266560937331 \zeta (13) \zeta (5)}{90092170240}+\frac{58925478183 \zeta (9)^2}{264665600}+\frac{6682311515022143 \zeta (7) \zeta (11)}{16216590643200}\right.\notag\\ 
                &\phantom{aaaaaaaa}\left.-\frac{2375425254570780269 \zeta (19)}{192530884853760}\right)+O\left(\kappa^{11}\right)
    \end{align}
    \paragraph{Model $\mathbf{E}$}
    \begin{align}
        \log & \, F^{\mathbf{E}\,(2)}(\kappa; 4) = -\frac{300 \kappa^3 \zeta (5)}{323}+\frac{3675 \kappa^4 \zeta (7)}{646}-\frac{11907 \kappa^5 \zeta (9)}{437}+\kappa^6 \left(\frac{4324320 \zeta (5)^2}{2399567}+\frac{899514 \zeta (11)}{7429}\right)\notag\\ 
            &+\kappa^7 \left(-\frac{56216160 \zeta (5) \zeta (7)}{2399567}-\frac{7791069 \zeta (13)}{14858}\right)+\kappa^8 \left(\frac{184459275 \zeta (7)^2}{2399567}+\frac{486486000 \zeta (5) \zeta (9)}{4093379}+\frac{31105689615 \zeta (15)}{13788224}\right)\notag\\
            &+\kappa^9 \left(-\frac{177866832000 \zeta (5)^3}{40987003927}-\frac{71209238100 \zeta (11) \zeta (5)}{126894749}-\frac{189189000 \zeta (7) \zeta (9)}{240787}-\frac{183399698911 \zeta (17)}{18857424}\right)\notag \\
            &+\kappa^{10} \left(\frac{186760173600 \zeta (7) \zeta (5)^2}{2157210733}+\frac{17296663956 \zeta (13) \zeta (5)}{6678671}+\frac{91270827648 \zeta (9)^2}{45179245}+\frac{24923233335 \zeta (7) \zeta (11)}{6678671}\right.\notag\\ &\phantom{aaaaaaaa}\left.+\frac{69791717567 \zeta (19)}{1654160}\right)+O\left(\kappa^{11}\right) \\
        \log & \, F^{\mathbf{E}\,(3)}(\kappa; 4) = -\frac{60 \kappa^3 \zeta (5)}{23}+\frac{735 \kappa^4 \zeta (7)}{46}-\frac{51597 \kappa^5 \zeta (9)}{667}+\kappa^6 \left(\frac{1840608 \zeta (5)^2}{475571}+\frac{7252014 \zeta (11)}{20677}\right)\notag \\
            &+\kappa^7 \left(-\frac{23927904 \zeta (5) \zeta (7)}{475571}-\frac{322623873 \zeta (13)}{206770}\right)+\kappa^8 \left(\frac{78513435 \zeta (7)^2}{475571}+\frac{123243120 \zeta (5) \zeta (9)}{475571}+\frac{9169675515 \zeta (15)}{1323328}\right)\notag \\
            &+\kappa^9 \left(-\frac{2226407040 \zeta (5)^3}{404710921}-\frac{22130316780 \zeta (11) \zeta (5)}{17596127}-\frac{814773960 \zeta (7) \zeta (9)}{475571}-\frac{1136214410617 \zeta (17)}{36722352}\right)\notag \\
            &+\kappa^{10} \left(\frac{44416820448 \zeta (7) \zeta (5)^2}{404710921}+\frac{529221183348 \zeta (13) \zeta (5)}{87980635}+\frac{11411072264592 \zeta (9)^2}{2551438415}+\frac{147166606587 \zeta (7) \zeta (11)}{17596127}\right. \notag\\
                &\phantom{aaaaaaaa}\left.+\frac{74162582637 \zeta (19)}{532208}\right)+O\left(\kappa^{11}\right)
    \end{align}

\section{Resummation of  $\zeta(5)^{k}$ contributions in the two-point functions of the $SU(3)$ theories}
\la{app:zeta5}

It seems very interesting to explore the large $\kappa$ behaviour of the   expansions (\ref{3.35}) and (\ref{3.36}). As we 
recalled in the introduction, this is possible in the $SU(2)$ model $\mathbf A$, see (\ref{1.8}).
For $SU(N)$  with $N>2$ this seems a very hard task since multiple products of $\zeta$-number appear
even after taking the logarithm of $F$ or $G$. Nevertheless, let us show how to resum all terms
proportional to $\zeta(5)^{k}$ in the two $SU(3)$ theories. For the $SU(N)$ 
 $\mathbf{A}$ and $\mathbf{E}$ models
the $\zeta(3)$ terms are already resummed, i.e. they appear as a single term in $\log F$ and $\log G$.
This is not true in the other $\mathbf{BCD}$ models. For $SU(3)$, this is the case thanks to the identifications (\ref{1.5}).
Of course, we are not claiming that this partial resummation is dominant in any sense. We just show that such
contributions may be  resummed and this might hopefully hint at some
structure or generalizations.

\paragraph{$F^{(2)}$ scaling function}

Let us begin with the scaling function $F^{(2)}$
in the $\mathbf{B}$ model, 
that turns out to be the simplest. We can write the function $\widetilde{f}^{(2)}(\kappa; 3)$ in  (\ref{3.30}) as 
\begin{align}
\exp \widetilde f^{\mathbf{B}\,(2)}(\kappa; 3) |_{\zeta(5)} & = \sum_{n=0}^{\infty}\frac{1}{n!}(-10\,\zeta(5)\,\kappa^{3})^{n}
\,\llangle [\tr(a^{3})]^{n}[\tr(a^{5})]^{n}\rrangle.
\end{align}
In $SU(3)$ we have $\tr(a^{5}) = \frac{5}{6}\,\tr(a^{2})\tr(a^{3})$. Hence, cf. the first relation in (\ref{B.4}), 
\begin{align}
\la{E.2}
\exp \widetilde f^{\mathbf{B}\,(2)}(\kappa; 3) |_{\zeta(5)} & = \sum_{n=0}^{\infty}\frac{1}{n!}\left(-\frac{5\,\zeta(5)\kappa^{3}}{3}
\right)^{n}\,\llangle [\tr(a^{3})]^{2n}\rrangle.
\end{align}
For general $N$ the expression of 
\be
t_{n} = \llangle [\tr(a^{3})]^{2n}\rrangle,
\ee
has not a simple dependence on $n$. Indeed
\begin{align}
t_{1} &= \frac{3 (N-2) (N+2)}{N (N^2+1) (N^2+3)},\notag \\
t_{2} &= \frac{27 (N-2) (N+2) 
(N^4+19 N^2-140)}{N^2 (N^2+1) (N^2+3) (N^2+5) (N^2+7) 
(N^2+9)}\notag \\
t_{3} &= \frac{405 (N-2) (N+2) (N^8+62 N^6+969 N^4-20632 
N^2+80080)}{N^3 (N^2+1) (N^2+3) (N^2+5) (N^2+7) (N^2+9) (N^2+11) 
(N^2+13) (N^2+15)}.
\end{align}
Nevertheless, for $N=3$ one has a simpler result (that may also be obtained by an explicit angular parametrization of the matrix $a$)
\be
\la{E.5}
SU(3):\qquad t_{n} = \frac{\Gamma(n+\frac{1}{2})}{6^{n}\,\sqrt\pi\,(n+1)!}.
\ee
Plugging this in (\ref{E.2}) gives 
\begin{align}
\exp \widetilde f^{\mathbf{B}\,(2)}(\kappa; 3) |_{\zeta(5)} &= \exp\left(-\frac{5\,\zeta(5)}{36}\,\kappa^{3}\right)\,\left[I_{0}\left(\frac{5\,\zeta(5)}{36}
\,\kappa^{3}\right)+I_{1}\left(\frac{5\,\zeta(5)}{36}\,\kappa^{3}\right)\right],
\end{align}
where $I_{n}$ are modified Bessel functions of the first kind. To get $\log F$ from (\ref{3.31}) we have to solve the  problem of expressing $F$ in (\ref{3.31}) in terms of $\widetilde f$ in (\ref{3.30}).
This problem will re-appear in every model and its solution is model independent. From the Mellin transform of the Mar\v{c}enko-Pastur weight we find the following relation (we omit the $N$ variable that 
plays no role in the relation)
\be
\log \left. F^{(2)}(\kappa)\right|_{\zeta(5)} = \int_{0}^{1}dt\,\left. \widetilde{f}(4\kappa\,t^{1/3})\right|_{\zeta(5)}\,\frac{t^{-5/6}}{3\,\pi\,\sqrt{1-t^{1/3}}}.
\ee
Hence, in the $\mathbf{B}$ model we have 
\be
\la{E.8}
\log \left. F^{\mathbf B\,(2)}(\kappa; 3)\right|_{\zeta(5)} = \frac{1}{3\pi}\int_{0}^{1}\frac{dt\,t^{-5/6}}{\sqrt{1-t^{1/3}}}\ \left\{
\log\left[I_{0}\left(\frac{80\,\zeta(5)}{9}\,\kappa^{3}\,t\right)
+I_{1}\left(\frac{80\,\zeta(5)}{9}\,\kappa^{3}\,t\right)\right]-\frac{80\,\zeta(5)}{9}\,\kappa^{3}\,t\right\}.
\ee
The small $\kappa$ expansion of this expressions indeed reproduces all the $\zeta(5)^{k}$ terms we already 
presented in (\ref{D.3}) and generates them for higher order
\begin{align}
& \log \left. F^{\mathbf B}(\kappa; 3)\right|_{\zeta(5)} = 
-\frac{25\, \zeta (5)}{18}\,\kappa ^3 + \frac{1925\,  
\zeta (5)^2}{864}\,\kappa ^6-\frac{1519375\, \zeta 
(5)^3}{559872}\, \kappa ^9\spek
+0\cdot\kappa^{12}+\frac{673253125\,  \zeta 
(5)^5}{80621568}\,\kappa ^{15}-\frac{11816582421875 \,
\zeta (5)^6}{835884417024}\,\kappa ^{18} +\mc O(\kappa ^{21}).
\end{align}
In the $\mathbf A$ model, one finds that in $SU(3)$
\begin{align}
\exp \widetilde f^{\mathbf{A}\,(2)}(\kappa; 3) |_{\zeta(5)} & = \sum_{n=0}^{\infty}\frac{1}{n!}\left(\frac{5\,\zeta(5)}{8}\,\kappa^{3}\right)^{n}
\,\llangle \left(1-[\tr(a^{3})]^{2}\right)^{n}\rrangle.
\end{align}
Expanding and using (\ref{E.5}) we find
\begin{align}
\la{E.11}
\exp & \widetilde f^{\mathbf{A}\,(2)}(\kappa; 3) |_{\zeta(5)}  = \sum_{n=0}^{\infty}\frac{1}{n!}\left(\frac{5\,\zeta(5)}{8}\,\kappa^{3}\right)^{n}\,{}_{2}F_{1}\left(\frac{1}{2}, -n, 2; \frac{2}{9}\right) \notag \\
&= \exp\left(\frac{5\,\zeta(5)}{9}\,\kappa^{3}\right)\,\left[I_{0}\left(\frac{5\,\zeta(5)}{72}
\,\kappa^{3}\right)+I_{1}\left(\frac{5\,\zeta(5)}{72}\,\kappa^{3}\right)\right].
\end{align}
To prove this one needs to compute the series
\be
f(x) = \sum_{n=0}^{\infty}\frac{x^{n}}{n!}\,{}_{2}F_{1}(1/2, -n, 2, 2/9) = \sum_{n=0}^{\infty }c_{n}x^{n}.
\ee
Using the recursion properties of the hypergeometric function, one shows that the sequence $\{c_{n}\}$ obeys 
\be
9\,(n+2)(n+3)\,c_{n+2}-(16 n+33)\,c_{n+1}+7\,c_{n}=0,\qquad c_{0}=1,\quad c_{1}=\frac{17}{18}.
\ee
Thus, $f(x) = x^{-1}\,g(x)$ where $g(x)$ solves the differential problem
\be
9\,x\,g''(x)-16\,x\,g'(x)+(7x-1)\,g(x)=0,\qquad g(0)=0,\quad g'(0)=1.
\ee
The solution is 
\be
g(x) = e^{\frac{8x}{9}}\,x\,\left[I_{0}\left(\frac{x}{9}\right)+I_{1}\left(\frac{x}{9}\right)\right],
\ee
and this gives the result in (\ref{E.11}). So, similarly to (\ref{E.8}), this gives
\be
\la{E.16}
\log \left. F^{\mathbf A\,(2)}(\kappa; 3)\right|_{\zeta(5)} = \frac{1}{3\pi}\int_{0}^{1}\frac{dt\,t^{-5/6}}{\sqrt{1-t^{1/3}}}\ \left\{
\log\left[I_{0}\left(\frac{40\,\zeta(5)}{9}\,\kappa^{3}\,t\right)
+I_{1}\left(\frac{40\,\zeta(5)}{9}\,\kappa^{3}\,t\right)\right]+\frac{320\,\zeta(5)}{9}\,\kappa^{3}\,t\right\}.
\ee
The first terms of the expansion at small $\kappa$ are 
\begin{align}
& \log \left. F^{\mathbf A\,(2)}(\kappa; 3)\right|_{\zeta(5)} = 
\frac{425\, \zeta (5)}{36}\,\kappa ^3 + \frac{1925\,  
\zeta (5)^2}{3456}\,\kappa ^6-\frac{1519375\, \zeta 
(5)^3}{4478976}\, \kappa ^9\spek
+0\cdot\kappa^{12}+\frac{673253125\,  \zeta 
(5)^5}{2579890176}\,\kappa ^{15}-\frac{11816582421875 \,
\zeta (5)^6}{53496602689536}\,\kappa ^{18} +\mc O(\kappa ^{21}).
\end{align}
Comparing (\ref{E.8}) and (\ref{E.16}) one notices the nice relation 
\be
\log \left. F^{\mathbf A\,(2)}(2^{1/3}\kappa; 3)\right|_{\zeta(5)}-
\log \left. F^{\mathbf B\,(2)}(\kappa; 3)\right|_{\zeta(5)} = 25\,\zeta(5)\,\kappa^{3}.
\ee

\paragraph{$F^{(3)}$ scaling function} The analysis of the $F^{(3)}$ scaling function is very similar. Our results
are
\begin{align}
\log \left. F^{\mathbf A\,(3)}(\kappa; 3)\right|_{\zeta(5)} &= 
 \frac{1}{3\pi}\int_{0}^{1}\frac{dt\,t^{-5/6}}{\sqrt{1-t^{1/3}}}\ \left\{
\log\left[\left(\frac{20\zeta(5)}{9}\,\kappa^{3}\,t\right)^{-1}\,
I_{1}\left(\frac{40\zeta(5)}{9}\,\kappa^{3}\,t\right)\right]+\frac{320\zeta(5)}{9}\,\kappa^{3}\,t\right\}\notag \\ 
&=\frac{100\, \zeta (5)}{9}\,\kappa^{3}+\frac{1925 \,\zeta(5)^2}{3456}\,\kappa^{6}+0\cdot\kappa^{9}
-\frac{422524375 \, \zeta(5)^4}{2579890176}\,\kappa^{12}+0\cdot \kappa^{15}\spek
+\frac{11816582421875 \, \zeta(5)^6}{106993205379072}\,\kappa^{18}-\frac{69981778826171875 \,\zeta(5)^8}{739537035580145664}\,\kappa^{24}+\mc O(\kappa^{27}),
\end{align}
and, for the $\mathbf{B}$ model, 
\begin{align}
\log \left. F^{\mathbf B\,(3)}(\kappa; 3)\right|_{\zeta(5)} &=
 \frac{1}{3\pi}\int_{0}^{1}\frac{dt\,t^{-5/6}}{\sqrt{1-t^{1/3}}}\ \left\{
\log\left[\left(\frac{40\zeta(5)}{9}\,\kappa^{3}\,t\right)^{-1}\,
I_{1}\left(\frac{80\zeta(5)}{9}\,\kappa^{3}\,t\right)\right]-\frac{80\zeta(5)}{9}\,\kappa^{3}\,t\right\}\notag \\ 
&=-\frac{25 \,\zeta (5)}{9}\,\kappa^{3}+\frac{1925 \, \zeta(5)^2}{864}\,\kappa^{6}+0\cdot\kappa^{9}
-\frac{422524375\, \zeta(5)^4}{161243136}\,\kappa^{12}+0\cdot \kappa^{15}\spek
+\frac{11816582421875 \,\zeta(5)^6}{1671768834048}\,\kappa^{18}
-\frac{69981778826171875 \, \zeta(5)^8}{2888816545234944}\,\kappa^{24}+\mc O(\kappa^{27}),
\end{align}
and, remarkably, we have again
\be
\log \left. F^{\mathbf A\,(3)}(2^{1/3}\kappa; 3)\right|_{\zeta(5)}-
\log \left. F^{\mathbf B\,(3)}(\kappa; 3)\right|_{\zeta(5)} = 25\,\zeta(5)\,\kappa^{3}.
\ee

\section{Asymptotics of one-point Wilson functions}
\la{simple_observable_appendix}

In this technical Appendix, we collect various results that are needed in the proof of  (\ref{6.13}) and (\ref{6.14}).

    We start by proving a bound that we will use to to ensure the validity of approximating one-point functions in the presence of Wilson loop with a sequence of two-point functions. Let $\Omega_{n}$ be a sequence of operators with $R$-charge growing linearly with $n$ and moreover $\Omega_{n}$ is orthogonal to all operators with lesser $R$-charge than it. Let $\mathcal{O}$ be an operator have a series expansion:
    \begin{align}
        \mathcal{O}(a) = \sum_{r}T_{r}(a) .    
    \end{align}  
    Where $T_{n}$ has the same $R$-charge as $\Omega_{n}$. Let us assume that there exist $n_{0}$, such that for $n \ge m \ge n_{0}$
    \begin{align}
        \label{F.2}
        \abs{\frac{\ev{T_{n+1}\,\Omega_{m}}}{\ev{T_{n}\,\Omega_{m}}}} \le \frac{c}{n}, 
    \end{align}
    with $c$ a constant, then for $n > n_{0}$
    \begin{align}
        \abs{\ev{T_{n+k}\,\Omega_{n}}} &\le \abs{\ev{T_{n}\,\Omega_{n}}}\frac{\abs{c}^{k}n!}{(n+k)!}
        \Rightarrow \sum_{k=1}^{\infty}\abs{\ev{T_{n+k}\,\Omega_{n}}} \le \abs{\ev{T_{n}\,\Omega_{n}}}\sum_{k=1}^{\infty} \frac{\abs{c}^{k}n!}{(n+k)!}.
    \end{align}
    In the  $n \to \infty$ limit we have $\frac{n!}{(n+k)!} \to n^{-k}$ so that
    \begin{align}
        \lim_{n \to \infty} \sum_{k=1}^{\infty}\abs{\ev{T_{n+k}\,\Omega_{n}}} &\le \lim_{ n \to \infty} \abs{\ev{T_{n}\,\Omega_{n}}}\frac{\abs{c}}{n}\sum_{k=0}^{\infty} \frac{\abs{c}^{k}}{n^{k}}.
    \end{align}
    The geometric series in the expression above converges to $1$, so that
    \begin{align}
        \lim_{n \to \infty} \sum_{k=1}^{\infty}\abs{\ev{T_{n+k}\,\Omega_{n}}} &\le \lim_{ n \to \infty} \abs{\ev{T_{n}\,\Omega_{n}}}\frac{\abs{c}}{n}.
    \end{align}
    This equation tells us that if \eqref{F.2} is satisfied, then  in the large $n$ limit,  the sum of contributions of $T_{r}$ with $r > n$ in $\ev{\mathcal{O}\, \Omega_{n}}$ is suppressed by $\frac{1}{n}$ compared to that of $T_{n}$. In addition $T_{r}$ with $r < n$ are orthogonal to $\Omega_{n}$ so  we can treat the large $n$ limit of one-point function $\ev{\Omega_{n}\,\mathcal{O}}$ as the limit of a sequence of two-point functions, i.e.
    \begin{align}
      \lim_{n \to \infty}  \ev{\Omega_{n}\,\mathcal{O}} = \lim_{n\to \infty} \ev{\Omega_{n}\, T_{n}}.
    \end{align}
  \subsection{Wilson one-point functions for $SU(2)$}
  \label{subsec:SU2_bound}
    Now we verify that this bound is satisfied in the case of Wilson loop for $SU(2)$. In this case $T_{n}(a) = \frac{(2a)^{2n}}{(2n)!}$ 
    and $\Omega_{n} = \phi_{n} = :(\tr a^{2})^{n}: = (2a^{2})^{n}+\cdots$. So,
    \begin{align}
        \abs{\frac{\ev{T_{n+k+1}\,\Omega_{n}}}{\ev{T_{n+k}\,\Omega_{n}}}} &= 
        \frac{1}{(2n+2k+1)(2n+2k+2)}\frac{\ev{(2a^{2})^{n+k+1}\,\phi_{n}}}{\ev{(2a^{2})^{n+k}\,\phi_{n}}},
    \end{align}
    where we recall that $\pm a$ are the two eigenvalues of the $SU(2)$ matrix.
    Although it is not possible to evaluate $\ev{(2a)^{2n+2k}\,\Omega_{n}}$ exactly, 
    with some effort we can extract the qualitative large $n$ behavior, which is all we need to verify \eqref{F.2}. The leading term in $\phi_{n}$ is just $a^{2n}$. 
    As a result we are looking for the large $n$ behavior of $\ev{a^{4n+2k}}$. This is still a hard quantity to compute in a $\mathcal{N}=2$ theory. 
    But this problem can be neatly sidestepped in the double scaling limit where we are taking the 
    large $n$-limit while simultaneously dialing down the coupling $g^{2}\sim \frac{1}{\Im \tau}$ . 
    In this regime the leading $n$ behavior is entirely determined by the $\mathcal{N}=4$ theory. As a result
      \begin{align}
          \frac{\ev{(2a^{2})^{n+k+1}\,\Omega_{n}}}{\ev{(2a^{2})^{n+k}\,\Omega_{n}}} \sim \frac{\ev{(2a^{2})^{2n+k+1}}_{\mathcal{N}=4}}{\ev{(2a^{2})^{2n+k}}_{\mathcal{N}=4}} \sim \frac{(4n+2k)}{\Im \tau}.
      \end{align}
    Hence,
    \begin{align}
      \abs{\frac{\ev{T_{n+k+1}\,\Omega_{n}}}{\ev{T_{n+k}\,\Omega_{n}}}} \sim \frac{1}{(n+k)\Im \tau}.
    \end{align}
    Which is exactly the bound on growth we need to satisfy \eqref{F.2}.  
    
  \subsection{$SU(N)$ generalization}
  \label{subsec:SUN_bound}
    Now we consider the case of $SU(N)$. Since the odd terms in the expansion of Wilson loop don't contribute to the two-point function with $\phi_{n}$ we can safely ignore those and consider $\mathcal{O}$ be the sum of even terms in Wilson loop. So, $T_{n} = \frac{\tr a^{2n}}{(2n)!}$. Hence,
    \begin{align}
      \abs{\frac{\ev{T_{n+k+1}\,\Omega_{n}}}{\ev{T_{n+k}\,\Omega_{n}}}} &= \frac{1}{(2n+2k+1)(2n+2k+2)}\frac{\ev{\tr(a^{2n+2k+2})\,\phi_{n}}}{\ev{\tr (a^{2n+2k})\,\phi_{n}}}.
    \end{align}
    The r.h.s. can be dealt using the same saddle point approximation that we employed in Sec.~(\ref{subsec:saddle_point}). After keeping only the leading term in $n$ the result is
    \begin{align}
        \ev{T_{n+k}\,\Omega_{n}} \sim c_{n+k}\int \dd r\, r^{N^{2}-2 + 4n + 2k}\exp(-4\pi\Im\tau r^{2})Z_{\mbox{\scriptsize 1-loop}}(r\tr a_{0}).
    \end{align}
    Once again the result is an integral whose behavior the weak coupling limit we can estimate by ignoring the $Z_{\mbox{\scriptsize 1-loop}}$, hence
    \begin{align}
      \abs{\frac{\ev{\tr (a^{2n + 2k + 2}) \,\Omega_{n}}}{\ev{\tr (a^{2n + 2k}) \,\Omega_{n}}}} &\sim \frac{c_{n+k+1}}{c_{n+k}}\frac{N^{2}-2 + 4n+2k}{\Im \tau}.
    \end{align}
    In the large $n$ limit, $c_{n+k}$ and $c_{n+k+1}$ contribute at the same order in $n$, while the $N^{2}$ in the expression above can be ignored. Hence we get the same result as in the $SU(2)$ case:
    \begin{align}
      \abs{\frac{\ev{T_{n+k+1}\,\Omega_{n}}}{\ev{T_{n+k}\,\Omega_{n}}}} \sim \frac{1}{(n+k)\Im \tau}.
    \end{align}
    Which justifies our approximation of one-point Wilson function by a sequence of two-point functions in the main text.
     
\section{Constrained extrema of $\tr a^{2n}$ for the $SU(N)$ dual matrix model}
\label{appendix_extrema}

    To obtain an effective matrix model \eqref{6.11}, we need to do a saddle point integral around the maxima of $\tr a^{2n}$ subject to the two constraints $\tr a = 0$ and $\tr a^{2} = 1$. Using Lagrange multiplier $\sigma$ and $\lambda$ respectively for the two constraints we find that the extrema of $\tr a^{2n}$ are given by
    \begin{align}
        \label{G.1}
        a_{\mu}^{2n-1} - 2\lambda\, a_{\mu} - \sigma = 0.      
    \end{align}  

    This equation tell us that $a_{\mu}$ are all roots of the same degree $2n-1$ polynomial. This polynomial has only three non-zero term in degree $2n-1 , 1$ and $0$.  As a result we can use Descartes' rule of signs to conclude that at most three of $a_{\mu}$ are distinct. Since $\tr a = 0$ constraint imposes that at least two of them have to be different, the possible number of distinct $a_{\mu}$ is $2$ or $3$. We will deal with both these cases separately, but before that we eliminate $\lambda$ and $\sigma$ from \eqref{G.1}. Summing over $\mu$ in \eqref{G.1} gives us $\sigma = \frac{\tr a^{2n-1}}{N}$. Multiplying by $a_{\mu}$ and then summing over $\mu$ results in $2\lambda = \tr a^{2n}$. Hence we need to solve
    \begin{align}
        \label{G.2}
       a_{\mu}^{2n-1} - \tr a^{2n}\, a_{\mu} - \frac{\tr a^{2n-1}}{N} = 0.
    \end{align} 
    Now we consider the case of two distinct $a_{\mu}$. Up to a permutation of coordinates we can write:
    \begin{align}
        \label{G.3}
        a_{\mu} = \alpha \,\,\,\mbox{for  } 1 \le \mu \le k, && a_{\mu} = \beta \,\,\,\mbox{for  } k+1 \le \mu \le N.    
    \end{align} 
    Imposing $\tr a = 0$ and $\tr a^{2} = 1$ results in 
    \begin{align}
         \alpha = \pm \sqrt{\frac{(N-k)}{k N}}, && \beta = \mp\sqrt{\frac{k}{N (N-k)}}.
    \end{align} 
    These $\alpha$ and $\beta$ solve the extremization equation \eqref{G.3} for any $k$. The resulting $\tr a^{2 n}$ is
    \begin{align}
        \label{G.5}
        \tr a^{2n} &= k \left(\sqrt{\frac{(N-k)}{k N}}\right)^{2n}+(N-k) \left(\sqrt{\frac{k}{N (N-k)}}\right)^{2n} \ .
    \end{align}
    Notice that the expression above is invariant under $k \to N-k$. Between the possible values of $k$ i.e. $1 \le k \le N-1$, $\tr a^{2n}$ takes the highest value at $k = 1$ or $k = N-1$. These correspond to the $\pm a_{0}$ used for the computation of one-point function in the presence of Wilson loop. Taking into account the $N$ extrema obtained from each of them by permutation of coordinates there are $2N$ such candidates for the global maxima. So at this stage our candidate for the maximum value of $\tr a^{2n}$ is
    \begin{align}
        \label{G.6}
        \tr (a_{0}^{2n}) =  [N(N-1)]^{-n}((N-1) + (1-N)^{2n}). 
     \end{align} 
   
    To prove that it is indeed the global maximum we need to consider the extrema with three distinct $\alpha_{\mu}$ 
    and show that for them $\tr a^{2n}$ does not exceed \eqref{G.6} . To this aim let us label the distinct values of 
    $a_{\mu}$ by $\alpha  > \beta >  \gamma$ and let $k$ and $l$ be the multiplicities of $\alpha$ and $\gamma$. 
    Some algebra shows that as a result of \eqref{G.2}
    we must have $\beta = 0$. So $\alpha = -\gamma$ and $k = l$. Imposing $\tr a^2 = 1$ gives us
    \begin{align}
        \alpha = \frac{1}{\sqrt{2k}} 
    \end{align}
    The resulting $a$ again satisfies \eqref{G.2} for any integer $1 \le k \le \frac{N}{2}$. The resulting $\tr a^{2n}$ is
    \begin{align}
        \tr a^{2 n} = (2k)^{1-n}    
    \end{align} 
    Which is indeed less than the maximum we found earlier in \eqref{G.6} for $N > 3$. For $N = 3$, it less than \eqref{G.6} for $n > 2$ and again we have the same result in large $n$ limit.

\bibliography{BT-Biblio}
\bibliographystyle{JHEP}
\end{document}